\documentclass[11pt,a4paper]{report}

\usepackage[dvips]{epsfig}
\usepackage{feynmf}

\begin{document}
%---------------------------Frontpage, temporary---------------------------------------------
\thispagestyle{empty}
\begin{flushleft}
UME\r{A} UNIVERSITY\\
Department of Physics
\\[60mm]
\end{flushleft}

\begin{center}

\LARGE{Detection of}
\\[1mm]

\Huge{Elastic Photon-Photon Scattering}
\\[2mm]

\LARGE{through}
\\[1mm]

\Huge{Four-Wave Coupling}
\\[25mm]
\bfseries\Large{Erik Lundstr\"{o}m}
\\[55mm]
\normalfont\normalsize Master of Science Diploma Thesis, 20p\\
Non-Linear Physics Group\\
Department of Physics\\
Ume\r{a} University, Sweden
\\[5mm]
Ume\r{a}, May 2005
\end{center}
\newpage
\thispagestyle{empty}
The frontpage is not the original one.

Appendix C, Geometric factors in hep-ph/0510076, was added in November 2005.
\newpage
%----------------------------------------------------------------------------------------
\begin{abstract}
According to the theory of quantum electrodynamics, photon-photon scattering can take place via exchange of virtual electron-positron pairs. Effectively, the interaction can be formulated in terms of non-linear corrections to Maxwell's equations, and hence may be treated by classical non-linear electrodynamics. Due to the strong electromagnetic fields needed to reach any noticeable effect, photon-photon scattering has not yet been observed experimentally, but recent improvements in laser technology have increased the possibility of direct detection. A verification of the phenomena would be of great scientific value as a confirmation of quantum electrodynamics.

In this thesis the possibility of direct detection of elastic photon-photon scattering through four-wave coupling is investigated, both for current and future systems. It is shown how three colliding laser pulses satisfying certain matching conditions, can generate scattered radiation in a fourth resonant direction. The interaction is modeled, and the number of scattered photons is estimated for optimized configurations of incoming pulses, both for the Vulcan laser at the Rutherford Appleton Laboratory as well as for the planned XFEL at DESY. An experiment using the Vulcan laser is predicted to produce 22 scattered photons per shot, an amount which is definitely detectable. The usefulness of the XFEL highly depends on future developments in laser technology, but a realistic estimation gives a result of $30\ \!000$ scattered photons per second.
\end{abstract}
\tableofcontents
\newpage
\section{Preface}
This master thesis originates from a suggestion set out by my supervisor Mattias Marklund in springtime 2004. Although some preparations started in August, the main work begun in late October the same year, and has been carried out within the Non-linear physics group at the Department of Physics, Ume\r{a} University.

I became interested in QED-effects thanks to my participation in a graduate course on QFT, which provided me with a basic understanding of photon-photon scattering. However, the calculations in this work are mainly performed within the framework of classical non-linear electrodynamics, and that is why I initially had to put some efforts in learning about non-linear interactions. Thereafter a time of experimentation with different configurations and models followed, which finally lead to the results presented here.

Chapter 1 is an introduction containing a general description of photon-photon scatter\-ing and the purpose and method of this work. There is also a survey of the unit systems and conventions used later on.

In Chapter 2 a short overview of QED is presented, and the lowest order Feynman diagram corresponding to photon-photon scattering is pictured.

The effective field theory outlined in Chapter 3 allows us to express the QED-effects giving rise to photon-photon interactions in terms of non-linear corrections to Maxwell's equations of classical electrodynamics.

Chapter 4 covers the theory of non-linear interactions needed for the calculations in this work, including coupling equations and matching conditions.

In Chapter 5 non-linear terms are calculated for two dimensional as well as three dimensional configurations.

In Chapter 6 the scattered intensity is determined for and compared between different models of the interaction.

Numerical values are inserted in Chapter 7. The number of scattered photons for optimal configurations is estimated and discussed, both for the existing Vulcan laser at the Rutherford Appleton Laboratory and for the planned XFEL at DESY.

\newpage
\section{Acknowledgments}
First of all I would like to thank my supervisor Mattias Marklund, not only for always being there to kindly answer all sorts of questions, but also for all his encouraging words and constant belief in me. A lot of thanks also to my examiner Gert Brodin, whose genuine engagement and interest in discussing any arising problems or possibilities have been of greatest help. I am very greatful for being given the opportunity to work with two such eminent scientists. Thanks also to Michael Bradley, for introducing me to quantum field theory and being inspiring in general, and to Daniel Eriksson for all help with the \LaTeX-typesetting.
\newpage
%------------------Introduction-----------------------------
\chapter{Introduction}
In classical theory, electromagnetic waves propagating in vacuum cannot, due to the linear\-ity of Maxwell's vacuum equations, interact with each other, and colliding light will not give rise to any scattering. However, the theory of quantum electrodynamics (QED) opens for the possibility of photon-photon scattering via virtual electron-positron pairs in the vacuum.

It is possible to formulate an effective field theory containing only the electromagnetic fields, valid for field strengths below the critical field $10^{18}\!~\mathrm{V}/\mathrm{m}$ and wavelengths larger than the Compton wavelength $10^{\!-12}\!~\mathrm{m}$, in which the interaction shows up as non-linear corrections to Maxwell's equations. Hence this pure QED-effect may instead be treated with classical non-linear electrodynamics.

In order to reach any noticeable effects, the fields need to be quite strong, and this is the main reason why photon-photon scattering has not yet been detected experimentally. Recent improvements in laser technology however make the situation look more promising.
\section{Purpose and methods}
\subsection{Purpose}
The aim of this thesis is to investigate, using parameters for current and future systems, the possibility of direct detection of elastic photon-photon scattering. Since it seems like today's technology is just on the edge to make it work, an important task is to carefully optimize the experiments.

Photon-photon scattering is a phenomena significant for QED, and direct detection would be of great scientific value as a confirmation of the theory, comparable to the detection of the Casimir effect \cite{Casimir}. Hence the purpose of this thesis is in first place of fundamental nature. Of course future fields of application may arise from the knowledge of how to experimentally produce and control this scattering.
\subsection{Earlier detection schemes}
To simply collide laser beams has not really been a realistic way to detect photon-photon scattering before the recent development of petawatt lasers, but during the last decades a number of other detection schemes, based on imposing boundary conditions, have been proposed.

The correction terms vanish in the limit of parallel propagating free waves, but if instead waves trapped between regions of conducting material are used, there will be a non-zero effect. One suggestion is detection by wave mixing inside a laser produced vacuum channel in an overdense plasma \cite{Shen}. Another uses the idea of non-linear excitations of new modes inside waveguides \cite{Eriksson}. However, there are some problems with this approach related to the properties of the conducting material, and it may be hard to separate the disturbances it gives rise to from the photon-photon scattering effects. In order to rule out any other scattering processes it is also important that the vacuum is almost perfect, which can be technically hard to achieve here.

The ultimate detection scheme would be when there is no disturbing matter at all close to the interaction region, and that is the case investigated in this thesis. 
\subsection{Method of this work}
The method of this work is based on the simple idea of colliding laser pulses in vacuum, and trying to get some scattered photons out of it. To increase our chances we will send in three carefully chosen laser pulses satisfying certain matching conditions, such that some scattering is, due to the non-linearities, generated in a fourth direction, i.e.$\!$ we will deal with four-wave coupling.

This scheme obviously does not suffer from any problems with properties of guiding material. It also mirrors the common picture of scattering as bouncing particles in a nice way: Instead of just exciting a new mode inside some guiding structure, we can really see that the light changes direction.

In order to get as nice result as possible, we will try to find an optimal configuration of incoming beams, and the resulting number of scattered photons will be calculated for two interesting lasers: The Vulcan laser and the XFEL at DESY.
\section{Units and conventions}
\subsection{Unit systems}
This work deals with three different unit systems. Instead of sticking to a single one, we will use units appropriate for the situation we are considering.

QED is with advantage formulated in natural units, where the vacuum speed of light and Planck's constant simply are put as $c=\hbar=\!1$. Hence this is the unit system chosen for Chapter 2.

Maxwell's equations are especially nice to work with in Gaussian units, where the electric and magnetic fields are of the same dimensions. Gaussian units are used from Chapter 3 up to and including  Eq.$\ $(6.12), after which we go over to SI-units to make the insertion of numerical values somewhat simpler. More information about how Gaussian and SI-units are related, and how they easily may be translated into each other, is for example to be found in Jackson \cite{Jackson}.
\newpage
\subsection{Conventions}
Some important conventions and notations used in this thesis are lined up below:
\begin{itemize}
	\item Four-vector components are labeled by Greek indices, running from $0$ to $3$.
	\item The metric signature is $\left(+---\right)$, i.e.$\!$ the non-vanishing components of the flat spacetime metric tensor $\eta_{\mu\nu}$ are given by $\eta_{00}\!=\!1$, $\ \eta_{11}\!=\eta_{22}\!=\eta_{33}\!=\!-1$.
	\item The completely antisymmetric symbol $\epsilon^{\mu\nu\sigma\rho}$ is defined to equal $+1$ for $\left(\mu,\ \!\nu,\ \!\sigma,\ \!\rho\right)$ an even permutation of $\left(0,\ \!\!1,\ \!2,\ \!3\right)$, $-1$ for an odd permutation, and $0$ otherwise.
	\item When writing scalar products of four-vectors the indices are sometimes suppressed, e.g. $px\equiv p^{\mu}x_{\mu}$. The spacetime four-vector $x^{\mu}$ is usually only written as $x$.
	\item The four-gradient $\frac{\partial}{\partial x^{\mu}}$ may be written like $\partial_{\mu}$ or simply just as $_{,\mu}$.
	\item The commutator of two operators A and B is denoted by $\left[A,B\right]\equiv AB-BA$, while the anticommutator is written as $\left[A,B\right]_{+}\!\equiv AB+BA$.
	\item Complex conjugation is denoted with a star ($^{*}$) and Hermitian conjugation with a dagger ($^{\dagger}$). Vectors with complex components are equipped with a tilde ($\tilde{~}$). 
\end{itemize}

%---------------Summary of QED------------------------------
\chapter{QED: A short overview}
 
This chapter is intended to give a short introduction to quantum field theory (QFT) in general, and quantum electrodynamics (QED) in specific. Although the main calculations in this thesis are carried out within the framework of classical non-linear electrodynamics, a brief introduction of the underlying theory will be outlined here. After all, it is a process predicted by QED which is considered, and for the reader not familiar with this theory a quick overview should be in its place. However, anyone who wishes to skip this part and directly proceed to Chapter 3, may do so without getting into problems.

It must be pointed out that the treatment given below is anything but complete. Innumer\-able things have been left out, both details as well as fundamentals. More profound texts are for instance given by Mandl\&Shaw \cite{Mandl} and Zee \cite{Zee}. Note that natural units ($c=\hbar=\!1$) are used in this chapter.
\section{Quantum field theory}
\subsection{Foundation and motivation}
At the turn of the 20th century some remarkable progresses were made in the science of physics. In 1900, Planck's explanation of the black-body radiation provided the basic ideas that would finally lead to the formulation of quantum theory. As an attempt to fit theory to experimental data, he proposed that the process of emission and absorption of radiation by atoms occurs discontinuously in quanta. Einstein went even a step longer when he in order to explain the photoelectric effect, concluded that the electromagnetic field itself consists of quanta, so called photons. About the same time, in 1905, Einstein also formulated his famous special theory of relativity (SR), forever changing our notions of space and time.

Both non-relativistic quantum mechanics (i.e. $\!\!$the quantum theory of classical mechan\-ics) and special relativity have turned out to be very accurate within their respective d\-omain of validity. But how to describe a particle moving at high speed, where both quantum theory and special relativity seem to be needed? The Schr\"{o}dinger equation deter\-mining the time evolution in non-relativistic quantum mechanics is not Lorentz covariant, and hence needs to be modified in order to be consistent with SR. Several candidates of covariant relativistic wave equations describing different kinds of particles exist, but suffer from difficulties in their single particle interpretation.

By quantizing the fields obeying those equations, and thereby going over from s\-ingle particles to a many particle theory, one can get around these problems in a smooth way. Anyway, when dealing with the interaction between radiation and atoms in quantum mechanics, the Fourier components of the electromagnetic field are often subjected to a quantization, giving rise to the photons. By symmetry reasons it then seems natural to quantize also all other particles. The many particle view allows for the creation and annihilation of different particles, something unachievable in the single particle interpretation.

We have here given a very rough overview of the utility of a quantum field theory, but the adequateness of it must of course utterly be determined by experiments. Let us now have a look at how the quantization of fields is performed.
\subsection{Formulations}
There exist two main approaches to quantum field theory: The canonical formulation and the path integral formulation.  These formalisms often appear to be complementary, and a problem hard to solve using one of them may be a lot simpler in the other, and vice versa. A characteristic difference between the  approaches is the presence of operators in one but not the other.

We will for both formalisms outline the quantization procedure of any classical field theory derivable from an action integral
\begin{eqnarray}
	S\!\left(\Sigma\right)=\int_{\Sigma}d^{4}x{\cal L}\left(\phi_{r},\phi_{r,\mu}\right),
\end{eqnarray}
by demanding the variation $\delta S\!\left(\Sigma\right)=0$ for any arbitrary spacetime region $\Sigma$ when variating the fields $\phi_{r}$, $r=\!1,\ldots,N$, needed to specify the system. That is, the equations of motion of the fields can be derived from the Lagrangian density ${\cal L}\left(\phi_{r},\phi_{r,\mu}\right)$ by using the Euler-Lagrange equations
\begin{eqnarray}
	\frac{\partial}{\partial x^{\mu}}\left(\frac{\partial{\cal L}}{\partial\phi_{r,\mu}}\right)-\frac{\partial{\cal L}}{\partial\phi_{r}}=0.
\end{eqnarray}

In the canonical formulation of quantum field theory the usual canonical quantization procedure is generalized to work on fields as well. In Heisenberg's approach to non-relativistic quantum mechanics, a system with configuration space variables $q_{j}\!\left(t\right)$ is quantized by identifying the canonical momenta $p_{j}\!\left(t\right)$, and then turning those conjugate coordinates and momenta into Heisenberg operators subjected to the commutation relations
\begin{eqnarray}
	\left.\begin{array}{l}\left[q_{j}\!\left(t\right),p_{k}\!\left(t\right)\right]=i\delta_{jk}\\ \left[q_{j}\!\left(t\right),q_{k}\!\left(t\right)\right]=\left[p_{j}\!\left(t\right),p_{k}\!\left(t\right)\right]=0\end{array}\right\}.
\end{eqnarray}

When dealing with fields, the dynamical quantities corresponding to the $q_{j}\!\left(t\right)$ are just the field amplitudes $\phi_{r}\!\left(\textbf{x},t\right)$ at each point in space. Hence the system has a continuously infinite number of degrees of freedom. By going over from the discrete to the continuous system, and by defining the fields conjugate as $\pi_{r}\!\left(\textbf{x},t\right)\equiv\!\partial{\cal L}/\partial\dot{\phi}_{r}$, the corresponding commutation relations which have to be imposed on the operator versions of the fields and their conjugate in order to quantize the theory, are seen to be
\begin{eqnarray}
\left.\begin{array}{l}	\left[\phi_{r}\!\left(\textbf{x},t\right),\pi_{s}\!\left(\textbf{x}',t\right)\right]=i\delta_{rs}\delta\!\left(\textbf{x}-\textbf{x}'\right)\\ \left[\phi_{r}\!\left(\textbf{x},t\right),\phi_{s}\!\left(\textbf{x}',t\right)\right]=\left[\pi_{r}\!\left(\textbf{x},t\right),\pi_{s}\!\left(\textbf{x}',t\right)\right]=0\end{array}\right\}.
\end{eqnarray}

The path integral formulation of quantum field theory is an extension of Feynman's approach to quantum mechanics: The amplitude $Z$ of propagation from a point $q_{I}$ to a point $q_{F}$ is equal to the sum of $e^{iS(q)}$ for all possible paths $q$ between the points in configuration space. Here, $S(q)$ stands for the classical action of the path $q$. By denoting the summation over all paths starting at $q_{I}$ and ending at $q_{F}$ by an integral over $Dq$, we may write the amplitude as
\begin{eqnarray}
	Z=\int Dq\left.e^{iS(q)}\right..
\end{eqnarray}
Of course the summation over all paths is not a trivial thing to get a grasp of, but we will not go deeper into any mathematical details in this short review.

The extension to fields is now quite obvious: The amplitude $Z$ for a system initially in state $I$ to finally end up in the state $F$ is given by
\begin{eqnarray}
	Z=\int D\phi\left.e^{iS(\phi)}\right.=\int D\phi\left.e^{i\int d^{4}x{\cal L}\left(\phi\right)}\right.,
\end{eqnarray}
where $D\phi$ tells us to sum over all possible field configurations smoothly connecting the initial and final states, and $S(\phi)=\int d^{4}x{\cal L}\left(\phi\right)$ is the action corresponding to the configuration $\phi$ of such type. Note that this summation is over a mindboggling set of possibilities, and the only integral that can be solved analytically is the Gaussian, corresponding to the free field case.

The path integral formulation is indeed the computationally more interesting nowadays, much thanks to its suitability to treat nonperturbative effects, but some features like the appearance of particle-like properties are easier to see in the canonical formalism. The Lorentz invariance is obviously built in from the start in the path integral formalism, while it may be harder to see in the canonical since the imposed commutation relations single out equal times.
\section{Quantum electrodynamics}
\subsection{Quantization of the Dirac field}
We now turn to quantum electrodynamics, the quantum field theory of the electromagnetic and the electron-positron fields
and their mutual interaction.

A relativistic wave equation describing spin $1/2$ fermions, including the electron-positron, with mass $m$ was found by Dirac in 1928. This so called Dirac equation can in a representation-free way be written
\begin{eqnarray}
	i\gamma^{\mu}\frac{\partial\psi(x)}{\partial x^{\mu}}-m\psi(x)=0,
\end{eqnarray}
where $\gamma^{\mu}$ are 4$\times$4 matrices satisfying the anticommutation relations
\begin{eqnarray}
	\left[\gamma^{\mu},\gamma^{\nu}\right]_{+}=2\eta^{\mu\nu}
\end{eqnarray}
as well as the Hermicity conditions
\begin{eqnarray}
	\gamma^{\mu\dagger}=\gamma^{0}\gamma^{\mu}\gamma^{0},
\end{eqnarray}
and where $\psi(x)$ accordingly is a four component spinor wavefunction $\psi_{\alpha}(x)$, $\alpha=\!1,\ldots,4$, but we choose to suppress all matrix and spinor indices.

The adjoint field $\bar{\psi}(x)\equiv\psi^{\dagger}(x)\gamma^{0}$ then satisfies the adjoint Dirac equation
\begin{eqnarray}
	i\frac{\partial\bar{\psi}(x)}{\partial x^{\mu}}\gamma^{\mu}-m\bar{\psi}(x)=0,
\end{eqnarray}
and Eqs.$\ $(2.7) and (2.10) may be derived from the Lagrangian density
\begin{eqnarray}
	{\cal L}=\bar{\psi}(x)\left(i\gamma^{\mu}\partial_{\mu}-m\right)\psi(x)
\end{eqnarray}
by taking $\psi_{\alpha}(x)$ and $\bar{\psi}_{\alpha}(x)$ as the independent fields in Euler-Lagrange equations (2.2).

Unfortunately, the quantization procedures outlined above really only works for bosons, and some adjustments have to be made for fermions in order to respect the antisymmetry of the wave function.

The expansion of a canonically quantized bosonic field in a complete set of s\-olutions leads to the identification of annihilation and creation operators satisfying certain commuta\-tion relations. For spin $1/2$ fermions we do a similar expansion in a complete set of solution\-s to the Dirac equation, but now instead impose anticommutation relations on the expansion coefficients.

A complete set of plane wave solutions to the Dirac equation (2.7) can be found from consideration of a cubic enclosure of volume $V$ with periodic boundary conditions. For each allowed momentum $\textbf{p}$, with corresponding energy $p^{0}\!\!=\!\!\sqrt{m^{2}+\textbf{p}^{2}}$, there are four independent solutions:
\begin{eqnarray}
u_{r}\!\left(\textbf{p}\right)\frac{e^{-ipx}}{\sqrt{V}},\left.\right.\left.\right.\left.\right.v_{r}\!\left(\textbf{p}\right)\frac{e^{ipx}}{\sqrt{V}},\left.\right.\left.\right.\left.\right.r=\!1,2
\end{eqnarray}
where the constant four component spinors $u_{r}\!\left(\textbf{p}\right)$ and $v_{r}\!\left(\textbf{p}\right)$ satisfy
\begin{eqnarray}
	\left(\gamma^{\mu}p_{\mu}-m\right)u_{r}\!\left(\textbf{p}\right)=0,\left.\right.\left.\right.\left.\right.\left(\gamma^{\mu}p_{\mu}+m\right)v_{r}\!\left(\textbf{p}\right)=0
\end{eqnarray}
and are chosen to be normalized according to
\begin{eqnarray}
u^{\dagger}_{r}\!\left(\textbf{p}\right)u_{r}\!\left(\textbf{p}\right)=v^{\dagger}_{r}\!\left(\textbf{p}\right)v_{r}\!\left(\textbf{p}\right)=\frac{p^{0}}{m}.
\end{eqnarray}
The quantized versions of the Dirac field and its adjoint field can be written as
\begin{eqnarray}
	\psi(x)=\sum_{p,r}\sqrt{\frac{m}{Vp^{0}}}\left[c_{r}\!\left(\textbf{p}\right)u_{r}\!\left(\textbf{p}\right)e^{-ipx}+d^{\dagger}_{r}\!\left(\textbf{p}\right)v_{r}\!\left(\textbf{p}\right)e^{ipx}\right],
\end{eqnarray}
\begin{eqnarray}
	\bar{\psi}(x)=\sum_{p,r}\sqrt{\frac{m}{Vp^{0}}}\left[d_{r}\!\left(\textbf{p}\right)\bar{v}_{r}\!\left(\textbf{p}\right)e^{-ipx}+c^{\dagger}_{r}\!\left(\textbf{p}\right)\bar{u}_{r}\!\left(\textbf{p}\right)e^{ipx}\right],
\end{eqnarray}
where the summations are over all allowed momenta $\textbf{p}$ and both spin states (corresponding to $r=\!1,2$), $\bar{v}_{r}\equiv{v}^{\dagger}_{r}\gamma^{0}$, $\bar{u}_{r}\equiv{u}^{\dagger}_{r}\gamma^{0}$, and the expansion coefficients are subjected to
\begin{eqnarray}
	\left[c_{r}\!\left(\textbf{p}\right),c^{\dagger}_{s}\!\left(\textbf{p}'\right)\right]_{+}=\left[d_{r}\!\left(\textbf{p}\right),d^{\dagger}_{s}\!\left(\textbf{p}'\right)\right]_{+}=\delta_{rs}\delta_{pp'}
\end{eqnarray}
while all the other combinations anticommute.

It follows from (2.17) that we can interpret $c_{r}$, $c^{\dagger}_{r}$ and $d_{r}$, $d^{\dagger}_{r}$ as absorption and creation operators of two different kinds of fermions, and physical properties of these particles may be deduced from the constants of motion derivable from the Lagrangian density (2.11). The conclusion is that the $c$- and $d$-operators corresponds to electrons and positrons respectively. For example, with the vacuum state $\left|0\right\rangle$ defined by $c_{r}\!\left(\textbf{p}\right)\left|0\right\rangle=d_{r}\!\left(\textbf{p}\right)\left|0\right\rangle=0$ for all $\textbf{p}$ and $r=\!1,2$, the state $d^{\dagger}_{r}\!\left(\textbf{p}\right)\left|0\right\rangle$ consists of one positron with momentum $\textbf{p}$ in spin state $r$, while the state $c^{\dagger}_{s}\!\left(\textbf{p}'\right)d^{\dagger}_{r}\!\left(\textbf{p}\right)\left|0\right\rangle$ also includes an electron with momentum $\textbf{p}'$ in spin state $s$ as well.

In the path integral formalism, the quantization of fermions can be performed by introducing so called Grassmann numbers. However, that is nothing which will be investigated here, and we choose to be satisfied with the canonical approach.
\subsection{Quantization of the electromagnetic field}
Expressed in terms of the four-vector potential $A^{\mu}\!\left(x\right)$, Maxwell's equations for the free electromagnetic field become (see Section 3.2 for definitions)
\begin{eqnarray}
	\partial^{\nu}\partial_{\nu}A^{\mu}\!\left(x\right)-\partial^{\mu}\left(\partial_{\nu}A^{\nu}\!\left(x\right)\right)=0.
\end{eqnarray}

The quantization can unfortunately not be performed using the usual Lagrangian density (3.23) due to problems with conjugate fields being identically zero, but another better suited one is given by
\begin{eqnarray}
	{\cal L}=-\frac{1}{2}\left(\partial_{\nu}A_{\mu}\!\left(x\right)\right)\left(\partial^{\nu}A^{\mu}\!\left(x\right)\right).
\end{eqnarray}
However, the field equations this Lagrangian density gives rise to are only equivalent to Eq.$\ $(2.18) in a Lorentz gauge, i.e.$\!$ we have to add the subsidiary condition
\begin{eqnarray}
	\partial_{\mu}A^{\mu}\!\left(x\right)=0.
\end{eqnarray}
On identifying the fields conjugate to $A^{\mu}\!\left(x\right)$, carrying out the canonical quantization procedure, and finally expanding the free electromagnetic field in a complete set of solutions to the field equations (2.18) in a Lorentz gauge (i.e.$\!$ to $\partial^{\nu}\partial_{\nu}A^{\mu}\!\left(x\right)\!=\!0$), one gets the quantized version
\begin{eqnarray}
	A^{\mu}\!\left(x\right)=\sum_{k,r}\sqrt{\frac{1}{2Vk^{0}}}\left[\epsilon^{\mu}_{r}\!\left(\textbf{k}\right)a_{r}\!\left(\textbf{k}\right)e^{-ikx}+\epsilon^{\mu}_{r}\!\left(\textbf{k}\right)a^{\dagger}_{r}\!\left(\textbf{k}\right)e^{ikx}\right],
\end{eqnarray}
with summation over all allowed wave vectors $\textbf{k}$ and four polarization states $r=0,\ldots,3$ with corresponding polarization vectors $\epsilon^{\mu}_{r}$, the energy $k^{0}\!=|\textbf{k}|$, and expansion coefficients as operators subjected to the commutation relations
\begin{eqnarray}
\left.\begin{array}{l}
\left[a_{r}\!\left(\textbf{k}\right),a^{\dagger}_{s}\!\left(\textbf{k}'\right)\right]=\zeta_{r}\delta_{rs}\delta_{kk'}\\ \left[a_{r}\!\left(\textbf{k}\right),a_{s}\!\left(\textbf{k}'\right)\right]=\left[a^{\dagger}_{r}\!\left(\textbf{k}\right),a^{\dagger}_{s}\!\left(\textbf{k}'\right)\right]=0\end{array}\right\},
\end{eqnarray}
where $\zeta_{0}\!=\!-1$ and $\zeta_{1}\!=\!\zeta_{2}\!=\!\zeta_{3}\!=\!1$. We are thus lead to the interpretation of $a_{r}\!\left(\textbf{k}\right)$ and $a^{\dagger}_{r}\!\left(\textbf{k}\right)$ as absorption and creation operators respectively for both transverse photons ($r\!=\!\!1,2$) and longitudinal photons ($r\!=\!3$). The minus sign present for scalar photons ($r\!=\!0$) brings about some problems, but can be handled by imposing the Lorentz gauge (2.20) (or really an equivalent condition). This will lead to the conclusion that only transverse photons may be real, and hence the number of independent polarizations is reduced to two, something familiar from classical electrodynamics.
\subsection{The interaction and Feynman diagrams}
After this compressed presentation of the quantized free electron-positron and electro\-magnetic fields, we turn to their mutual interaction, something which allows for annihilation, creation and scattering processes to take place.

Before doing so, we will rewrite the Lagrangian densities of the free fields a little bit. It turns out that for the Lagrangian densities (2.11) and (2.19), the energy of the vacuum state is not finite. However, since only differences in energy are observable, we can redefine the vacuum energy such that all other energies are measured with respect to it. The removing of the infinities can be done by normal ordering of the operators, that is moving all annihilation operators to the right of the creation operators, treating them as if they all commute (bosons) or anticommute (fermions). Hence we will redefine the Lagrangian densities as normal products (denoted by N).

The form of the interaction Lagrangian density ${\cal L}_{I}$ can be found following the procedure of incorporating an electromagnetic field into non-relativistic quantum mechanics. There are good reasons for this form to be the correct one, but ultimately that is of course up to experiments to determine. The total Lagrangian density in QED can be written as ${\cal L}={\cal L}_{0}+{\cal L}_{I}$, with free and interaction parts given by
\begin{eqnarray}
	  {\cal L}_{0}\!\left(x\right)=\mathrm{N}\left[\bar{\psi}(x)\left(i\gamma^{\mu}\partial_{\mu}-m\right)\psi(x)-\frac{1}{2}\left(\partial_{\nu}A_{\mu}\!\left(x\right)\right)\left(\partial^{\nu}A^{\mu}\!\left(x\right)\right)\right]
\end{eqnarray}
and
\begin{eqnarray}
	{\cal L}_{I}\!\left(x\right)=\mathrm{N}\left[e\bar{\psi}(x)\gamma^{\mu}A_{\mu}\!\left(x\right)\psi(x)\right],
\end{eqnarray}
where $\left(\!-e\right)$ is the charge of the electron. In the same way, the total Hamiltonian may be divided into a free and an interaction part, $H=H_{0}+H_{I}$.

Due to the non-linearity of the interaction, the coupled field equations become really hard to solve, and we have to rely on perturbation theory, which is justified thanks to the weak coupling between the free fields. To treat this situation, we have to go over from our Heisenberg picture to the interaction picture, where the operators continue to satisfy the same equations of motion and the same commutation relations as the free fields in the Heisenberg picture, but where the state vectors $\left|\Phi\!\left(t\right)\right\rangle$ carry a time dependence and evolve in time in accordance with
\begin{eqnarray}
	i\frac{d}{dt}\left|\Phi\!\left(t\right)\right\rangle=H_{I}\!\left(t\right)\left|\Phi\!\left(t\right)\right\rangle.
\end{eqnarray}
In the interaction picture the interaction Hamiltonian, $H_{I}\!\left(t\right)=\int d^{3}\textbf{x}{\cal H}_{I}\!\left(x\right)$, also becomes time dependent, and in QED it is given through the interaction Hamiltonian density
\begin{eqnarray}
	{\cal H}_{I}\!\left(x\right)=-{\cal L}_{I}\!\left(x\right)=\mathrm{N}\!\left[-e\bar{\psi}(x)\gamma^{\mu}A_{\mu}\!\left(x\right)\psi(x)\right].
\end{eqnarray}

For a scattering process we are interested in the states a long time before ($t=\!-\infty$) and a long time after ($t=\!\infty$) the interaction takes place. These states are related through the $S$-matrix defined by $\left|\Phi\!\left(\infty\right)\right\rangle=S\left|\Phi\!\left(-\infty\right)\right\rangle$. The amplitude for an initial state $\left|i\right\rangle$ to end up in a final state $\left|f\right\rangle$ is correspondingly $\left\langle f\right|S\left|i\right\rangle$. By solving Eq.$\ $(2.25) iteratively, the $S$-matrix is identified as the expansion
\begin{eqnarray}
	S=\sum^{\infty}_{n=0}\frac{\left(-i\right)^n}{n!}\int\ldots\int d^{4}x_{1}d^{4}x_{2}\ldots d^{4}x_{n}\mathrm{T}\!\left\{{\cal H}_{I}\left(x_{1}\right){\cal H}_{I}\left(x_{2}\right)\ldots{\cal H}_{I}\left(x_{n}\right)\right\}
\end{eqnarray}
where the integrations are over all space and time. The time-ordered product T orders the operators in chronological order with earlier times to the right.

The $S$-matrix expansion may be written as a sum of normal products (Wick's theorem), so when calculating an amplitude $\left\langle f\right|S\left|i\right\rangle$ we only need to pick out the parts that are able to first absorb the initial particles, and then create the final ones. When the initial and final states are specified by particles with given momenta and spin (polarization), it is found that the scattering amplitude can be pictured as a set of so called Feynman diagrams in momentum space. Each diagram consists of external lines corresponding to the initial and final particles of the process, with internal lines and vertices in between. The internal lines may be seen as virtual particles present only during the interaction, and the number of vertices corresponds to the perturbation order. Often it is accurate enough to perform the calculations to first non-vanishing order, and hence neglect all diagrams with a larger number of vertices. Photons are usually pictured by wavy lines, while electrons (positrons) are drawn as straight lines with arrows pointing along (against) the time direction, which we here choose to be from the left to the right.

The correspondence between scattering amplitudes and Feynman diagrams may be summarized in a set of rules, the Feynman rules for QED. Following those we can in principle calculate the amplitude for any scattering process to any given order. We just draw all topologically different Feynmann diagrams of current interest, and then via the Feynman rules translate those into mathematical expressions. Hence we need not deal with the equivalent, but less practical, method of identifying terms directly from the normal products following from Wick's theorem. It may be pointed out that the same Feynman rules are obtained by using the path integral formulation.

As an example, the lowest order Feynman diagrams for electron-positron pair creation is shown in Figure 2.1. The diagram represents two initial real photons, an intermediate virtual fermion, and a final electron-positron pair.
\begin{figure}
\centering
\unitlength = 1mm
\begin{fmffile}{simple100}
\begin{fmfgraph}(50,40)
\fmfleft{i1,i2}
\fmfright{o1,o2}
\fmf{photon}{i1,v1}
\fmf{photon}{i2,v2}
\fmf{fermion}{o1,v1,v2,o2}
\end{fmfgraph}
\end{fmffile}
\caption{\emph{The lowest order Feynman diagram for electron-positron pair creation}.}
\end{figure}
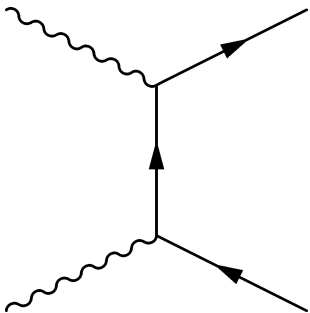

When going to higher orders one encounters problems with infinities from divergent integrals, threatening to ruin the complete theory. Luckily these infinities can be handled through the methods of regularization and renormalization, where the infinities are, loosely spoken, put into the relations between bare and physical particles instead. This procedure works to all orders of perturbation, and hence QED is called a renormalizable theory.

It is a general feature, following from the form of the interaction Hamiltonian density (2.26), that every vertex is built up by two fermion lines and one photon line. Thus we have to add more vertices in order to find a diagram which describes photon-photon scattering, that is a diagram with two photons both initially and finally. The lowest order photon-photon scattering diagram is pictured in Figure 2.2. It shows that the interaction takes place via virtual electron-positron pairs in the intermediate stage.
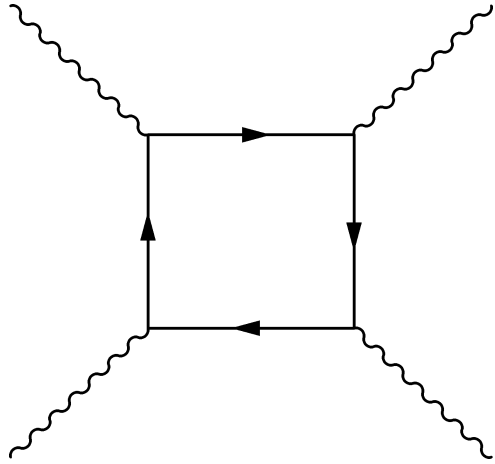
\begin{figure}
\centering
\unitlength = 1mm
\begin{fmffile}{simple_box100}
\begin{fmfgraph}(80,60)
\fmfleft{i1,i2}
\fmfright{o1,o2}
\fmf{fermion}{v1,v2,v4,v3,v1}
\fmf{photon,tension=1.5}{i1,v1}
\fmf{photon,tension=1.5}{i2,v2}
\fmf{photon,tension=1.5}{v3,o1}
\fmf{photon,tension=1.5}{v4,o2}
\end{fmfgraph}
\end{fmffile}
\caption{\emph{The lowest order Feynman diagram for photon-photon scattering}.}
\end{figure}
%---------------------Derivation of the non-linear wave equations----------------------------
\chapter{Effective field theory}
It may often be useful to relate the properties of the quantum vacuum to properties of a medium in classical theory. This can be done by integrating out all high-energy degrees of freedom, which are invisible in the classical low-energy domain, leaving us with an effective action from which classical equations of motion, effectively incorporating the quantum effects, can be derived. QED is a particularly nice candidate for this treatment, since it is easy to identify and separate the high energy degrees of freedom, corresponding to the massive electrons, from the low-energy massless photons.
\section{Effective Lagrangian density}
\subsection{Quantum vacuum}
The vacuum of quantum field theory differs quite remarkably from the classical view, and is sometimes pictured as a stormy sea of fluctuations, with virtual particles constantly popping up just to quickly disappear again. Since virtual particles cannot be measured, a fact making any discussion about whether they really exist or not meaningless, this is of course just a mental picture. The vacuum becomes interesting when we disturb it with an external field, and observe how it responds.

For external electromagnetic fields stronger than the critical field strength of $10^{18}~\mathrm{V}/\mathrm{m}$, QED accounts for a significant creation of real electron-positron pairs out of the vacuum. This phenomena, as well as the detailed structure of the quantum vacuum in general, is carefully treated by Greiner {\itshape et al.} $\!$\cite{Greiner}. Photon-photon scattering arises in not so strong fields, where the electron-positron pairs only appear as virtual particles in intermediate stages.
\subsection{Constant electromagnetic field}
The effective Lagrangian density for a vacuum perturbed with a constant external electro\-magnetic field was presented by Heisenberg and Euler in 1936 \cite{Heisenberg}, and is often referred to as the Euler-Heisenberg Lagrangian density. The situation is of course an idealization, because of the infinite extension of the field, but is a very useful approximation if one deals with electromagnetic fields that vary little over an electron Compton wavelength. This means that the Euler-Heisenberg Lagrangian density may without problems be used for plane electromagnetic waves with wavelengths larger than $10^{\!-12}\!~\mathrm{m}$. 

The formal derivation was given by Schwinger in 1951 \cite{Schwinger}, where he used the so called proper-time method, a particularly elegant regularization procedure which explicitly deals only with gauge-covariantly objects, and thereby conserves formal gauge invariance throughout the calculations. The vacuum energy may be represented in terms of an integral with respect to a scalar variable which has many similarities with the proper time, and thereof the name of the method. The full derivation is beyond the scope of this thesis, but a nice review is given by Dittrich\&Gies \cite{Dittrich}. Technically speaking we integrate out the electrons and positrons coupled to the external field, i.e.$\!$ we take care of all the internal fermion lines in Feynman diagrams like Figure 2.2. This finally leads to an expression of the effective Lagrangian density ${\cal L}$ for constant external electromagnetic fields, which for relatively weak fields can be expanded as
\begin{eqnarray}
	{\cal L}={\cal L}_{0}+\delta{\cal L}+\ldots=\frac{1}{8\pi}\left(E^{2}\!-B^{2}\right)+\frac{\xi}{8\pi}\left[\left(E^{2}\!-B^{2}\right)^{2}+7\left(\textbf{E}\cdot\textbf{B}\right)^{2}\right]+\ldots\ ,
\end{eqnarray}
where $\xi=\hbar e^{4}/45\pi m^{4}c^{7}$. Observe that the expression is given in Gaussian units, which are used from now on. ${\cal L}_{0}$ is, as will be shown in the next section, the classical Lagrangian density, while $\delta{\cal L}$ represents the non-linear QED-correction. Since $\xi$ is very small, the correction is negligible for weak fields. Also note the vanishing of $\delta{\cal L}$ for parallel propagating waves. We repeat that Eq.$\ $(3.1) is valid for field strengths below $10^{18}\!~\mathrm{V}/\mathrm{m}$ and wavelengths larger than $10^{\!-12}\!~\mathrm{m}$, conditions which are met for all electromagnetic waves considered in this work.

A new quasi-classical theory, describing the quantum effects in a classical language, may now be derived with the effective Lagrangian density ${\cal L}$ as starting-point. Both classical electrodynamics and QED are characterized by Lorentz and gauge invariance, but while the original classical theory is linear, quantum effects introduces non-linear self-interaction terms. These may be identified with a polarization and magnetization of the vacuum, as shown below.
\section{Effective field equations}
\subsection{Maxwell's equations}
In the case of no free charges or free currents, but including possible polarization $\textbf{P}$ and magnetization $\textbf{M}$, Maxwell's equations can be written
\begin{eqnarray}
	\nabla\cdot\textbf{E}=-4\pi\nabla\cdot\textbf{P},
\end{eqnarray}
\begin{eqnarray}
	\nabla\times\textbf{E}+\frac{1}{c}\frac{\partial\textbf{B}}{\partial t}=0,
\end{eqnarray}
\begin{eqnarray}
	\nabla\cdot\textbf{B}=0,
\end{eqnarray}
\begin{eqnarray}
	\nabla\times\textbf{B}-\frac{1}{c}\frac{\partial\textbf{E}}{\partial t}=4\pi\frac{1}{c}\frac{\partial\textbf{P}}{\partial t}+4\pi\nabla\times\textbf{M}.
\end{eqnarray}
From these we can directly derive the wave equations
\begin{eqnarray}
	\left(\frac{\partial^{2}}{\partial t^{2}}-c^{2}\nabla^{2}\right)\textbf{E}\!\!\!\!&=&\!\!\!\!\frac{\partial^{2}\textbf{E}}{\partial t^{2}}-c^{2}\nabla^{2}\textbf{E}+c^{2}\nabla\left(\nabla\cdot\textbf{E}\right)+4\pi c^{2}\nabla\left(\nabla\cdot\textbf{P}\right) \nonumber \\ \!\!\!\!&=&\!\!\!\! \frac{\partial^{2}\textbf{E}}{\partial t^{2}}+c^{2}\nabla\times\nabla\times\textbf{E}+4\pi c^{2}\nabla\left(\nabla\cdot\textbf{P}\right)\nonumber \\ \!\!\!\!&=&\!\!\!\!c\frac{\partial}{\partial t}\left(\frac{1}{c}\frac{\partial\textbf{E}}{\partial t}-\nabla\times\textbf{B}\right)+4\pi c^{2}\nabla\left(\nabla\cdot\textbf{P}\right)\nonumber \\ \!\!\!\!&=&\!\!\!\!4\pi c^{2}\!\left[\nabla \left(\nabla\cdot\textbf{P}\right)-\frac{1}{c}\frac{\partial}{\partial t}\left(\frac{1}{c}\frac{\partial\textbf{P}}{\partial t}+\nabla\times\textbf{M}\right)\right],
\end{eqnarray}
\begin{eqnarray}	
	\left(\frac{\partial^{2}}{\partial t^{2}}-c^{2}\nabla^{2}\right)\textbf{B}\!\!\!\!&=&\!\!\!\!\frac{\partial^{2}\textbf{B}}{\partial t^{2}}-c^{2}\nabla^{2}\textbf{B}+c^{2}\nabla\left(\nabla\cdot\textbf{B}\right)\nonumber \\
	\!\!\!\!&=&\!\!\!\! \frac{\partial^{2}\textbf{B}}{\partial t^{2}}+c^{2}\nabla\times\nabla\times\textbf{B}\nonumber \\ \!\!\!\!&=&\!\!\!\!c^{2}\nabla\times\left(-\frac{1}{c}\frac{\partial\textbf{E}}{\partial t}+\nabla\times\textbf{B}\right)\nonumber \\ \!\!\!\!&=&\!\!\!\! 4\pi c^{2}\nabla\times\left(\frac{1}{c}\frac{\partial\textbf{P}}{\partial t}+\nabla\times\textbf{M}\right).
\end{eqnarray}
In (classical) vacuum, where $\textbf{P}\!=\!\textbf{M}\!=\!\textbf{0}$, Maxwell's equations reduces to
\begin{eqnarray}
	\nabla\cdot\textbf{E}=0,
\end{eqnarray}
\begin{eqnarray}
	\nabla\times\textbf{E}+\frac{1}{c}\frac{\partial\textbf{B}}{\partial t}=0,
\end{eqnarray}
\begin{eqnarray}
	\nabla\cdot\textbf{B}=0,
\end{eqnarray}
\begin{eqnarray}
	\nabla\times\textbf{B}-\frac{1}{c}\frac{\partial\textbf{E}}{\partial t}=\textbf{0}.
\end{eqnarray}
These equations can be written in covariant form as
\begin{eqnarray}
	F^\mu{}^\nu{}_{,\nu}=0,
\end{eqnarray}
\begin{eqnarray}
	F_{\mu\nu,\tau}+F_{\nu\tau,\mu}+F_{\tau\mu,\nu}=0,
\end{eqnarray}
where the antisymmetric electromagnetic field tensors $F^\mu{}^\nu{}$ and $F_{\mu\nu}=\eta_{\mu\sigma}\eta_{\nu\rho}F^\sigma{}^\rho{}$ are given by
\begin{eqnarray}
	F^\mu{}^\nu{} = \left( \begin{array}{cccc}0 & -E_{x} & -E_{y} & -E_{z} \\ 
	E_{x} & 0 & -B_{z} & B_{y} \\ 
	E_{y} & B_{z} & 0 & -B_{x} \\ 
	E_{z} & -B_{y} & B_{x} & 0\end{array} \right),
\end{eqnarray}
\begin{eqnarray}
	F_{\mu\nu}= \left( \begin{array}{cccc}0 & E_{x} & E_{y} & E_{z} \\ 
	-E_{x} & 0 & -B_{z} & B_{y} \\ 
	-E_{y} & B_{z} & 0 & -B_{x} \\ 
	-E_{z} & -B_{y} & B_{x} & 0\end{array} \right).
\end{eqnarray}
It can easily be checked that Eq.$\ $(3.12) gives us Eqs.$\ $(3.8) and (3.11), while Eq.$\ $(3.13) corresponds to Eqs.$\ $(3.9) and (3.10).

The existence of scalar and vector potentials, $\Phi$ and $\textbf{A}$, defined through
\begin{eqnarray}
	\textbf{B}=\nabla\times\textbf{A}, 
\end{eqnarray}
\begin{eqnarray}
	\textbf{E}=-\frac{1}{c}\frac{\partial\textbf{A}}{\partial t}-\nabla\Phi,
\end{eqnarray}
is implied by Eqs.$\ $(3.9) and (3.10). By defining the four-vector potential as
\begin{eqnarray}
	A^{\mu}\equiv\left(\Phi,\textbf{A}\right),
\end{eqnarray}
it is straight forward to verify that the field tensor can be expressed as
\begin{eqnarray}
	F_{\mu\nu}=A_{\nu,\mu}\!-A_{\mu,\nu}.
\end{eqnarray}
In terms of the potentials, Eq.$\ $(3.13) is identically satisfied.

Further on we notice the identities
\begin{eqnarray}
	E^{2}\!-B^{2}=-\frac{1}{2}F_{\mu\nu}F^\mu{}^\nu{},
\end{eqnarray}
\begin{eqnarray}
	\textbf{E}\cdot\textbf{B}=-\frac{1}{4}F_{\mu\nu}\cal F^\mu{}^\nu{},
\end{eqnarray}
where the so called dual field tensor $\cal{F}^\mu{}^\nu{}$ is defined as
\begin{eqnarray}
	{\cal F}^\mu{}^\nu{}\equiv\frac{1}{2}\epsilon^{\mu\nu\sigma\rho}F_{\sigma\rho}=\left( \begin{array}{cccc}0 & -B_{x} & -B_{y} & -B_{z} \\ 
	B_{x} & 0 & E_{z} & -E_{y} \\ 
	B_{y} & -E_{z} & 0 & E_{x} \\ 
	B_{z} & E_{y} & -E_{x} & 0\end{array} \right).
\end{eqnarray}
This explicitly shows the Lorentz invariance of both $E^{2}\!-B^{2}$ and $\textbf{E}\cdot\textbf{B}$.

We will now show that the field equations (3.12) can be derived from the Lagrangian density
\begin{eqnarray}
	{\cal L}_{0} = \frac{1}{8\pi}\left(E^{2}\!-B^{2}\right)=-\frac{1}{16\pi}F_{\mu\nu}F^\mu{}^\nu{},
\end{eqnarray}
by treating the components of $A^{\mu}$ as the independent fields in the variational principle, that is solving Euler-Lagrange equations
\begin{eqnarray}
	\frac{\partial}{\partial x^{\mu}}\left(\frac{\partial{\cal L}_{0}}{\partial A_{\nu,\mu}}\right)-\frac{\partial{\cal L}_{0}}{\partial A_{\nu}}=0.
\end{eqnarray}
We have
\begin{eqnarray}
	\frac{\partial{\cal L}_{0}}{\partial A_{\nu,\mu}}\!\!\!\!&=&\!\!\!\!-\frac{1}{16\pi}\frac{\partial F_{\alpha\beta}}{\partial A_{\nu,\mu}}F^{\alpha\beta}-\frac{1}{16\pi}\frac{\partial F^{\alpha\beta}}{\partial A_{\nu,\mu}}F_{\alpha\beta}=-\frac{1}{8\pi}\frac{\partial F_{\alpha\beta}}{\partial A_{\nu,\mu}}F^{\alpha\beta}\nonumber \\
\!\!\!\!&=&\!\!\!\!-\frac{1}{8\pi}\frac{\partial \left(A_{\beta,\alpha}\!-A_{\alpha,\beta}\right)}{\partial A_{\nu,\mu}}F^{\alpha\beta}=-\frac{1}{8\pi}\left(\delta^{\nu}_{\beta}\delta^{\mu}_{\alpha}-\delta^{\nu}_{\alpha}\delta^{\mu}_{\beta}\right)F^{\alpha\beta}\nonumber \\ \!\!\!\!&=&\!\!\!\!-\frac{1}{8\pi}\left(F^{\mu\nu}\!-F^{\nu\mu}\right)=\frac{1}{8\pi}\left(F^{\nu\mu}\!+F^{\nu\mu}\right)=\frac{1}{4\pi}F^{\nu\mu}
\end{eqnarray}
giving
\begin{eqnarray}
	\frac{\partial}{\partial x^{\mu}}\left(\frac{\partial{\cal L}_{0}}{\partial A_{\nu,\mu}}\right)=\frac{1}{4\pi}F^\nu{}^\mu{}_{,\mu}\ ,
\end{eqnarray}
and since
\begin{eqnarray}
	\frac{\partial{\cal L}_{0}}{\partial A_{\nu}}=0
\end{eqnarray}
we finally end up with Eq.$\ $(3.12).
\subsection{Effective polarization and magnetization}
Now let us investigate what the equations will look like when we consider the total effective Lagrangian density from Eq.$\ $(3.1):
\begin{eqnarray}
	{\cal L}={\cal L}_{0}+\delta{\cal L}\!\!\!\!&=&\!\!\!\!\frac{1}{8\pi}\left(E^{2}\!-B^{2}\right)+\frac{\xi}{8\pi}\left[\left(E^{2}\!-B^{2}\right)^{2}+7\left(\textbf{E}\cdot\textbf{B}\right)^{2}\right] \nonumber \\ \!\!\!\!&=&\!\!\!\!-\frac{1}{16\pi}F_{\mu\nu}F^\mu{}^\nu{}+\frac{\xi}{8\pi}\left(\frac{1}{4}F_{\mu\nu}F^\mu{}^\nu{}F_{\sigma\rho}F^\sigma{}^\rho{}+\frac{7}{16}F_{\mu\nu}{\cal F}^\mu{}^\nu{}F_{\sigma\rho}{\cal F}^\sigma{}^\rho{}\right).\ \ \ \ 
\end{eqnarray}
We have
\begin{eqnarray}
	\frac{\partial\left(F_{\alpha\beta}F^\alpha{}^\beta{}F_{\sigma\rho}F^\sigma{}^\rho{}\right)}{\partial A_{\nu,\mu}}\!\!\!\!&=&\!\!\!\!2\frac{\partial\left(F_{\alpha\beta}F^\alpha{}^\beta{}\right)}{\partial A_{\nu,\mu}}F_{\sigma\rho}F^\sigma{}^\rho{}\nonumber \\ \!\!\!\!&=&\!\!\!\!-32\pi\frac{\partial{\cal L}_{0}}{\partial A_{\nu,\mu}}F_{\sigma\rho}F^\sigma{}^\rho{}=16\left(E^{2}\!-B^{2}\right)F^\nu{}^\mu{},
\end{eqnarray}
and
\begin{eqnarray}
	\frac{\partial\left(F_{\mu\nu}{\cal F}^\mu{}^\nu{}F_{\sigma\rho}{\cal F}^\sigma{}^\rho{}\right)}{\partial A_{\nu,\mu}}\!\!\!\!&=&\!\!\!\!2\frac{\partial\left(F_{\alpha\beta}{\cal F}^\alpha{}^\beta{}\right)}{\partial A_{\nu,\mu}}F_{\sigma\rho}{\cal F}^\sigma{}^\rho{}=\epsilon^{\alpha\beta\gamma\delta}\frac{\partial\left(F_{\alpha\beta}F_{\gamma\delta}\right)}{\partial A_{\nu,\mu}}F_{\sigma\rho}{\cal F}^\sigma{}^\rho{}\nonumber  \\ \!\!\!\!&=&\!\!\!\!\epsilon^{\alpha\beta\gamma\delta}\left[\frac{\partial\left(A_{\beta,\alpha}\!-A_{\alpha,\beta}\right)}{\partial A_{\nu,\mu}}F_{\gamma\delta}+\frac{\partial\left(A_{\delta,\gamma}\!-A_{\gamma,\delta}\right)}{\partial A_{\nu,\mu}}F_{\alpha\beta}\right]F_{\sigma\rho}{\cal F}^\sigma{}^\rho{}\nonumber  \\ \!\!\!\!&=&\!\!\!\!\epsilon^{\alpha\beta\gamma\delta}\left[\left(\delta^{\nu}_{\beta}\delta^{\mu}_{\alpha}-\delta^{\nu}_{\alpha}\delta^{\mu}_{\beta}\right)F_{\gamma\delta}+\left(\delta^{\nu}_{\delta}\delta^{\mu}_{\gamma}-\delta^{\nu}_{\gamma}\delta^{\mu}_{\delta}\right)F_{\alpha\beta}\right]F_{\sigma\rho}{\cal F}^\sigma{}^\rho{}\nonumber \\ \!\!\!\!&=&\!\!\!\!\left[\left(\epsilon^{\mu\nu\gamma\delta}\!-\epsilon^{\nu\mu\gamma\delta}\right)F_{\gamma\delta}+\left(\epsilon^{\alpha\beta\mu\nu}\!-\epsilon^{\alpha\beta\nu\mu}\right)F_{\alpha\beta}\right]F_{\sigma\rho}{\cal F}^\sigma{}^\rho{}\nonumber \\ \!\!\!\!&=&\!\!\!\!4\epsilon^{\mu\nu\alpha\beta}F_{\alpha\beta}F_{\sigma\rho}{\cal F}^\sigma{}^\rho{}=32\left(\textbf{E}\cdot\textbf{B}\right){\cal F}^\nu{}^\mu{},
\end{eqnarray}
so Eqs.$\ $(3.29) and (3.30) together with (3.28) yields
\begin{eqnarray}
	\frac{\partial}{\partial x^{\mu}}\left(\frac{\partial\left(\delta{\cal L}\right)}{\partial A_{\nu,\mu}}\right)=\frac{\xi}{4\pi}\left[2\left(E^{2}\!-B^{2}\right)F^\nu{}^\mu{}+7\left(\textbf{E}\cdot\textbf{B}\right){\cal F}^\nu{}^\mu{}\right]_{,\mu}.
\end{eqnarray}
We also have
\begin{eqnarray}
	\frac{\partial\left(\delta{\cal L}\right)}{\partial A_{\nu}}=0,
\end{eqnarray}
so by putting Eqs.$\ $(3.26), (3.27), (3.31) and (3.32) into Euler-Lagrange equations (3.24), but now with the total effective Lagrangian density ${\cal L}=\!{\cal L}_{0}+\delta{\cal L}$ instead of just ${\cal L}_{0}$, we arrive at the effective field equations
\begin{eqnarray}
	F^\nu{}^\mu{}_{,\mu}=\xi\left[-2\left(E^{2}\!-B^{2}\right)F^\nu{}^\mu{}-7\left(\textbf{E}\cdot\textbf{B}\right){\cal F}^\nu{}^\mu{}\right]_{,\mu}.
\end{eqnarray}
It is straight forward to check that $\nu=0$ corresponds to
\begin{eqnarray}
\nabla\cdot\textbf{E}=-\xi\nabla\cdot\left[2\left(E^{2}\!-B^{2}\right)\textbf{E}+7\left(\textbf{E}\cdot\textbf{B}\right)\textbf{B}\right],
\end{eqnarray}
and $\nu=1,2,3$ is equivalent to
\begin{eqnarray}
	\nabla\times\textbf{B}-\frac{1}{c}\frac{\partial\textbf{E}}{\partial t}\!\!\!\!&=&\!\!\!\!\xi\frac{1}{c}\frac{\partial}{\partial t}\left[2\left(E^{2}\!-B^{2}\right)\textbf{E}+7\left(\textbf{E}\cdot\textbf{B}\right)\textbf{B}\right]\nonumber \\ &&\!\!\!\!+\ \xi\nabla\times\left[-2\left(E^{2}\!-B^{2}\right)\textbf{B}+7\left(\textbf{E}\cdot\textbf{B}\right)\textbf{E}\right].
\end{eqnarray}
Comparing these results with Eqs.$\ $(3.2) and (3.5) reveals that we can express the correc\-tions to Maxwell's equations, arising from photon-photon scattering, as an effective polarization and magnetization given by
\begin{eqnarray}
	\textbf{P}=\frac{\xi}{4\pi}\left[2\left(E^{2}\!-B^{2}\right)\textbf{E}+7\left(\textbf{E}\cdot\textbf{B}\right)\textbf{B}\right],
\end{eqnarray}
\begin{eqnarray}
\textbf{M}=\frac{\xi}{4\pi}\left[-2\left(E^{2}\!-B^{2}\right)\textbf{B}+7\left(\textbf{E}\cdot\textbf{B}\right)\textbf{E}\right].
\end{eqnarray}
\subsection{Effective wave equations}
Thus the wave equations (3.6) and (3.7) can be written
\begin{eqnarray}
	\left(\frac{\partial^{2}}{\partial t^{2}}-c^{2}\nabla^{2}\right)\textbf{E}=-c^{2}\left[\nabla (N1)+\frac{1}{c}\frac{\partial}{\partial t}\overline{(N234)}\right], 
\end{eqnarray}
\begin{eqnarray}
	\left(\frac{\partial^{2}}{\partial t^{2}}-c^{2}\nabla^{2}\right)\textbf{B}=c^{2}\nabla\times\overline{(N234)},
\end{eqnarray}
where $(N1)$ and $\overline{(N234)}$ are defined as
\begin{eqnarray}
	(N1)\equiv-\xi\nabla\cdot\left[2\left(E^{2}\!-B^{2}\right)\textbf{E}+7\left(\textbf{E}\cdot\textbf{B}\right)\textbf{B}\right],
\end{eqnarray}
\begin{eqnarray}
	\overline{(N234)}\!\!\!\!&\equiv&\!\!\!\!\xi\frac{1}{c}\frac{\partial}{\partial t}\left[2\left(E^{2}\!-B^{2}\right)\textbf{E}+7\left(\textbf{E}\cdot\textbf{B}\right)\textbf{B}\right]\nonumber \\&&\!\!\!\!+\ \xi\nabla\times\left[-2\left(E^{2}\!-B^{2}\right)\textbf{B}+7\left(\textbf{E}\cdot\textbf{B}\right)\textbf{E}\right].
\end{eqnarray}

%--------------------------------

\chapter{Non-linear interactions}
The interest in non-linear effects has been steadily increasing during the last four decades, and it has been especially important for the development of fields such as plasma physics and modern optics. Here the theory will only be worked out to the extent that it covers what is needed for coming chapters.

Non-linear interactions may be classified into weak and strong non-linearities. If the variation of the wave amplitude due to non-linearity is slow compared to the harmonic oscillation we have a weak non-linearity, while it is said to be strong otherwise.

There are two main methods for describing weak non-linear wave-wave interactions, namely the coherent-phase description and the random-phase description. In the coherent-phase description the coherence time is assumed to be much longer than the time of interaction, while the opposite assumption is made in the random-phase description. Therefore phase effects play an important role in the first but not in the later case.

In this work weak interactions between laser beams are considered, and thus we are interested in weak coherent non-linear wave-wave interactions. A more complete treatment of this area can be found in Weiland\&Wilhelmsson \cite{Weiland}. However, we will just restrict ourselves to the case of four plane waves, of which one is much weaker than the other three, interacting in a small region of spacetime. By carefully choosing wave vectors to satisfy certain matching conditions, the weak wave can be driven by the others, that is its amplitude increases due to the non-linear interaction. Let us see how this comes about.
\section{Four-wave coupling}
\subsection{Coupling equations}
As starting point we consider four plane waves, with amplitudes allowed to have a weak spacetime dependence due to interactions,
\begin{eqnarray}
\textbf{E}_{j}\!\left(\textbf{r},t\right)\!\!\!\!&=&\!\!\!\!\frac{1}{2}\left(\tilde{\textbf{E}}_{j}\!\left(\textbf{r},t\right)e^{i\left(\textbf{k}_{j}\cdot\textbf{r}-\omega_{j}t\right)}+\tilde{\textbf{E}}^{*}_{j}\!\left(\textbf{r},t\right)e^{-i\left(\textbf{k}_{j}\cdot\textbf{r}-\omega_{j}t\right)}\right)\ \ ,\ j=1,\ldots,4
\end{eqnarray}
together with a non-linear wave equation of the form
\begin{eqnarray}
	\left(\frac{\partial^{2}}{\partial t^{2}}-c^{2}\nabla^{2}\right)\textbf{E}\!\left(\textbf{r},t\right)\!\!\!\!&=&\!\!\!\!\left(\frac{\partial^{2}}{\partial t^{2}}-c^{2}\nabla^{2}\right)\left(\sum^{4}_{j=1}\textbf{E}_{j}\!\left(\textbf{r},t\right)\right)\nonumber\\ =\sum^{8}_{l=1}\ \!\sum^{8}_{m=l}\sum^{8}_{n=m}\!\!\!\!\!\!\!\!&&\!\!\!\!\!\!\!\!A_{l,m,n}\tilde{E}_{l}\tilde{E}_{m}\tilde{E}_{n}e^{i\left[\left(\textbf{k}_{l}+\textbf{k}_{m}+\textbf{k}_{n}\right)\cdot\textbf{r}-\left(\omega_{l}+\omega_{m}+\omega_{n}\right)t\right]}\hat{\textbf{u}}_{l,m,n},
\end{eqnarray}
where for $i=5,\ldots,8$ we have defined $\tilde{E}_{i}=\!\tilde{E}^{*}_{i-4}$, $\ \textbf{k}_{i}=\!-\textbf{k}_{i-4}$, $\ \omega_{i}=\!-\omega_{i-4}$, $\ $and $A_{l,m,n}$ are constants small enough to ensure weak non-linearity, while $\hat{\textbf{u}}_{l,m,n}$ are unit vectors. As will be apparent, this is in fact the form of our wave equation of interest. We also assume that the wave vectors and frequencies satisfy the matching conditions
\begin{eqnarray}
	\textbf{k}_{1}+\textbf{k}_{2}=\textbf{k}_{3}+\textbf{k}_{4},
\end{eqnarray}
\begin{eqnarray}
	\omega_{1}+\omega_{2}=\omega_{3}+\omega_{4}.
\end{eqnarray}
Due to the weak non-linearity, space and time derivatives of the amplitudes are much smaller than of the harmonic parts. Taking the dispersion relation $\omega_{j}=ck_{j}$ into account, we thus have to lowest non-vanishing order
\begin{eqnarray}
	\left(\frac{\partial^{2}}{\partial t^{2}}-c^{2}\nabla^{2}\right)\left[\tilde{\textbf{E}}_{j}\!\left(\textbf{r},t\right)e^{i\left(\textbf{k}_{j}\cdot\textbf{r}-\omega_{j}t\right)}\right]\!\!\!\!&\approx&\!\!\!\!\left(\!-\omega^{2}_{j}+c^{2}k^{2}_{j}\right)\tilde{\textbf{E}}_{j}\!\left(\textbf{r},t\right)e^{i\left(\textbf{k}_{j}\cdot\textbf{r}-\omega_{j}t\right)}\nonumber  \\ &&\!\!\!\!-2i\omega_{j}\left[\left(\frac{\partial}{\partial t}+c\hat{\textbf{k}}_{j}\cdot\nabla\right)\tilde{\textbf{E}}_{j}\!\left(\textbf{r},t\right)\right]e^{i\left(\textbf{k}_{j}\cdot\textbf{r}-\omega_{j}t\right)}\nonumber  \\\!\!\!\!&=&\!\!\!\!-2i\omega_{j}\frac{d\tilde{\textbf{E}}_{j}\!\left(\textbf{r},t\right)}{dt}e^{i\left(\textbf{k}_{j}\cdot\textbf{r}-\omega_{j}t\right)},
\end{eqnarray}
and clearly
\begin{eqnarray}
	\left(\frac{\partial^{2}}{\partial t^{2}}-c^{2}\nabla^{2}\right)\left[\tilde{\textbf{E}}^{*}_{j}\!\left(\textbf{r},t\right)e^{-i\left(\textbf{k}_{j}\cdot\textbf{r}-\omega_{j}t\right)}\right]\approx2i\omega_{j}\frac{d\tilde{\textbf{E}}^{*}_{j}\!\left(\textbf{r},t\right)}{dt}e^{-i\left(\textbf{k}_{j}\cdot\textbf{r}-\omega_{j}t\right)},
\end{eqnarray}
where $\frac{d}{dt}=\!\left(\frac{\partial}{\partial t}+\frac{\partial\textbf{r}}{\partial t}\cdot\nabla\right)=\!\left(\frac{\partial}{\partial t}+c\hat{\textbf{k}}_{j}\cdot\nabla\right)$ is the convective derivative along the wave in question.

From Eqs.$\ $(4.1), (4.2), (4.5) and (4.6) we have
\begin{eqnarray}
\sum^{4}_{j=1}i\omega_{j}\!\!\!\!\!\!&&\!\!\!\!\!\!\left(\frac{d\tilde{\textbf{E}}_{j}\!\left(\textbf{r},t\right)}{dt}e^{i\left(\textbf{k}_{j}\cdot\textbf{r}-\omega_{j}t\right)}-\frac{d\tilde{\textbf{E}}^{*}_{j}\!\left(\textbf{r},t\right)}{dt}e^{-i\left(\textbf{k}_{j}\cdot\textbf{r}-\omega_{j}t\right)}\right)\nonumber \\ \!\!\!\!\!\!&&\!\!\!\! \approx -\sum^{8}_{l=1}\ \!\sum^{8}_{m=l}\sum^{8}_{n=m}\!\!\ A_{l,m,n}\tilde{E}_{l}\tilde{E}_{m}\tilde{E}_{n}e^{i\left[\left(\textbf{k}_{l}+\textbf{k}_{m}+\textbf{k}_{n}\right)\cdot\textbf{r}-\left(\omega_{l}+\omega_{m}+\omega_{n}\right)t\right]}\hat{\textbf{u}}_{l,m,n}.
\end{eqnarray}
Let us multiply both sides of Eq.$\ $(4.7) by $e^{-i\left(\textbf{k}_{1}\cdot\textbf{r}-\omega_{1}t\right)}$ and thereafter average over many wavelengths and period times. Due to rapid oscillations, all terms with non-zero exponents will then average to zero and be negligible. Thanks to the matching conditions (4.3) and (4.4), there is a term that will survive on the right hand side, namely the one including the product $\tilde{E}^{*}_{2}\tilde{E}_{3}\tilde{E}_{4}$. Thus there is no need to ever consider the other terms, an insight which definitely makes the calculations easier. When evaluating the right hand side of a wave equation like (4.2) we only have to keep track of the resonant terms, that is only determine one of the in total 120 different  $A_{l,m,n}$. Similar arguments of course works for the other three waves. So by neglecting all non-resonant terms, we end up with wave equations looking like
\begin{eqnarray}
	\left(\frac{\partial^{2}}{\partial t^{2}}-c^{2}\nabla^{2}\right)\tilde{\textbf{E}}_{1}\!\left(\textbf{r},t\right)=C_{1}\tilde{E}^{*}_{2}\tilde{E}_{3}\tilde{E}_{4}\hat{\textbf{u}}_{1},
\end{eqnarray}
\begin{eqnarray}
	\left(\frac{\partial^{2}}{\partial t^{2}}-c^{2}\nabla^{2}\right)\tilde{\textbf{E}}_{2}\!\left(\textbf{r},t\right)=C_{2}\tilde{E}^{*}_{1}\tilde{E}_{3}\tilde{E}_{4}\hat{\textbf{u}}_{2},
\end{eqnarray}
\begin{eqnarray}
	\left(\frac{\partial^{2}}{\partial t^{2}}-c^{2}\nabla^{2}\right)\tilde{\textbf{E}}_{3}\!\left(\textbf{r},t\right)=C_{3}\tilde{E}_{1}\tilde{E}_{2}\tilde{E}^{*}_{4}\hat{\textbf{u}}_{3},
\end{eqnarray}
\begin{eqnarray}
	\left(\frac{\partial^{2}}{\partial t^{2}}-c^{2}\nabla^{2}\right)\tilde{\textbf{E}}_{4}\!\left(\textbf{r},t\right)=C_{4}\tilde{E}_{1}\tilde{E}_{2}\tilde{E}^{*}_{3}\hat{\textbf{u}}_{4},
\end{eqnarray}
where $C_{j}$ are the coupling coefficients, and $\hat{\textbf{u}}_{j}$ are unit vectors.

Next assume that initially one field is much weaker than the others, say for instance $E_{4}\!\!\ll\!E_{1}, E_{2}, E_{3}$. From Eqs.$\ $(4.8)-(4.11) we then see that, for interactions occupying small regions of spacetime, the driving of the strong field amplitudes may be neglected, strengthening the point of view to treat those fields as unaffected by the interaction. Only the driving of the weak field amplitude will be of importance.
\subsection{Generated field}
Now consider the case when the initial weak field in fact does not exist, but the strong fields $\textbf{E}_{1}\!\left(\textbf{r},t\right)$, $\textbf{E}_{2}\!\left(\textbf{r},t\right)$ and $\textbf{E}_{3}\!\left(\textbf{r},t\right)$ still satisfy the matching conditions (4.3) and (4.4) for some $\textbf{k}_{4}$. In other words the wave vectors satisfy the equation
\begin{eqnarray}
\left|\textbf{k}_{1}+\textbf{k}_{2}-\textbf{k}_{3}\right|=k_{1}+k_{2}-k_{3},
\end{eqnarray}
and we make the definitions $\textbf{k}_{4}\!\equiv\textbf{k}_{1}+\textbf{k}_{2}-\textbf{k}_{3}$ and $k_{4}\!\equiv k_{1}+k_{2}-k_{3}$.
From the above discussion it is clear we expect a wave to be generated in the $\hat{\textbf{k}}_{4}$-direction, but how about directions differing only slightly from it? To clear this out we write the generated field $\textbf{E}_{g}\!\left(\textbf{r},t\right)$ as
\begin{eqnarray}
\textbf{E}_{g}\!\left(\textbf{r},t\right)=\frac{1}{2}\left(\tilde{\textbf{E}}_{g}\!\left(\textbf{r},t\right)+\tilde{\textbf{E}}^{*}_{g}\!\left(\textbf{r},t\right)\right),
\end{eqnarray}
where $\tilde{\textbf{E}}_{g}\!\left(\textbf{r},t\right)$ is the part corresponding to $\tilde{\textbf{E}}_{4}\!\left(\textbf{r},t\right)e^{i\left(\textbf{k}_{4}\cdot\textbf{r}-\omega_{4}t\right)}$. Following the procedure of just keeping resonant terms, we will end up with an equation like
\begin{eqnarray}
	\left(\frac{\partial^{2}}{\partial t^{2}}-c^{2}\nabla^{2}\right)\tilde{\textbf{E}}_{g}\!\left(\textbf{r},t\right)=C_{4}\tilde{E}_{1}\tilde{E}_{2}\tilde{E}^{*}_{3}\hat{\textbf{u}}_{4}e^{i\left(\textbf{k}_{4}\cdot\textbf{r}-\omega_{4}t\right)},
\end{eqnarray}
having the solution \cite{Byron}
\begin{eqnarray}
	\tilde{\textbf{E}}_{g}\!\left(\textbf{r},t\right)=\frac{1}{4\pi c^{2}}C_{4}\hat{\textbf{u}}_{4}\int\tilde{E}_{1}\tilde{E}_{2}\tilde{E}^{*}_{3}|_{t_{R}}\frac{e^{i\left(\textbf{k}_{4}\cdot\textbf{r}-\omega_{4}t_{R}\right)}}{R}dV',
\end{eqnarray}
where the retarded time is defined as $t_{R}\equiv t-\frac{R}{c}$, $R\equiv\left|\textbf{r}-\textbf{r}'\right|$ and integration is over all space. The term $\tilde{E}_{1}\tilde{E}_{2}\tilde{E}^{*}_{3}|_{t_{R}}$ is kept inside the integral since the fields will only occupy a limited region of spacetime. The subscript indicates that the term really should be evaluated at $t_{R}$ instead of $t$. For now we model the amplitudes to be constants inside some interaction region, and zero outside. This is to be seen as a source generating the fourth wave.

Assume that this source region is constituted out of the volume $V'$ in space, and infinite in time. The solution may then be written
\begin{eqnarray}
	\tilde{\textbf{E}}_{g}\!\left(\textbf{r},t\right)=\frac{1}{4\pi c^{2}}C_{4}\tilde{E}_{1}\tilde{E}_{2}\tilde{E}^{*}_{3}\hat{\textbf{u}}_{4}\int_{V'}\frac{e^{i\left(\textbf{k}_{4}\cdot\textbf{r}-\omega_{4}t_{R}\right)}}{R}dV'.
\end{eqnarray}
In the radiation zone, $r\gg r'$, we have
\begin{eqnarray}
	R=\left|\textbf{r}-\textbf{r}'\right|=\sqrt{r^{2}+(r')^{2}-2\textbf{r}\cdot\textbf{r}'}\approx r\sqrt{1-2\frac{\textbf{r}\cdot\textbf{r}'}{r^{2}}}\approx r-\hat{\textbf{r}}\cdot\textbf{r}'.
\end{eqnarray}
In the denominator of the integrand in Eq.$\ $(4.16) it is enough to just use $R\approx r$, but since $k_{4}$ may be large, the full expression (4.17) will be used in the exponent, giving us the radiation zone approximation
\begin{eqnarray}
	\tilde{\textbf{E}}_{g}\!\left(\textbf{r},t\right)\!\!\!\!&\approx &\!\!\!\!\frac{1}{4\pi c^{2}}C_{4}\tilde{E}_{1}\tilde{E}_{2}\tilde{E}^{*}_{3}\hat{\textbf{u}}_{4}\int_{V'}\frac{e^{i\left(\textbf{k}_{4}\cdot\textbf{r}'-\omega_{4}t+k_{4}r-k_{4}\hat{\textbf{r}}\cdot\textbf{r}'\right)}}{r}\ dV' \nonumber  \\ \!\!\!\!&=&\!\!\!\!\frac{1}{4\pi rc^{2}}C_{4}\tilde{E}_{1}\tilde{E}_{2}\tilde{E}^{*}_{3}e^{i\left(k_{4}r-\omega_{4}t\right)}\hat{\textbf{u}}_{4}\int_{V'}e^{i{k}_{4}\left(\hat{\textbf{k}}_{4}-\hat{\textbf{r}}\right)\cdot\textbf{r}'}dV'.
\end{eqnarray}
The fact that $\textbf{k}_{4}\!={k}_{4}\hat{\textbf{k}}_{4}$, used in the last line, follows directly from the imposed matching conditions (4.12). Note that the solution takes the form of an outgoing spherical wave, multiplied by a direction dependent factor peaked for $\hat{\textbf{r}}=\hat{\textbf{k}}_{4}$. 
%--------------------------Matching conditions--------------------------------------
\section{Matching conditions}
\subsection{Interpretation}
In order to take advantage of the resonance phenomena discussed in the last section, we need to find a configuration of plane waves with wave vectors satisfying the matching conditions (4.3) and (4.4), or equivalently
\begin{eqnarray}
	\textbf{k}_{1}+\textbf{k}_{2}=\textbf{k}_{3}+\textbf{k}_{4},
\end{eqnarray}
\begin{eqnarray}
	k_{1}+k_{2}=k_{3}+k_{4}.
\end{eqnarray}
These conditions correspond to momentum and energy conservations respectively. (Compare with the momentum $\hbar\textbf{k}$ and energy $\hbar\omega$ of a photon.)

Let us first discuss some questions that may arise if one is not familiar with this kind of wave interactions. The form of the matching conditions screams for the interpretation of wave one and two as incoming photons, while wave three and four should represent the scattered ones. But in the experiment we will have three incoming waves generating a fourth one, so how does this then make sense? Well, the problem has already been translated from QED into classical electrodynamics via the effective Lagrangian, and we should in fact no longer have to interpret the waves as particles. We are free to choose whatever matching conditions we like to, and as long as the calculations end up with something useful, one should in principle be satisfied. However, there is a simple way of thinking which may help. We can see the third wave as an additional external contribution to the particles scattered in the $\hat{\textbf{k}}_{3}$-direction, with the purpose of stimulating the scattering in the $\hat{\textbf{k}}_{4}$-direction, i.e.$\!$ making the right hand side of Eq.$\ $(4.11) larger. Another question is why it is not enough with just three waves, two incoming and a third being generated? Since the non-linear terms are cubic in the fields, the wave vectors then have to satisfy matching conditions like $	2\textbf{k}_{1}\!+\textbf{k}_{2}\!=\textbf{k}_{3}$, $\ 2k_{1}\!+k_{2}\!=k_{3}$ or $2\textbf{k}_{1}\!-\textbf{k}_{2}\!=\textbf{k}_{3}$, $\ 2k_{1}\!-k_{2}\!=k_{3}$. For these to hold the three vectors need to point along the same line, i.e.$\!$ we are confined to a one dimensional configuration, making the experimental set up and measuring very hard. (This can with advantage be compared to the classical elastic two particle collision.) Following the same line of reasoning, we can also rule out the choice of the matching conditions $\textbf{k}_{1}\!+\textbf{k}_{2}\!+\textbf{k}_{3}\!=\textbf{k}_{4}$, $\ k_{1}\!+k_{2}\!+k_{3}\!=k_{4}$.
\subsection{Two dimensional configurations}
Putting focus on finding configurations satisfying (4.19) and (4.20), we write the wave vectors as
\begin{eqnarray}
\textbf{k}_{j}=k_{j}\sin\theta_{j}\cos\phi_{j}\hat{\textbf{x}}+k_{j}\sin\theta_{j}\sin\phi_{j}\hat{\textbf{y}}+k_{j}\cos\theta_{j}\hat{\textbf{z}},
\end{eqnarray}
where $\theta_{j}$ and $\phi_{j}$ are defined in the usual way, as in Figure 4.1. In total we have twelve unknown parameters ($k_{j},\ \!\theta_{j},\ \!\phi_{j}$) and four equations, (4.19) and (4.20), so given eight parameters the remaining four can in principle be determined.
%---------------------------------------------------
\begin{figure}
	\centering
	\includegraphics[width=0.6\textwidth]{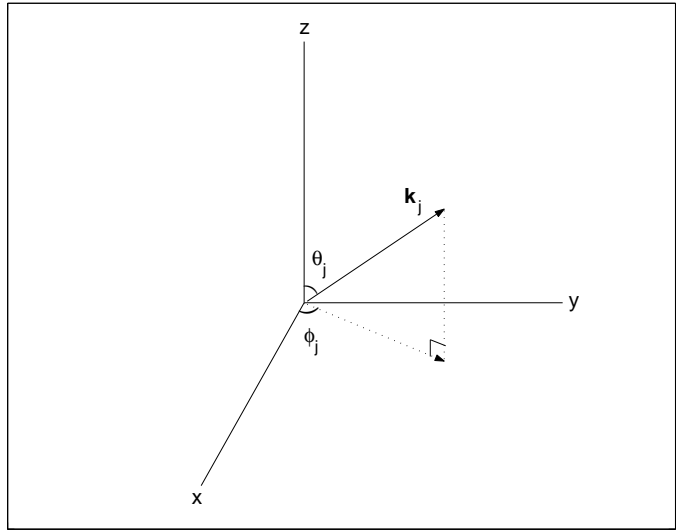}	
	\caption{\emph{Definition of $\ \theta_{j}$ and $\phi_{j}$}.}
	\label{fig:coordinatesystem10}
\end{figure}
%---------------------------------------------------

For a two dimensional configuration, which we choose to be in the x-y-plane so that $\theta_{1}\!=\theta_{2}\!=\theta_{3}\!=\pi/2\ \Rightarrow\ \theta_{4}\!=\pi/2$ giving
\begin{eqnarray}
\textbf{k}_{j}=k_{j}\cos\phi_{j}\hat{\textbf{x}}+k_{j}\sin\phi_{j}\hat{\textbf{y}},
\end{eqnarray}
the system is reduced to eight unknown ($k_{j},\ \!\phi_{j}$) and three equations
\begin{eqnarray}
	k_{1}\cos\phi_{1}+k_{2}\cos\phi_{2}=k_{3}\cos\phi_{3}+k_{4}\cos\phi_{4}
\end{eqnarray}
\begin{eqnarray}
	k_{1}\sin\phi_{1}+k_{2}\sin\phi_{2}=k_{3}\sin\phi_{3}+k_{4}\sin\phi_{4}
\end{eqnarray}
\begin{eqnarray}
	k_{1}+k_{2}=k_{3}+k_{4},
\end{eqnarray}
leaving us with five independent parameters.

As a first approach we consider the case when $k_{1}\!=\!k_{2}\!=\!k_{3}\!\equiv k$, corresponding to using three laser beams of the same frequency. According to Eq.$\ $(4.25) we will then get $k_{4}\!=\!k$, so Eqs.$\ $(4.23) and (4.24) reduces to
\begin{eqnarray}
	\cos\phi_{1}+\cos\phi_{2}=\cos\phi_{3}+\cos\phi_{4}
\end{eqnarray}
\begin{eqnarray}
	\sin\phi_{1}+\sin\phi_{2}=\sin\phi_{3}+\sin\phi_{4},
\end{eqnarray}
which are satisfied for $\phi_{1}\!=\!\phi_{3}$, $\ \phi_{2}\!=\!\phi_{4}$ and also when $\phi_{1}\!=\!\phi_{2}+\pi$, $\ \phi_{3}\!=\!\phi_{4}+\pi$.
The first case where two of the incoming waves must have the same wave vector, may be seen mentally as a rhombus with sides as in Figure 4.2(a). The second case with two pairs of head-on colliding beams can be pictured as in Figure 4.2(b). As we should expect, and as can be seen from the coupling coefficients derived in the next chapter, the last solution is indeed the most interesting. When modeling the interaction we will examine the two extreme cases $\phi_{1}\!=\pi/2$, $\ \phi_{2}\!=\!-\pi/2$, $\ \phi_{3}\!=\pi$, $\ \phi_{4}\!=0\ $ and $\ \phi_{1}\!=\!\phi_{3}\!=\pi$, $\ \phi_{2}\!=\!\phi_{4}\!=0$, visualized in Figure 4.3. It will be shown that the later will give rise to a larger generated intensity, and therefore is the favoured configuration.
%---------------------------------------------------------------------------------------
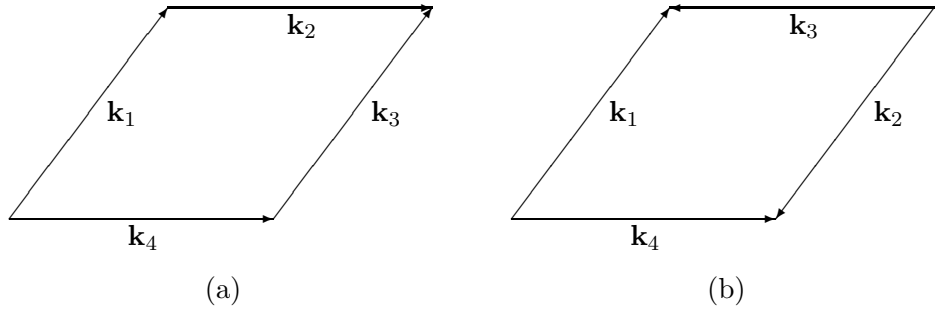
\begin{figure}
\begin{center}
		\begin{picture}(350,120)
										
		\put(0,30){\vector(3,4){60}}
		\put(60,110){\vector(1,0){100}}
		\put(0,30){\vector(1,0){100}}
		\put(100,30){\vector(3,4){60}}
		
		\put(74,0){\textrm{(a)}}
		\put(37,67){\textrm{$\textbf{k}_{1}$}}
		\put(105,100){\textrm{$\textbf{k}_{2}$}}
		\put(137,67){\textrm{$\textbf{k}_{3}$}}
		\put(45,20){\textrm{$\textbf{k}_{4}$}}
		
		\put(190,30){\vector(3,4){60}}
		\put(350,110){\vector(-1,0){100}}
		\put(190,30){\vector(1,0){100}}
		\put(350,110){\vector(-3,-4){60}}
		
		\put(264,0){\textrm{(b)}}
		\put(227,67){\textrm{$\textbf{k}_{1}$}}
		\put(295,100){\textrm{$\textbf{k}_{3}$}}
		\put(327,67){\textrm{$\textbf{k}_{2}$}}
		\put(235,20){\textrm{$\textbf{k}_{4}$}}
	\end{picture}
\end{center}
\caption{\emph{Diagrammatic representations of the solutions to the two dimensional matching conditions for equal frequencies, Eqs.$\ $(4.26), (4.27)}: (a) $\phi_{1}\!=\!\phi_{3}$, $\ \phi_{2}\!=\!\phi_{4}$; (b) $\phi_{1}\!=\!\phi_{2}+\pi$, $\ \phi_{3}\!=\!\phi_{4}+\pi$.}
\end{figure}

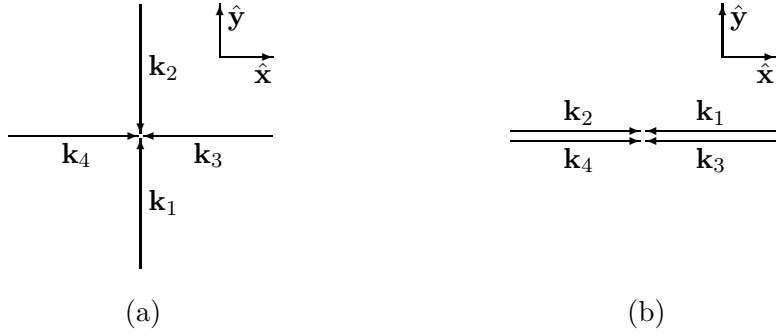
\begin{figure}
\begin{center}
		\begin{picture}(350,120)
								
		\put(80,20){\vector(0,1){49}}
		\put(30,70){\vector(1,0){49}}
		\put(80,120){\vector(0,-1){49}}
		\put(130,70){\vector(-1,0){49}}
		
		\put(74,0){\textrm{(a)}}
		\put(83,42){\textrm{$\textbf{k}_{1}$}}
		\put(83,93){\textrm{$\textbf{k}_{2}$}}
		\put(100,60){\textrm{$\textbf{k}_{3}$}}
		\put(50,60){\textrm{$\textbf{k}_{4}$}}
		
		\put(110,100){\vector(1,0){20}}
		\put(110,100){\vector(0,1){20}}
		
		\put(123,90){\textrm{$\hat{\textbf{x}}$}}
		\put(113,112){\textrm{$\hat{\textbf{y}}$}}
		
		\put(220,72){\vector(1,0){49}}
		\put(220,68){\vector(1,0){49}}
		\put(320,72){\vector(-1,0){49}}
		\put(320,68){\vector(-1,0){49}}
		
		\put(264,0){\textrm{(b)}}
		\put(240,76){\textrm{$\textbf{k}_{2}$}}
		\put(240,58){\textrm{$\textbf{k}_{4}$}}
		\put(290,76){\textrm{$\textbf{k}_{1}$}}
		\put(290,58){\textrm{$\textbf{k}_{3}$}}
		
		\put(300,100){\vector(1,0){20}}
		\put(300,100){\vector(0,1){20}}
		
		\put(313,90){\textrm{$\hat{\textbf{x}}$}}
		\put(303,112){\textrm{$\hat{\textbf{y}}$}}
		
	\end{picture}
\end{center}
\caption{\emph{Two specific configurations satisfying the matching conditions (4.26) and (4.27)}: (a) $\phi_{1}\!=\pi/2$, $\ \phi_{2}\!=\!-\pi/2$, $\ \phi_{3}\!=\pi$, $\ \phi_{4}\!=0$; (b) $\ \phi_{1}\!=\!\phi_{3}\!=\pi$, $\ \phi_{2}\!=\!\phi_{4}\!=0$.}	
\end{figure}
%---------------------------------------------------------------------------------------
\subsection{Practical adjustments}
There is however some problems with this latest choice of angles (as there is also with all other alternatives discussed so far, but let us focus on this case now). For the first, the $\hat{\textbf{k}}_{4}$- and $\hat{\textbf{k}}_{2}$-direction coincides, making it really hard to separate the small amount of scattered photons from the intense external beam. Secondly, the generated wave will have its peak straight into the sources of the two opposite traveling waves, yielding problems with the placing of the detector. What we want to do is thus to find a configuration looking almost as the original (to keep as much of the advantages of the model as possible) but where the $\hat{\textbf{k}}_{4}$-direction is slightly changed to make sure none of the above problems arise.

\begin{figure}[b]
\begin{center}
		\begin{picture}(90,65)
		
		\put(5,10){\dashbox(0,60)}
		\put(85,10){\dashbox(0,60)}		
		\put(85,70){\vector(-4,-3){80}}
		\put(5,70){\vector(4,-3){80}}
		\put(5,70){\vector(1,0){80}}
		\put(85,10){\vector(-1,0){80}}
		
		\put(42,0){\textrm{$\textbf{k}_{3}$}}
		\put(42,60){\textrm{$\textbf{k}_{2}$}}
		\put(56,20){\textrm{$\textbf{k}_{4}$}}
		\put(32,20){\textrm{$\textbf{k}_{1}$}}
		
	\end{picture}
\end{center}
\caption{\emph{Picture of the adjusted wave vectors}.}	
\end{figure}
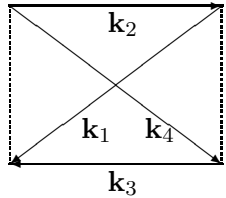
Say we want to displace $\hat{\textbf{k}}_{4}$ at least an angle $\delta$ from both $\hat{\textbf{k}}_{2}$ and the sources of the opposite directed waves. What is needed to obtain such a configuration is a small shifting of frequencies. To get a feeling for how this can be done we draw a rectangle in which the diagonals represent ${\textbf{k}}_{1}$ and ${\textbf{k}}_{4}$, while two of the sides correspond to ${\textbf{k}}_{2}$ and ${\textbf{k}}_{3}$, all in accordance with Figure 4.4. Guided by this picture we now choose $\phi_{1}\!=\pi+2\delta$, $\ \phi_{2}\!=\delta$, $\ \phi_{3}\!=\pi+\delta$, $\ \phi_{4}=0$, $\ k_{1}\!=\!k_{4}\!\equiv k$, $\ k_{2}\!=\!k_{3}\!=\!k\cos\delta$, and show that with these choices the matching conditions (4.23)-(4.25) are satisfied. Since
\begin{eqnarray}
	\frac{1}{k}\!\!\!\!\!&&\!\!\!\!\!\left[k_{1}\cos\phi_{1}+k_{2}\cos\phi_{2}-k_{3}\cos\phi_{3}-k_{4}\cos\phi_{4}\right]\nonumber \\ &&\ \ \ \ \ \ \ \ \ \ \ \ \ \ \ \ =\cos\left(\pi+2\delta\right)+\cos\delta\cos\delta-\cos\delta\cos\left(\pi+\delta\right)-\cos0\nonumber \\ &&\ \ \ \ \ \ \ \ \ \ \ \ \ \ \ \ =-\cos2\delta+2\cos^{2}\delta-1=\sin^{2}\delta+\cos^{2}\delta-1=0,
\end{eqnarray}
\begin{eqnarray}
	\frac{1}{k}\!\!\!\!\!&&\!\!\!\!\!\left[k_{1}\sin\phi_{1}+k_{2}\sin\phi_{2}-k_{3}\sin\phi_{3}-k_{4}\sin\phi_{4}\right]\nonumber \\ &&\ \ \ \ \ \ \ \ \ \ \ \ \ \ \ \ =\sin\left(\pi+2\delta\right)+\cos\delta\sin\delta-\cos\delta\sin\left(\pi+\delta\right)-\sin0\nonumber \\ &&\ \ \ \ \ \ \ \ \ \ \ \ \ \ \ \ =-\sin2\delta+2\cos\delta\sin\delta=0
\end{eqnarray}
and
\begin{eqnarray}
	k_{1}+k_{2}-k_{3}-k_{4}=k+k\cos\delta-k\cos\delta-k=0,
\end{eqnarray}
this is clearly the case. Consequently we must be able to shift two frequencies from $k$ to $k\cos\delta$ in order to follow this recipe.
\subsection{Three dimensional configurations}
The restriction to two dimensional configurations is not as limited as it first may sound. As soon as two of the waves are heads on, the four wave vectors will always lie in a plane, which we of course are free to call the x-y-plane. Although one might expect this to be the optimal situation even when considering three dimensions, we have to investigate if there could be something favourable with choosing a configuration not confined to a plane. It turns out that for a general case the calculations of the coupling coefficients are quite hard to perform, and therefore we will just work out one specific three dimensional configuration.

What should be most interesting is the extreme case when the incoming waves form straight angles to each other, one along each of the cartesian coordinate axises. This takes us as far as possible from a plane. Let us now see if we can find such wave vectors satisfying (4.19) and (4.20). To start with we choose $\theta_{1}\!=\pi/2$, $\ \phi_{1}\!=0$, $\ \theta_{2}\!=\pi/2$, $\ \phi_{1}\!=\pi/2$, $\ \theta_{3}\!=0$, i.e.$\!$ the vectors lie along the x-, y- and z-axis respectively ($\phi_{3}$ has no meaning in this case). To specify the situation completely two more parameters need to be given, so why not choose $k_{1}\!=\!k_{2}\!\equiv k$. Left to be determined are four parameters which from imposing the matching conditions easily can be found as $k_{3}\!=\!k/2$, $\ k_{4}\!=3k/2$, $\ \theta_{4}\!=\pi/2+\arctan \ \!\!(1/\ \!\!2\!\sqrt{2})$ and $\phi_{4}\!=\pi/4$. 
We now have a specific three dimensional configuration, pictured in Figure 4.5, ready to be investigated later on.
\begin{figure}
	\centering
	\includegraphics[width=0.6\textwidth]{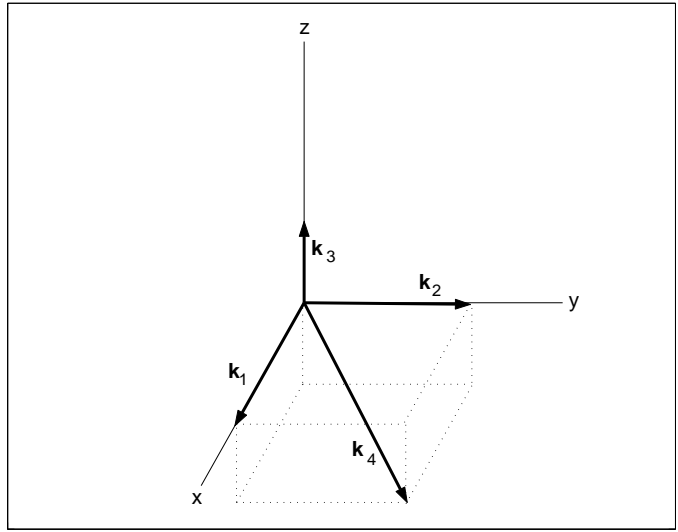}	
	\caption{\emph{Specific three dimensional configuration}.}
	\label{fig:3dconfiguration2}
\end{figure}
%------------------------Calculations of non-linear terms-----------------------------------------
\chapter{Calculations of non-linear terms}
In this chapter we derive expressions for the non-linear terms in the effective wave equation (3.38). It will be assumed the wave vectors of the incoming plane waves ($\textbf{k}_{1}, \ \!\textbf{k}_{2}, \ \!\textbf{k}_{3}$) have been chosen in such way that the matching conditions (4.19) and (4.20) are fulfilled for some $\textbf{k}_{4}$. Thus we can use the method of only saving resonant terms as discussed in Section 4.1. For every specific wave configuration the calculations hence end up with an expression like (4.14). However, since we will consider more general cases than just single ones, the results will by convenience be written on the form
\begin{eqnarray}
	\left(\frac{\partial^{2}}{\partial t^{2}}-c^{2}\nabla^{2}\right)\tilde{\textbf{E}}_{g}\!\left(\textbf{r},t\right)=4\xi\omega^{2}_{4}\textbf{G}\tilde{E}_{1}\tilde{E}_{2}\tilde{E}^{*}_{3}e^{i\left(\textbf{k}_{4}\cdot\textbf{r}-\omega_{4}t\right)},
\end{eqnarray}
where the vector $\textbf{G}$ is a geometric factor depending only on the directions of wave vectors and polarization vectors. As before the plane waves will be written on complex form such that
\begin{eqnarray}
\textbf{E}_{j}\!\left(\textbf{r},t\right)=\frac{1}{2}\left(\tilde{\textbf{E}}_{j}\!\left(\textbf{r},t\right)e^{i\left(\textbf{k}_{j}\cdot\textbf{r}-\omega_{j}t\right)}+\tilde{\textbf{E}}^{*}_{j}\!\left(\textbf{r},t\right)e^{-i\left(\textbf{k}_{j}\cdot\textbf{r}-\omega_{j}t\right)}\right).
\end{eqnarray}

An attempt to work out the calculations for the most general case of three dimensional geometry and arbitrary polarization soon becomes quite messy. Therefore we put focus on a general two dimensional geometry, and only work out the calculations for a specific example in three dimensions. As discussed in Section 4.2, this is really not such a limited restriction since the optimal situations seem to be when we have head-on collisions, in which the wave vectors always form a plane. But at least we should check if there could be some advantages with a truly three dimensional configuration.
\section{Two dimensional geometries}
\subsection{Arbitrary incoming angles and polarizations}
Wave vectors confined to the x-y-plane can be written
\begin{eqnarray}
	\textbf{k}_{j}=k_{j}\cos\phi_{j}\hat{\textbf{x}}+k_{j}\sin\phi_{j}\hat{\textbf{y}},
\end{eqnarray}
where $\phi_{j}$ is defined in accordance with Figure 4.1.

With intentions of being general, the calculations will be done for arbitrary linear polar\-izations of the plane waves. Let $\gamma_{j}$ be the angle between the z-axis and the polarization vector of the $E_{j}$-field, in such way it is positive in clockwise direction when looking in the $\textbf{k}_{j}$-direction, as shown in Figure 5.1. From inspection of the picture it can be seen that the electric fields then are given by
\begin{eqnarray}
\textbf{E}_{j}\!\left(\textbf{r},t\right)=E_{j}\!\left(\textbf{r},t\right)\left[\sin\gamma_{j}\sin\phi_{j}\hat{\textbf{x}}-\sin\gamma_{j}\cos\phi_{j}\hat{\textbf{y}}+\cos\gamma_{j}\hat{\textbf{z}}\right],
\end{eqnarray}
where obviously
\begin{eqnarray}
{E}_{j}\!\left(\textbf{r},t\right)=\frac{1}{2}\left(\tilde{E}_{j}\!\left(\textbf{r},t\right)e^{i\left(\textbf{k}_{j}\cdot\textbf{r}-\omega_{j}t\right)}+\tilde{E}^{*}_{j}\!\left(\textbf{r},t\right)e^{-i\left(\textbf{k}_{j}\cdot\textbf{r}-\omega_{j}t\right)}\right).
\end{eqnarray}
To avoid any possible confusion about the notation it should be pointed out that $|\tilde{E}_{j}\!\left(\textbf{r},t\right)|=|\tilde{E}^{*}_{j}\!\left(\textbf{r},t\right)|$ is the amplitude of the oscillating function ${E}_{j}\!\left(\textbf{r},t\right)$, and nothing else.
\begin{figure}
	\centering
	\includegraphics[width=0.6\textwidth]{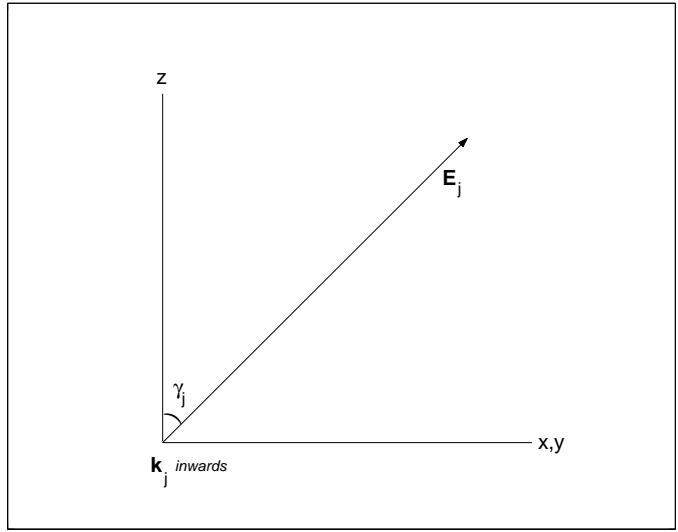}	
	\caption{\emph{Definition of $\gamma_{j}$}.}
	\label{fig:gamma6}
\end{figure}

The magnetic fields may be expressed in terms of the electric ones using (5.3) and (5.4) in the well-known (and from Maxwell's equations easily derivable) relation for plane electromagnetic waves
\begin{eqnarray}
\textbf{B}_{j}\!\left(\textbf{r},t\right)=\hat{\textbf{k}}_{j}\times\textbf{E}_{j}\!\left(\textbf{r},t\right)=E_{j}\!\left(\textbf{r},t\right)\left[\cos\gamma_{j}\sin\phi_{j}\hat{\textbf{x}}-\cos\gamma_{j}\cos\phi_{j}\hat{\textbf{y}}-\sin\gamma_{j}\hat{\textbf{z}}\right].
\end{eqnarray}

We are now ready to work out the right hand side of Eq.$\ $(3.38). Due to the weak non-linearity the amplitude variations are much slower than the harmonic oscillations, and hence the space and time derivatives is taken to operate only on the exponential parts. This just brings down some factors in front of the different terms in the expression, so we can conclude it really takes the form of Eq.$\ $(4.2). Using Eqs.$\ $(5.4)-(5.6) and just keeping the resonant parts, the non-linear expression may in a tedious way be calculated term by term, leading to the final result
\begin{eqnarray}
	\left(\frac{\partial^{2}}{\partial t^{2}}-c^{2}\nabla^{2}\right)\tilde{\textbf{E}}_{g}\!\left(\textbf{r},t\right)=4\xi\omega^{2}_{4}\textbf{G}_{2d}\tilde{E}_{1}\tilde{E}_{2}\tilde{E}^{*}_{3}e^{i\left(\textbf{k}_{4}\cdot\textbf{r}-\omega_{4}t\right)},
\end{eqnarray}
where the geometric factor is given by
\begin{eqnarray}
	\textbf{G}_{2d}\!\!\!\!&=&\!\!\!\!+\frac{1}{2}\left\{\ \left[\frac{7}{4}\cos\gamma_{3}\sin\left(\gamma_{1}+\gamma_{2}\right)-\sin\gamma_{3}\cos\left(\gamma_{1}+\gamma_{2}\right)\right]\right.\nonumber \\ &&\ \ \ \ \ \ \times\left[\sin\left(\phi_{3}-\phi_{4}\right)\cos\phi_{4}-\left(\sin\phi_{3}-\sin\phi_{4}\right)\right]\sin^{2}\left(\frac{\phi_{1}-\phi_{2}}{2}\right)\nonumber \\ &&\ \ \ + \left[\frac{7}{4}\cos\gamma_{2}\sin\left(\gamma_{1}+\gamma_{3}\right)-\sin\gamma_{2}\cos\left(\gamma_{1}+\gamma_{3}\right)\right]\nonumber \\ &&\ \ \ \ \ \ \times\left[\sin\left(\phi_{2}-\phi_{4}\right)\cos\phi_{4}-\left(\sin\phi_{2}-\sin\phi_{4}\right)\right]\sin^{2}\left(\frac{\phi_{1}-\phi_{3}}{2}\right)\nonumber \\ &&\ \ \ + \left[\frac{7}{4}\cos\gamma_{1}\sin\left(\gamma_{2}+\gamma_{3}\right)-\sin\gamma_{1}\cos\left(\gamma_{2}+\gamma_{3}\right)\right]\nonumber \\ &&\ \ \ \ \ \ \times\left.\left[\sin\left(\phi_{1}-\phi_{4}\right)\cos\phi_{4}-\left(\sin\phi_{1}-\sin\phi_{4}\right)\right]\sin^{2}\left(\frac{\phi_{2}-\phi_{3}}{2}\right)\ \right\}\hat{\textbf{x}}\nonumber \\ &&\!\!\!\!+ \frac{1}{2}\left\{\ \left[\frac{7}{4}\cos\gamma_{3}\sin\left(\gamma_{1}+\gamma_{2}\right)-\sin\gamma_{3}\cos\left(\gamma_{1}+\gamma_{2}\right)\right]\right.\nonumber \\ &&\ \ \ \ \ \ \times\left[\sin\left(\phi_{3}-\phi_{4}\right)\sin\phi_{4}+\left(\cos\phi_{3}-\cos\phi_{4}\right)\right]\sin^{2}\left(\frac{\phi_{1}-\phi_{2}}{2}\right)\nonumber \\ &&\ \ \ + \left[\frac{7}{4}\cos\gamma_{2}\sin\left(\gamma_{1}+\gamma_{3}\right)-\sin\gamma_{2}\cos\left(\gamma_{1}+\gamma_{3}\right)\right]\nonumber \\ &&\ \ \ \ \ \ \times\left[\sin\left(\phi_{2}-\phi_{4}\right)\sin\phi_{4}+\left(\cos\phi_{2}-\cos\phi_{4}\right)\right]\sin^{2}\left(\frac{\phi_{1}-\phi_{3}}{2}\right)\nonumber \\ &&\ \ \ + \left[\frac{7}{4}\cos\gamma_{1}\sin\left(\gamma_{2}+\gamma_{3}\right)-\sin\gamma_{1}\cos\left(\gamma_{2}+\gamma_{3}\right)\right]\nonumber \\ &&\ \ \ \ \ \ \times\left.\left[\sin\left(\phi_{1}-\phi_{4}\right)\sin\phi_{4}+\left(\cos\phi_{1}-\cos\phi_{4}\right)\right]\sin^{2}\left(\frac{\phi_{2}-\phi_{3}}{2}\right)\ \right\}\hat{\textbf{y}}\nonumber \\ &&\!\!\!\!+ \left\{\ \left[\frac{7}{4}\sin\gamma_{3}\sin\left(\gamma_{1}+\gamma_{2}\right)+\cos\gamma_{3}\cos\left(\gamma_{1}+\gamma_{2}\right)\right]\right.\nonumber \\ &&\ \ \ \ \times\ \sin^{2}\left(\frac{\phi_{3}-\phi_{4}}{2}\right)\sin^{2}\left(\frac{\phi_{1}-\phi_{2}}{2}\right)\nonumber \\ &&\ + \left[\frac{7}{4}\sin\gamma_{2}\sin\left(\gamma_{1}+\gamma_{3}\right)+\cos\gamma_{2}\cos\left(\gamma_{1}+\gamma_{3}\right)\right]\nonumber \\ &&\ \ \ \ \times\ \sin^{2}\left(\frac{\phi_{2}-\phi_{4}}{2}\right)\sin^{2}\left(\frac{\phi_{1}-\phi_{3}}{2}\right)\nonumber \\ &&\ + \left[\frac{7}{4}\sin\gamma_{1}\sin\left(\gamma_{2}+\gamma_{3}\right)+\cos\gamma_{1}\cos\left(\gamma_{2}+\gamma_{3}\right)\right]\nonumber \\ &&\ \ \ \ \times\left.\ \sin^{2}\left(\frac{\phi_{1}-\phi_{4}}{2}\right)\sin^{2}\left(\frac{\phi_{2}-\phi_{3}}{2}\right)\ \right\}\hat{\textbf{z}}.
\end{eqnarray}
The complete calculations are to be found in Appendix A. Here we will just check that the expression at least satisfies some reasonable conditions.
\subsection{Checks of the result and a special case}
Since we are expecting a plane wave to be formed in the $\hat{\textbf{k}}_{4}$-direction, a basic property should be the orthogonality between the generated E-field and $\hat{\textbf{k}}_{4}$, that is the geometric factor should satisfy
\begin{eqnarray}
	\textbf{G}\cdot\hat{\textbf{k}}_{4}=0.
\end{eqnarray}
From substituting Eqs.$\ $(5.3) and (5.8) into (5.9), and using the identity
\begin{eqnarray}
	&&\sin\left(\phi_{l}-\phi_{k}\right)\cos^{2}\phi_{k}-\left(\sin\phi_{l}-\sin\phi_{k}\right)\cos\phi_{k}\nonumber \\ &&+\sin\left(\phi_{l}-\phi_{k}\right)\sin^{2}\phi_{k}+\left(\cos\phi_{l}-\cos\phi_{k}\right)\sin\phi_{k}\nonumber \\ &&\ \ \ \ \ \ \ \ \ \ \ \ \ \ \ \ \ \ \ \ \ \ \ \ =\sin\left(\phi_{l}-\phi_{k}\right)-\left(\sin\phi_{l}\cos\phi_{k}-\cos\phi_{l}\sin\phi_{k}\right)=0,
\end{eqnarray}
the orthogonality relation is in fact seen to hold.

A particularly nice form of $\textbf{G}_{2d}$ appears when all three incoming waves are polarized in the z-direction, i.e. $\gamma_{1}\!=\!\gamma_{2}\!=\!\gamma_{3}\!=0$. It then reduces to
\begin{eqnarray}
	\textbf{G}_{2d,0}\!\!\!\!&=&\!\!\!\!\left\{\ \sin^{2}\left(\frac{\phi_{3}-\phi_{4}}{2}\right)\sin^{2}\left(\frac{\phi_{1}-\phi_{2}}{2}\right)\right.\nonumber \\&&\!\!\!\!+ \sin^{2}\left(\frac{\phi_{2}-\phi_{4}}{2}\right)\sin^{2}\left(\frac{\phi_{1}-\phi_{3}}{2}\right)\nonumber \\&&\!\!\!\!+ \left.\sin^{2}\left(\frac{\phi_{1}-\phi_{4}}{2}\right)\sin^{2}\left(\frac{\phi_{2}-\phi_{3}}{2}\right)\ \right\}\hat{\textbf{z}}.
\end{eqnarray}
We will for this special case of the geometric factor check a relation corresponding to energy conservation. Since the energy density of a plane wave in vacuum is proportional to the square of the electric field amplitude, such a relation should be given by
\begin{eqnarray}
\frac{d}{dt}\left(\sum^{4}_{j=1}\tilde{E}_{j}\tilde{E}^{*}_{j}\right)=0.
\end{eqnarray}
Comparison of Eq.$\ $(4.5) and Eq.$\ $(5.1) allows us to draw the conclusion
\begin{eqnarray}
	\frac{d\tilde{E}_{4}}{dt}=2i\xi\omega_{4}G_{2d,z}\tilde{E}_{1}\tilde{E}_{2}\tilde{E}^{*}_{3},
\end{eqnarray}
and with that obviously
\begin{eqnarray}
	\frac{d\tilde{E}^{*}_{4}}{dt}=-2i\xi\omega_{4}G_{2d,z}\tilde{E}^{*}_{1}\tilde{E}^{*}_{2}\tilde{E}_{3},
\end{eqnarray}
where $G_{2d,z}$ as usual stands for the modulus of $\textbf{G}_{2d,z}$.
To evaluate (5.12) we also need the total derivatives of the three other amplitudes. Looking back on what has been done, these are quite trivial to find. We just have to make some modifications of Eq.$\ $(5.13). As an example, to get an expression for the derivative of $\tilde{E}_{3}$ we have to change $\omega_{4}$ to $\omega_{3}$, $\tilde{E}_{1}\tilde{E}_{2}\tilde{E}^{*}_{3}$ to $\tilde{E}_{1}\tilde{E}_{2}\tilde{E}^{*}_{4}$, and also interchange $\phi_{3}$ and $\phi_{4}$ in the geometric factor $G_{2d,z}$. The other derivatives are obtained in a similar way. Due to its nice form, $G_{2d,z}$ will however not be affected by the interchanging of angles (just perform it to see), and thus will be the same in each of the expressions. Hence we have
\newpage
\begin{eqnarray}
\frac{1}{2i\xi G_{2d,z}}\frac{d}{dt}\left(\sum^{4}_{j=1}\tilde{E}_{j}\tilde{E}^{*}_{j}\right)\!\!\!\!&=&\!\!\!\!\omega_{1}\tilde{E}^{*}_{1}\tilde{E}^{*}_{2}\tilde{E}_{3}\tilde{E}_{4}-\omega_{1}\tilde{E}_{1}\tilde{E}_{2}\tilde{E}^{*}_{3}\tilde{E}^{*}_{4}\nonumber \\&&\!\!\!\!+\omega_{2}\tilde{E}^{*}_{1}\tilde{E}^{*}_{2}\tilde{E}_{3}\tilde{E}_{4}-\omega_{2}\tilde{E}_{1}\tilde{E}_{2}\tilde{E}^{*}_{3}\tilde{E}^{*}_{4}\nonumber \\ &&\!\!\!\!+\omega_{3}\tilde{E}_{1}\tilde{E}_{2}\tilde{E}^{*}_{3}\tilde{E}^{*}_{4}-\omega_{3}\tilde{E}^{*}_{1}\tilde{E}^{*}_{2}\tilde{E}_{3}\tilde{E}_{4}\nonumber \\&&\!\!\!\!+\omega_{4}\tilde{E}_{1}\tilde{E}_{2}\tilde{E}^{*}_{3}\tilde{E}^{*}_{4}-\omega_{4}\tilde{E}^{*}_{1}\tilde{E}^{*}_{2}\tilde{E}_{3}\tilde{E}_{4}\nonumber \\ \!\!\!\!&=&\!\!\!\!\left(\omega_{1}+\omega_{2}-\omega_{3}-\omega_{4}\right)\tilde{E}^{*}_{1}\tilde{E}^{*}_{2}\tilde{E}_{3}\tilde{E}_{4}\nonumber \\&&\!\!\!\!-\left(\omega_{1}+\omega_{2}-\omega_{3}-\omega_{4}\right)\tilde{E}_{1}\tilde{E}_{2}\tilde{E}^{*}_{3}\tilde{E}^{*}_{4}=0,
\end{eqnarray}
where the last equality follows from the matching conditions (4.20). We conclude that everything is as it should be concerning energy conservation.

The verifications of conditions (5.9) and (5.12), the first for general polarizations and the second for a special case, indicates that the $\textbf{G}_{2d}$ we have found at least has some expected properties. This is of course nothing but a hint of being on the right track, and for the full proof of (5.8) one must turn to Appendix A.
\section{Three dimensional geometries}
\subsection{Fixed incoming angles and arbitrary polarizations}
As discussed above, it would be interesting to see what the geometric factor $\textbf{G}$ will look like in a truly three dimensional situation. We choose to work with a configuration which is, so to say, as far from a plane as possible. Such a set of wave vectors satisfying the matching conditions (4.19) and (4.20) was found at the end of Section 4.2, and may be written as
\begin{eqnarray}
	\left.\begin{array}{l}\textbf{k}_{1}=k\hat{\textbf{x}}\\ \textbf{k}_{2}=k\hat{\textbf{y}}\\ \textbf{k}_{3}=\frac{k}{2}\hat{\textbf{z}}\\ \textbf{k}_{4}=k\hat{\textbf{x}}+k\hat{\textbf{y}}-\frac{1}{2}k\hat{\textbf{z}}\end{array}\right\}.
\end{eqnarray}
Although the wave vectors now are fixed, we can still investigate the configuration for arbitrary polarizations. The polarization directions of the two waves lying in the x-y-plane can be defined exactly as before through the angles $\gamma_{1}$ and $\gamma_{2}$, but the same definition is obviously not possible for the third one as it is pointing in the z-direction. Instead we let $\beta_{3}$ be the angle between the x-axis and the polarization vector of the $E_{3}$-field, such that when looking in the $\textbf{k}_{3}$-direction it is positive in the clockwise direction, see Figure 5.2. The incoming electric fields will now be
\begin{figure}
	\centering
	\includegraphics[width=0.6\textwidth]{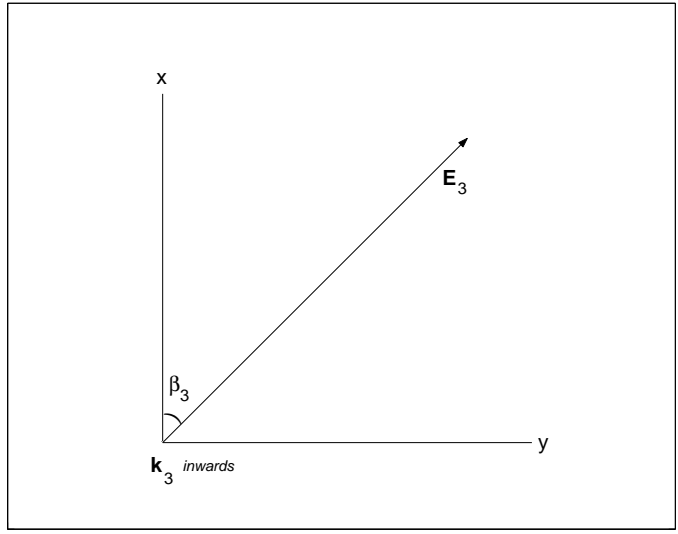}	
	\caption{\emph{Definition of $\beta_{3}$}.}
	\label{fig:beta}
\end{figure}
\begin{eqnarray}
\left.\begin{array}{l}\textbf{E}_{1}\!\left(\textbf{r},t\right)={E}_{1}\!\left(\textbf{r},t\right)\left[-\sin\gamma_{1}\hat{\textbf{y}}+\cos\gamma_{1}\hat{\textbf{z}}\right] \\ \textbf{E}_{2}\!\left(\textbf{r},t\right)={E}_{2}\!\left(\textbf{r},t\right)\left[\sin\gamma_{2}\hat{\textbf{x}}+\cos\gamma_{2}\hat{\textbf{z}}\right] \\ \textbf{E}_{3}\!\left(\textbf{r},t\right)={E}_{3}\!\left(\textbf{r},t\right)\left[\cos\beta_{3}\hat{\textbf{x}}+\sin\beta_{3}\hat{\textbf{y}}\right]\end{array}\right\},
\end{eqnarray}
and thus giving us the magnetic fields
\begin{eqnarray}
	\left.\begin{array}{l}\textbf{B}_{1}\!\left(\textbf{r},t\right)=\hat{\textbf{k}}_{1}\times\textbf{E}_{1}\!\left(\textbf{r},t\right)=E_{1}\!\left(\textbf{r},t\right)\left[-\cos\gamma_{1}\hat{\textbf{y}}-\sin\gamma_{1}\hat{\textbf{z}}\right] \\ \textbf{B}_{2}\!\left(\textbf{r},t\right)=\hat{\textbf{k}}_{2}\times\textbf{E}_{2}\!\left(\textbf{r},t\right)=E_{2}\!\left(\textbf{r},t\right)\left[\cos\gamma_{2}\hat{\textbf{x}}-\sin\gamma_{2}\hat{\textbf{z}}\right] \\ \textbf{B}_{3}\!\left(\textbf{r},t\right)=\hat{\textbf{k}}_{3}\times\textbf{E}_{3}\!\left(\textbf{r},t\right)=E_{3}\!\left(\textbf{r},t\right)\left[-\sin\beta_{3}\hat{\textbf{x}}+\cos\beta_{3}\hat{\textbf{y}}\right]\end{array}\right\}.
\end{eqnarray}

Following the same method as for the two dimensional case, that is derivatives only operate on the exponentials and we throw away all non-resonant terms, we finally arrive at the non-linear wave equation
\begin{eqnarray}
	\left(\frac{\partial^{2}}{\partial t^{2}}-c^{2}\nabla^{2}\right)\tilde{\textbf{E}}_{g}\!\left(\textbf{r},t\right)=4\xi\omega^{2}\textbf{G}_{3d}\tilde{E}_{1}\tilde{E}_{2}\tilde{E}^{*}_{3}e^{i\left(\textbf{k}_{4}\cdot\textbf{r}-\omega_{4}t\right)},
\end{eqnarray}
with the geometric factor 
\begin{eqnarray}
	\textbf{G}_{3d}\!\!\!\!&=&\!\!\!\!-\frac{1}{2}\left\{\ \left[\left(\frac{1}{2}\sin\beta_{3}-\cos\beta_{3}\right)\cos\left(\gamma_{1}+\gamma_{2}\right)\right.\right.\nonumber \\&&\ \ \ \ \ \  +\left(\frac{1}{8}\sin\gamma_{2}-\frac{1}{4}\cos\gamma_{2}\right)\cos\left(\gamma_{1}+\beta_{3}\right)\nonumber \\&&\ \ \ \ \ \ +\!\left.\left(\frac{1}{4}\sin\gamma_{1}+\frac{1}{8}\cos\gamma_{1}\right)\sin\left(\gamma_{2}+\beta_{3}\right)\right]\nonumber \\&&\ \ \ +\frac{7}{4}\left[\left(-\sin\beta_{3}-\frac{1}{2}\cos\beta_{3}\right)\sin\left(\gamma_{1}+\gamma_{2}\right)\right.\nonumber \\&&\ \ \ \ \ \ \ \  +\left(-\frac{1}{8}\cos\gamma_{2}-\frac{1}{4}\sin\gamma_{2}\right)\sin\left(\gamma_{1}+\beta_{3}\right)\nonumber \\&&\ \ \ \ \ \ \ \  +\!\!\left.\left.\left(\frac{1}{4}\cos\gamma_{1}-\frac{1}{8}\sin\gamma_{1}\right)\cos\left(\gamma_{2}+\beta_{3}\right)\right]\ \right\}\hat{\textbf{x}}\nonumber \\&&\!\!\!\! -\frac{1}{2}\left\{\ \left[\left(\frac{1}{2}\cos\beta_{3}-\sin\beta_{3}\right)\cos\left(\gamma_{1}+\gamma_{2}\right)\right.\right.\nonumber \\&&\ \ \ \ \ \ +\left(\frac{1}{8}\cos\gamma_{2}-\frac{1}{4}\sin\gamma_{2}\right)\cos\left(\gamma_{1}+\beta_{3}\right)\nonumber \\&&\ \ \ \ \ \  +\!\left.\left(-\frac{1}{4}\cos\gamma_{1}-\frac{1}{8}\sin\gamma_{1}\right)\sin\left(\gamma_{2}+\beta_{3}\right)\right]\nonumber \\&&\ \ \  +\frac{7}{4}\left[\left(\cos\beta_{3}+\frac{1}{2}\sin\beta_{3}\right)\sin\left(\gamma_{1}+\gamma_{2}\right)\right.\nonumber \\&&\ \ \ \ \ \ \ \  +\left(\frac{1}{8}\sin\gamma_{2}+\frac{1}{4}\cos\gamma_{2}\right)\sin\left(\gamma_{1}+\beta_{3}\right)\nonumber \\&&\ \ \ \ \ \ \ \  +\!\!\left.\left.\left(\frac{1}{4}\sin\gamma_{1}-\frac{1}{8}\cos\gamma_{1}\right)\cos\left(\gamma_{2}+\beta_{3}\right)\right]\ \right\}\hat{\textbf{y}}\nonumber \\&&\!\!\!\! +\left\{\ \left[\frac{1}{2}\left(\cos\beta_{3}+\sin\beta_{3}\right)\cos\left(\gamma_{1}+\gamma_{2}\right)\right.\right.\nonumber \\&&\ \ \ \ +\frac{1}{8}\left(\sin\gamma_{2}+\cos\gamma_{2}\right)\cos\left(\gamma_{1}+\beta_{3}\right)\nonumber \\ &&\ \ \ \ \!\left.+\frac{1}{8}\left(\cos\gamma_{1}-\sin\gamma_{1}\right)\sin\left(\gamma_{2}+\beta_{3}\right)\right]\nonumber \\&&\ +\frac{7}{4}\left[\frac{1}{2}\left(\sin\beta_{3}-\cos\beta_{3}\right)\sin\left(\gamma_{1}+\gamma_{2}\right)\right.\nonumber \\&&\ \ \ \ \ \ +\frac{1}{8}\left(\sin\gamma_{2}-\cos\gamma_{2}\right)\sin\left(\gamma_{1}+\beta_{3}\right)\nonumber \\&&\ \ \ \ \ \ +\!\!\left.\left.\frac{1}{8}\left(-\cos\gamma_{1}-\sin\gamma_{1}\right)\cos\left(\gamma_{2}+\beta_{3}\right)\right]\ \right\}\hat{\textbf{z}}.
\end{eqnarray}
The calculations are shown in Appendix B, and here we just verify that condition (5.9) is satisfied.
\subsection{Check of the result and a specific example}
Using Eqs.$\ $(5.16) and (5.20) we have
\begin{eqnarray}
	-2\textbf{G}_{3d}\cdot\hat{\textbf{k}}_{4}\!\!\!\!&=&\!\!\!\!\left(\frac{1}{2}\sin\beta_{3}-\cos\beta_{3}+\frac{1}{2}\cos\beta_{3}\right.\nonumber \\&&\!\!\!\!\left.-\sin\beta_{3}+\frac{1}{2}\cos\beta_{3}+\frac{1}{2}\sin\beta_{3}\right)\cos\left(\gamma_{1}+\gamma_{2}\right)\nonumber \\&&\!\!\!\! +\left(\frac{1}{8}\sin\gamma_{2}-\frac{1}{4}\cos\gamma_{2}+\frac{1}{8}\cos\gamma_{2}\right.\nonumber \\&&\ \left.-\frac{1}{4}\sin\gamma_{2}+\frac{1}{8}\sin\gamma_{2}+\frac{1}{8}\cos\gamma_{2}\right)\cos\left(\gamma_{1}+\beta_{3}\right)\nonumber \\&&\!\!\!\!+\left(\frac{1}{4}\sin\gamma_{1}+\frac{1}{8}\cos\gamma_{1}-\frac{1}{4}\cos\gamma_{1}\right.\nonumber \\&&\ \left.-\frac{1}{8}\sin\gamma_{1}+\frac{1}{8}\cos\gamma_{1}-\frac{1}{8}\sin\gamma_{1}\right)\sin\left(\gamma_{2}+\beta_{3}\right)\nonumber \\&&\!\!\!\! +\frac{7}{4}\left(-\sin\beta_{3}-\frac{1}{2}\cos\beta_{3}+\cos\beta_{3}\right.\nonumber \\&& \ \ \ \left.+\frac{1}{2}\sin\beta_{3}+\frac{1}{2}\sin\beta_{3}-\frac{1}{2}\cos\beta_{3}\right)\sin\left(\gamma_{1}+\gamma_{2}\right)\nonumber \\&&\!\!\!\! +\frac{7}{4}\left(-\frac{1}{8}\cos\gamma_{2}-\frac{1}{4}\sin\gamma_{2}+\frac{1}{8}\sin\gamma_{2}\right.\nonumber \\ && \ \ \ \left.+\frac{1}{4}\cos\gamma_{2}+\frac{1}{8}\sin\gamma_{2}-\frac{1}{8}\cos\gamma_{2}\right)\sin\left(\gamma_{1}+\beta_{3}\right)\nonumber \\&&\!\!\!\! +\frac{7}{4}\left(\frac{1}{4}\cos\gamma_{1}-\frac{1}{8}\sin\gamma_{1}+\frac{1}{4}\sin\gamma_{1}\right.\nonumber \\&& \ \ \ \left.-\frac{1}{8}\cos\gamma_{1}-\frac{1}{8}\cos\gamma_{1}-\frac{1}{8}\sin\gamma_{1}\right)\cos\left(\gamma_{2}+\beta_{3}\right)=0
\end{eqnarray}
since in each bracket the terms exactly adds up to zero. Hence we have checked the orthogonality condition (5.9) for this configuration.

As a specific example we may consider $\gamma_{1}\!=\!\gamma_{2}\!=\!\beta_{3}\!=0$, that is with wave one and two polarized in the z-direction and wave three along the x-direction. The geometric factor then becomes
\begin{eqnarray}
	\textbf{G}_{3d,0}=\frac{13}{32}\hat{\textbf{x}}-\frac{13}{64}\hat{\textbf{y}}+\frac{13}{32}\hat{\textbf{z}}.
\end{eqnarray}
%-----------------------Interaction models-------------------------------------------------
\chapter{Interaction models}
 
As discussed in Section 4.1, the interaction region needs to be specified in order to integrate out the generated fields from the derived wave equations. Any too complex model leads to great difficulties in the computation, and we will have to sacrifice some accuracy to be able to keep as much of the calculations analytical as possible. After all, we are just heading for a nice estimation of the scattered amount of photons, but we must of course watch out for getting too sloppy.

It seems reasonable, at least as a basic approach, to model the spatial form of the incoming laser pulses as parallel epipeds with quadratic cross sections, inside which the field amplitudes are constant. Although the pulses often are more of a Gaussian shape in reality, this should work quite well as long as the energies of them are correct. It could seem more natural to use a geometry of a cylinder instead of a parallel epiped, but our choice has the advantage of giving interaction volumes well adapted for carrying out the integrations in cartesian coordinates, something which simplifies the calculations a lot.
\section{Cubic interaction region}
\subsection{Formation}
Let us consider the case when the region of interaction takes the shape of a cube. This will for instance happen when $\phi_{1}\!=\pi/2$, $\ \phi_{2}\!=\!-\pi/2\ $ and $\phi_{3}=\pi$ in our two dimensional configuration discussed earlier, provided the beams collide at a single spot, which we choose to be the origin. When the beams pass through the origin, a cubic region containing all three waves will be formed, exist for a while, and ultimately disappear. The side of the cube will equal the side of the quadratic cross section of the beams, while the time of existence is determined by the pulse length. We will neglect any scattering during the construction and end phases, and only model the situation by an instantaneously created cube which after a while suddenly disappears again. Note that our three dimensional special case also may be treated by this model, due to the straight angles between the incoming waves.
\subsection{Generated electric field}
Following the theory outlined in Eqs.$\ $(4.14)-(4.18), the radiation zone solution to the wave equation (5.1) will be
\begin{eqnarray}
	\tilde{\textbf{E}}_{g}(\textbf{r},t)=\frac{\xi k^{2}_{4}}{\pi r}\textbf{G}\tilde{E}_{1}\tilde{E}_{2}\tilde{E}^{*}_{3}e^{i\left(k_{4}r-\omega_{4}t\right)}\int_{V'}e^{i{k}_{4}\left(\hat{\textbf{k}}_{4}-\hat{\textbf{r}}\right)\cdot\textbf{r}'}dV',
\end{eqnarray}
where $V'$ is the cubic region centred at the origin. Let us call the side length $b$. Consider the two dimensional case where we have
\begin{eqnarray}
\left(\hat{\textbf{k}}_{4}-\hat{\textbf{r}}\right)=\left(\cos\phi_{4}-\cos\phi\sin\theta\right)\hat{\textbf{x}}+\left(\sin\phi_{4}-\sin\phi\sin\theta\right)\hat{\textbf{y}}-\left(\cos\theta\right)\hat{\textbf{z}},
\end{eqnarray}
and thus for $\phi\neq\phi_{4}$ and $\theta\neq\pi/2$ the integral in (6.1) can be evaluated as
\begin{eqnarray}
\int_{V'}\!\!&&\!\!\!\!\!\!\!\!e^{i{k}_{4}\left(\hat{\textbf{k}}_{4}-\hat{\textbf{r}}\right)\cdot\textbf{r}'}dV'\nonumber \\ &&= \int^{\frac{b}{2}}_{-\frac{b}{2}}e^{ik_{4}\left(\cos\phi_{4}-\cos\phi\sin\theta\right)x'}dx'\int^{\frac{b}{2}}_{-\frac{b}{2}}e^{ik_{4}\left(\sin\phi_{4}-\sin\phi\sin\theta\right)y'}dy'\int^{\frac{b}{2}}_{-\frac{b}{2}}e^{-ik_{4}\left(\cos\theta\right) z'}dz'\nonumber \\ &&= \left[\frac{e^{ik_{4}\left(\cos\phi_{4}-\cos\phi\sin\theta\right)x'}}{ik_{4}\left(\cos\phi_{4}-\cos\phi\sin\theta\right)}\right]^{\frac{b}{2}}_{-\frac{b}{2}}\left[\frac{e^{ik_{4}\left(\sin\phi_{4}-\sin\phi\sin\theta\right)y'}}{ik_{4}\left(\sin\phi_{4}-\sin\phi\sin\theta\right)}\right]^{\frac{b}{2}}_{-\frac{b}{2}}\left[\frac{e^{-ik_{4}\left(\cos\theta\right) z'}}{-ik_{4}\left(\cos\theta\right)}\right]^{\frac{b}{2}}_{-\frac{b}{2}}\nonumber \\ &&= \frac{2\sin\left[k_{4}\frac{b}{2}\left(\cos\phi_{4}-\cos\phi\sin\theta\right)\right]}{k_{4}\left(\cos\phi_{4}-\cos\phi\sin\theta\right)}\frac{2\sin\left[k_{4}\frac{b}{2}\left(\sin\phi_{4}-\sin\phi\sin\theta\right)\right]}{k_{4}\left(\sin\phi_{4}-\sin\phi\sin\theta\right)}\nonumber \\ &&\ \ \ \ \!\times\frac{2\sin\left[k_{4}\frac{b}{2}\left(\cos\theta\right)\right]}{k_{4}\left(\cos\theta\right)}.
\end{eqnarray}
Plugging (6.3) into (6.1) gives us
\begin{eqnarray}
	\tilde{\textbf{E}}_{g}\!\left(\textbf{r},t\right)\!\!\!\!&=&\!\!\!\!\frac{8\xi}{k_{4}\pi r} \frac{\sin\left[k_{4}\frac{b}{2}\left(\cos\phi_{4}-\cos\phi\sin\theta\right)\right]}{\left(\cos\phi_{4}-\cos\phi\sin\theta\right)}\frac{\sin\left[k_{4}\frac{b}{2}\left(\sin\phi_{4}-\sin\phi\sin\theta\right)\right]}{\left(\sin\phi_{4}-\sin\phi\sin\theta\right)}\nonumber \\&&\!\!\!\!\times \frac{\sin\left[k_{4}\frac{b}{2}\left(\cos\theta\right)\right]}{\left(\cos\theta\right)}\textbf{G}_{2d}\tilde{E}_{1}\tilde{E}_{2}\tilde{E}^{*}_{3}e^{i\left(k_{4}r-\omega_{4}t\right)}.\left.\right.
\end{eqnarray}
The real generated electric field is hence given by
\begin{eqnarray}
	\textbf{E}_{g}\!\left(\textbf{r},t\right)\!\!\!\!&=&\!\!\!\!\frac{8\xi}{k_{4}\pi r} \frac{\sin\left[k_{4}\frac{b}{2}\left(\cos\phi_{4}-\cos\phi\sin\theta\right)\right]}{\left(\cos\phi_{4}-\cos\phi\sin\theta\right)}\frac{\sin\left[k_{4}\frac{b}{2}\left(\sin\phi_{4}-\sin\phi\sin\theta\right)\right]}{\left(\sin\phi_{4}-\sin\phi\sin\theta\right)}\nonumber \\&&\!\!\!\!\times \frac{\sin\left[k_{4}\frac{b}{2}\left(\cos\theta\right)\right]}{\left(\cos\theta\right)}\textbf{G}_{2d}|\tilde{E}_{1}||\tilde{E}_{2}||\tilde{E}_{3}|\cos\left(k_{4}r-\omega_{4}t+\delta\right),
\end{eqnarray}
where $|\tilde{E}_{j}|$ are the amplitudes of the incoming waves, and $\delta$ is the total phase from the complex amplitudes $\tilde{E}_{j}$.

In the resonant direction, $\phi=\phi_{4}$ and $\theta=\pi/2$, that is $\hat{\textbf{r}}=\hat{\textbf{k}}_{4}$, the integral simply becomes
\begin{eqnarray}
	\int_{V'}e^{i{k}_{4}\left(\hat{\textbf{k}}_{4}-\hat{\textbf{r}}\right)\cdot\textbf{r}'}dV'=\int_{V'}dV'=b^{3},
\end{eqnarray}
which leads to
\begin{eqnarray}
	\tilde{\textbf{E}}_{g,res}\!\left(\textbf{r},t\right)=\frac{\xi k^{2}_{4}b^{3}}{\pi r} \textbf{G}_{2d}\tilde{E}_{1}\tilde{E}_{2}\tilde{E}^{*}_{3}e^{i\left(k_{4}r-\omega_{4}t\right)},
\end{eqnarray}
and thereby we can express the real field as
\begin{eqnarray}
	\textbf{E}_{g,res}\!\left(\textbf{r},t\right)=\frac{\xi k^{2}_{4}b^{3}}{\pi r} \textbf{G}_{2d}|\tilde{E}_{1}||\tilde{E}_{2}||\tilde{E}_{3}|\cos\left(k_{4}r-\omega_{4}t+\delta\right).
\end{eqnarray}
Since $\frac{\sin ax}{x}\!\rightarrow a$ when $x\rightarrow 0$ it follows from Eq.$\ $(6.5) that
\begin{eqnarray}
	\lim_{\phi\rightarrow\phi_{4}\atop\theta\rightarrow\frac{\pi}{2}}\textbf{E}_{g}\!\left(\textbf{r},t\right)=\frac{\xi k^{2}_{4}b^{3}}{\pi r} \textbf{G}_{2d}|\tilde{E}_{1}||\tilde{E}_{2}||\tilde{E}_{3}|\cos\left(k_{4}r-\omega_{4}t+\delta\right)=\textbf{E}_{g,res}\!\left(\textbf{r},t\right),
\end{eqnarray}
as expected.

Plots of the generated field $\textbf{E}_{g}\!\left(\textbf{r},t\right)$ show a characteristic interference pattern with a maximum in the resonant direction. Such plots will be considered later on in the next chapter.

To summarize, the generated field may be written like
\begin{eqnarray}
	\textbf{E}_{g}\!\left(\textbf{r},t\right)={E}_{0}\!\left(r,\theta,\phi\right)\cos\left(k_{4}r-\omega_{4}t+\delta\right)\hat{\textbf{G}}_{2d},
\end{eqnarray}
where the direction dependent amplitude is given by
\begin{eqnarray}
	{E}_{0}\!\left(r,\theta,\phi\right)\!\!\!\!&=&\!\!\!\!\frac{8\xi}{k_{4}\pi r} \frac{\sin\left[k_{4}\frac{b}{2}\left(\cos\phi_{4}-\cos\phi\sin\theta\right)\right]}{\left(\cos\phi_{4}-\cos\phi\sin\theta\right)}\frac{\sin\left[k_{4}\frac{b}{2}\left(\sin\phi_{4}-\sin\phi\sin\theta\right)\right]}{\left(\sin\phi_{4}-\sin\phi\sin\theta\right)}\nonumber \\&&\!\!\!\!\times\frac{\sin\left[k_{4}\frac{b}{2}\left(\cos\theta\right)\right]}{\left(\cos\theta\right)}{G}_{2d}|\tilde{E}_{1}||\tilde{E}_{2}||\tilde{E}_{3}|,
\end{eqnarray}
with the resonant direction peaked value of
\begin{eqnarray}
	{E}_{0,res}\!\left(r,\theta,\phi\right)={E}_{0}\!\left(r,\frac{\pi}{2},\phi_{4}\right)=\frac{\xi k^{2}_{4}b^{3}}{\pi r}{G}_{2d}|\tilde{E}_{1}||\tilde{E}_{2}||\tilde{E}_{3}|.
\end{eqnarray}

Until now the equations have been expressed in Gaussian units, but the input of numerical values will get somewhat smoother if we translate our result into SI-units instead. This is done by following the prescriptions \cite{Jackson}
\begin{eqnarray}
	\left.\begin{array}{l}E\rightarrow\sqrt{4\pi\epsilon_{0}}\ E\\ 
	\ e\rightarrow\frac{1}{\sqrt{4\pi\epsilon_{0}}}\ e\end{array}\right\},
\end{eqnarray}
giving the SI-expressions
\begin{eqnarray}
	{E}_{0}\!\left(r,\theta,\phi\right)\!\!\!\!&=&\!\!\!\!\frac{1}{4\pi\epsilon_{0}}\frac{8\xi}{k_{4}\pi r} \frac{\sin\left[k_{4}\frac{b}{2}\left(\cos\phi_{4}-\cos\phi\sin\theta\right)\right]}{\left(\cos\phi_{4}-\cos\phi\sin\theta\right)}\frac{\sin\left[k_{4}\frac{b}{2}\left(\sin\phi_{4}-\sin\phi\sin\theta\right)\right]}{\left(\sin\phi_{4}-\sin\phi\sin\theta\right)}\nonumber \\&&\!\!\!\!\times\frac{\sin\left[k_{4}\frac{b}{2}\left(\cos\theta\right)\right]}{\left(\cos\theta\right)}{G}_{2d}|\tilde{E}_{1}||\tilde{E}_{2}||\tilde{E}_{3}|,
\end{eqnarray}
and
\begin{eqnarray}
	{E}_{0,res}\!\left(r\right)=\frac{1}{4\pi\epsilon_{0}}\frac{\xi k^{2}_{4}b^{3}}{\pi r}{G}_{2d}|\tilde{E}_{1}||\tilde{E}_{2}||\tilde{E}_{3}|.
\end{eqnarray}
\subsection{Generated intensity}
The omission of all non-resonant terms causes our calculated generated electric field to have the same polarization, independent of the traveling direction. This leads to the unpleasant property of $\textbf{E}_{g}\!\left(\textbf{r},t\right)$ and $\textbf{r}$ in general not being mutually orthogonal, as they should be for plane waves. However, for $\hat{\textbf{r}}\approx\hat{\textbf{k}}_{4}$ the orthogonality is almost fulfilled, and the plane wave intensity expression
\begin{eqnarray}
	I=\frac{1}{2}c\epsilon_{0}E^{2}_{0}
\end{eqnarray}
will be safe to use for all directions close to the resonant.
The generated intensity is accordingly
\begin{eqnarray}
	I_{g}\!\left(r,\theta,\phi\right)\!\!\!\!&=&\!\!\!\!\frac{1}{2}c\epsilon_{0}{E}^{2}_{0}\!\left(r,\theta,\phi\right)\nonumber \\ \!\!\!\!&=&\!\!\!\!\frac{16\xi^{2}}{\pi^{4}c^{2}\epsilon^{4}_{0}}\frac{1}{k^{2}_{4}r^{2}}\frac{\sin^{2}\left[k_{4}\frac{b}{2}\left(\cos\phi_{4}-\cos\phi\sin\theta\right)\right]}{\left(\cos\phi_{4}-\cos\phi\sin\theta\right)^{2}}\frac{\sin^{2}\left[k_{4}\frac{b}{2}\left(\sin\phi_{4}-\sin\phi\sin\theta\right)\right]}{\left(\sin\phi_{4}-\sin\phi\sin\theta\right)^{2}}\nonumber \\&&\!\!\!\!\times\frac{\sin^{2}\left[k_{4}\frac{b}{2}\left(\cos\theta\right)\right]}{\left(\cos\theta\right)^{2}}{G}^{2}_{2d}\left(\frac{1}{2}c\epsilon_{0}|\tilde{E}_{1}|^{2}\right)\left(\frac{1}{2}c\epsilon_{0}|\tilde{E}_{2}|^{2}\right)\left(\frac{1}{2}c\epsilon_{0}|\tilde{E}_{3}|^{2}\right)\nonumber \\ \!\!\!\!&=&\!\!\!\!\frac{16\xi^{2}}{\pi^{4}c^{2}\epsilon^{4}_{0}}\frac{1}{k^{2}_{4}r^{2}}\frac{\sin^{2}\left[k_{4}\frac{b}{2}\left(\cos\phi_{4}-\cos\phi\sin\theta\right)\right]}{\left(\cos\phi_{4}-\cos\phi\sin\theta\right)^{2}}\frac{\sin^{2}\left[k_{4}\frac{b}{2}\left(\sin\phi_{4}-\sin\phi\sin\theta\right)\right]}{\left(\sin\phi_{4}-\sin\phi\sin\theta\right)^{2}}\nonumber \\&&\!\!\!\!\times\frac{\sin^{2}\left[k_{4}\frac{b}{2}\left(\cos\theta\right)\right]}{\left(\cos\theta\right)^{2}}{G}^{2}_{2d}I_{1}I_{2}I_{3},
\end{eqnarray}
with the maximum at resonance
\begin{eqnarray}
	I_{g,res}\!\left(r\right)=\frac{\xi^{2}}{4\pi^{4}c^{2}\epsilon^{4}_{0}}\frac{k^{4}_{4}b^{6}}{r^{2}}{G}^{2}_{2d}I_{1}I_{2}I_{3},
\end{eqnarray}
where $I_{j}$ are the intensities of the incoming waves.

Following the same steps, a similar scattered intensity pattern is obtained for the three dimensional special case, with the peaked value in the $\hat{\textbf{k}}_4$-direction given by
\begin{eqnarray}
I_{g,res}\!\left(r\right)=\frac{\xi^{2}}{4\pi^{4}c^{2}\epsilon^{4}_{0}}\frac{k^{4}b^{6}}{r^{2}}{G}^{2}_{3d}I_{1}I_{2}I_{3}.
\end{eqnarray}
Note that the maximum intensity is determined by the power of the incoming waves $P_{j}=b^{2}I_{j}$, together with the frequency $\omega_{4}$ (or really for the three dimensional case, $\omega$) and the geometric factor ${G}^{2}_{2d}$ (${G}^{2}_{3d}$). However, the shape of the interference pattern depends on the size of the cubic interaction region, that is it depends on $b$. For large regions the generated intensity is more focused towards the resonant direction, while the fringes becomes more prominent for a small $b$.
\subsection{Two dimensions vs three dimensions}
Let us make a comparison between our two dimensional and three dimensional configurations. The expressions (6.18) and (6.19) reveals that the two cases should give scattered intensities of the same order, as long as ${G}^{2}_{2d}$ and ${G}^{2}_{3d}$ do not differ remarkably. As a quick check we compare ${G}^{2}_{2d,0}$ with ${G}^{2}_{3d,0}$. By taking $\phi_{1}\!=\pi/2$, $\ \phi_{2}\!=\!-\pi/2$, $\ \phi_{3}\!=\pi$ and $\phi_{4}\!=0$ it follows from Eq.$\ $(5.11) that in this case
\begin{eqnarray}
	{G}^{2}_{2d,0}=\left(1+\frac{1}{4}+\frac{1}{4}\right)^{2}=2.25,
\end{eqnarray}
and from Eq.$\ $(5.22) we have
\begin{eqnarray}
	{G}^{2}_{3d,0}=\left(\frac{13}{32}\right)^{2}+\left(\frac{13}{64}\right)^{2}+\left(\frac{13}{32}\right)^{2}\approx0.37.
\end{eqnarray}
Thus it seems like there really are no advantages with using the truly three dimensional configuration. Of course one should try to optimize the squared geometric factors by choosing other polarizations, but since they are built up by trigonometric functions there is in fact not much hope of finding something extraordinary. A glance at Eq.$\ $(5.8) gives us the insight that we should try to choose the $\gamma_{j}$ such that the terms including the factor 7/4 play a greater role. This can be done by taking $\gamma_{1}\!=0$ and $\gamma_{2}\!=\!\gamma_{3}\!=\pi/2$, and thereby increase the squared geometric factor to ${G}^{2}_{2d}\approx3.75$. Inspection of Eq.$\ $(5.20) inspires to choose $\gamma_{1}\!=0$ and $\gamma_{2}\!=\!\beta_{3}\!=\pi/2$, giving ${G}^{2}_{3d}\approx3.88$, so the two cases are still of the same order. Guided by this, we choose from this point to abandon all three dimensional configurations, and put focus on two dimensions only. If this does not seem well-reasoned, it is just because we have not considered the most interesting two dimensional case yet. Let us do that right now.
\section{Time dependent interaction region}
\subsection{Formation}
In the model discussed above, the side length of the cubic interaction region is, due to the chosen geometry, determined by the width $b$ of the incoming laser pulses. The peaked value of the generated intensity is directly proportional to $b^{6}$, i.e.$\!$ to the square of the interaction volume. However, since it is the power of the lasers which is limited, an increase of the beam cross section area must be accompanied by a corresponding reduction of intensity, keeping the peaked generated intensity unaffected. We may now ask ourselves if it is possible to find another configuration, with the same beam parameters used as previous, but with a considerably larger volume of interaction? If so, we expect the scattered intensity to be higher, at least as long as nothing dramatic at the same time happens with the geometric factor.

Since the pulse length often exceeds the pulse width with at least an order of magnitude, it would be nice if the size of the interaction region somehow could depend on it. This is indeed the case when the incoming pulses travel along the same line in space, two overlapping waves in one direction and a third in the opposite, eventually colliding head-on at some point (the origin). We choose to arrange our coordinate system such that this one dimensional configuration is described by $\theta_{1}\!=\!\theta_{2}\!=\!\theta_{3}\!=\pi/2$, $\ \phi_{1}\!=\!\phi_{3}\!=\pi$ and $\phi_{2}\!=0$. Moreover we assume that all three pulses are of the same kind, and in specific $k_{1}\!=\!k_{2}\!=\!k_{3}\equiv k$. The matching conditions (4.23)-(4.25) are then satisfied for $k_{4}\!=\!k$, $\ \theta_{4}\!=\pi/2$ and $\phi_{4}\!=0$.

Now try to imagine what this region of interaction will be like. Since wave one and three are overlapping, these can be pictured as a single parallel epiped with length $L$ and cross section area $b^{2}$, moving to the left along the x-axis. Consequently a similar parallel epiped traveling in the opposite direction then corresponds to wave two. As before, the interaction volume is defined as the part of space where all three incoming pulses are present, i.e.$\!$ the region where the two colliding parallel epipeds overlap. When passing through each other, the pulses give rise to an interaction volume of the form of a parallel epiped with width and hight $b$ in the y- and z-directions, but with a time dependent length $l\!\left(t\right)$ in the x-direction. Let $\tau\!=\!L/c$ denote the pulse duration, and the time $t\!=\!0$ to be fixed at the moment when the edges of the pulses first meet at the origin. The length $l\!\left(t\right)$ will thus increase linearly from $l\!\left(0\right)\!=\!0$ to its maximum $l\!\left(\tau/2\right)\!=\!L$, and then decrease linearly down to $l\!\left(\tau\right)\!=\!0$ again.
\subsection{Retarded time dependence}
From Eq.$\ $(4.15) it is clear that what we really are interested in is how the length depends on the retarded time $t_{R}$, and not on $t$. As we are about to measure the scattered intensity close to the $\phi_{4}$-direction out in the radiation zone, we investigate what $l\!\left(t_{R}\right)$ will look like when the observer is situated far away on the positive x-axis. At the very first moment when the incoming pulses overlap, some initially generated radiation will leave the region with the speed of light, going in the $\phi_{4}$-direction. But wave two is also traveling with the same speed in the same direction, so all radiation generated at later times at the edge of it will reach the observer exactly at the same moment as the portion initially generated. Since wave one and three both have the opposite velocity of the second, the front of wave two will be situated outside the interaction region as soon as it has crossed $x\!=\!L/2$, and hence no more radiation is generated from that edge. The first scattered radiation reaching the observer thus comes from a parallel epiped with length $L/2$ situated between $x\!=\!0$ and $x\!=\!L/2$. The scattering occurring at the end of pulse two is obviously the last portion arriving at the observer, and a similar mental picture tells us it comes from an equally sized parallel epiped, now located between $x\!=\!\!-L/2$ and $x\!=\!0$. In the same way it is easy to convince oneself that for any intermediate portion of radiation received by the observer, the size of the region it comes from is always the same, it is just moving continuously with half the speed of light in the negative x-direction. Hence we conclude that $l\!\left(t_{R}\right)\!=\!L/2$ during some time interval of length $\tau$, and zero otherwise.
\subsection{Generated intensity}
An expression for the generated intensity during this time interval is thus found in the same way as for the cubic interaction volume, provided we replace the limits of integration $x\!=\!\pm b/2$ with $x\!=\!\pm L/4$. The generated intensity for this configuration then becomes
\begin{eqnarray}
I_{g}\!\left(r,\theta,\phi\right)\!\!\!\!&=&\!\!\!\!\frac{16\xi^{2}}{\pi^{4}c^{2}\epsilon^{4}_{0}}\frac{1}{k^{2}_{4}r^{2}}\frac{\sin^{2}\left[k_{4}\frac{L}{4}\left(\cos\phi_{4}-\cos\phi\sin\theta\right)\right]}{\left(\cos\phi_{4}-\cos\phi\sin\theta\right)^{2}}\frac{\sin^{2}\left[k_{4}\frac{b}{2}\left(\sin\phi_{4}-\sin\phi\sin\theta\right)\right]}{\left(\sin\phi_{4}-\sin\phi\sin\theta\right)^{2}}\nonumber \\&&\!\!\!\!\times\frac{\sin^{2}\left[k_{4}\frac{b}{2}\left(\cos\theta\right)\right]}{\left(\cos\theta\right)^{2}}{G}^{2}_{2d}I_{1}I_{2}I_{3},
\end{eqnarray}
with the limit in the resonant direction
\begin{eqnarray}
I_{g,res}\!\left(r\right)=\frac{\xi^{2}}{4\pi^{4}c^{2}\epsilon^{4}_{0}}\frac{k^{4}_{4}b^{4}\left(\frac{L}{2}\right)^{2}}{r^{2}}{G}^{2}_{2d}I_{1}I_{2}I_{3}.
\end{eqnarray}
\subsection{Cube vs parallel epiped}
From Eqs.$\ $(6.18) and (6.23) we find the ratio between the peak intensities of the parallel epiped region and the cubic region to be
\begin{eqnarray}
	\frac{I_{g_{para},res}\!\left(r\right)}{I_{g_{cube},res}\!\left(r\right)}=\left(\frac{L}{2b}\right)^{2}\left(\frac{{G}_{2d_{para}}}{{G}_{2d_{cube}}}\right)^{2}.
\end{eqnarray}
For most laser pulses $L\!\gg\!b$, and hence the configuration giving rise to the parallel epiped interaction region seems to be the favourable. Let us just make a quick check to ensure nothing terrible happens with the geometric factor. For simplicity we choose the polarizations $\gamma_{1}\!=\!\gamma_{2}\!=\!\gamma_{3}\!=0$ (the optimization will be considered in the next chapter). Insertion of $\phi_{1}\!=\!\phi_{3}\!=\pi$, $\ \phi_{2}\!=\!\phi_{4}\!=0$ in Eq.$\ $(5.11) leads to the result
\begin{eqnarray}
	{G}^{2}_{2d,0_{para}}\!\!=4.
\end{eqnarray}
By comparing this to the geometric factor for a cubic interaction region (6.20), we see that in fact ${G}^{2}_{2d,0_{para}}\!\!>\!{G}^{2}_{2d,0_{cube}}$, so there is really nothing to worry about.
\subsection{Validity after practical adjustments}
As pointed out in Section 4.2, our choice of configuration brings about some practical problems concerning the detection of the scattered photons. This can, as discussed, be overcome by a shifting of some frequencies and angles, leaving us with a configuration slightly different from that investigated above. However, for the small changes we are going to make, it will be justified to continue using the results derived in this chapter.
%-------------------------Numerical examples----------------------------------------
\chapter{Numerical results}
The Vulcan laser at the Rutherford Appleton Laboratory, capable of delivering petawatt shots, is an interesting candidate to use for detection of photon-photon scattering. As is the planned X-ray free electron laser at DESY, based on a technique quite different from ordinary lasers. In this chapter we investigate and discuss the detection possibilities for both of these alternatives.
\section{Vulcan laser}
\subsection{Development of high power lasers}
Since the 1960s when the pulsed laser was invented, the peak power has increased by twelve orders of magnitude through a succession of leaps \cite{Joshi,Mourou,Pukhov}. After a few years the first free running lasers had been improved, first through the process of Q-switching and then by mode locking, such that the pulse duration could be cut down from nanoseconds to picoseconds for the same amount of energy. Consequently the power was pushed a million times from kilowatts to gigawatts. Although some minor amplifications were accomplished during the following two decades, the high intensities of the ultrashort pulses prevented any larger increase of the peak power. At the associated intensities of gigawatts per square centimeter, non-linear optical responses in the gain media imposed an intensity limit for the laser. This is due to that the material refractive index shows up a linear dependence on the intensity at such high values, and thus causes a deformation of the beam. Even if some progress were made in lowering the pulse duration down to femtoseconds, no increase in power could at the same time be achieved.

Thanks to the technique of chirped pulse amplification (CPA) demonstrated in the mid 1980s, this problem was mainly solved. In CPA one basically takes an ultrashort pulse, stretches it $10^{3}$--$10^{5}$ times and thereby reduces the intensity, amplifies it, and finally recompresses it back to its original shape again. By doing the amplification during the stretched (chirped) intermediate state of low intensity, one gets around the problem with non-linear optical effects. The stretching can be done by passing the pulse through a dispersive delay line, and thereby separate the spectral components of the wave packet. For ultrashort pulses the bandwidth gets large, and hence it is important that the amplification is the same over a large range of wavelengths. The compression, performed by a conjugate delay line, must be perfectly matched to the stretching in order to restore the shape of the pulse.

The CPA lead to the development of terawatt lasers, and in last years several petawatt upgrade projects have started. We will put our focus on the Vulcan laser at the Rutherford Appleton Laboratory, which fired its first petawatt shot as recent as 5th October 2004.
\subsection{Laser capacity}
The Vulcan Laser \cite{Vulcan} is capable of delivering $423\!~\mathrm{J}$ in a single $410\!~\mathrm{fs}$ pulse at wavelength $1054\!~\mathrm{nm}$, generating a power of $1.03\!~\mathrm{PW}$. The beam can be focused to an intensity of $1.06\!\cdot\!10^{25}\!~\mathrm{Wm}^{\!-2}$ on target, corresponding to a beam cross section area of $0.97\!\cdot\!10^{\!-10}\!~\mathrm{m}^{\!-2}$. The repetition rate takes on the moderate value of one shot per hour.

In reality the shape of the pulse is of course more complex than a parallel epiped of constant intensity, but in order to get something out in a fairly simple way we will stick to the model in Chapter 6. Anyway, we are not aiming for any detailed calculation of the scattered number of photons, but only a nice estimation.
\subsection{Scattered intensity distribution}
To carry out the experiment, three incoming beams are needed. Thus we have to split the petawatt pulse into three parts, each with a third of the original energy. This may for instance be done with the help of two beam splitters, but the exact procedure will be left for the experimentalists. We hence start out with three beams, each of the same spatial shape and wavelength as the original, but with energy, power and intensity cut down by a factor one third.

In Chapter 6 the most promising geometric configuration was found to be described by the parameters $\theta_{1}\!=\!\theta_{2}\!=\!\theta_{3}\!=\!\theta_{4}\!=\pi/2$, $\ \phi_{1}\!=\!\phi_{3}\!=\pi$, $\ \phi_{2}\!=\!\phi_{4}\!=0$, $\ k_{1}\!=\!k_{2}\!=\!k_{3}\!=\!k_{4}$. On arranging the incoming beams like this, we expect a generated intensity as in the expressions (6.22) and (6.23). The pulse length is now given by $L\!=\!c\tau\!\approx\!120\!~\mathrm{\mu m}$, while the width is from the cross section area found to be $b\!\approx\!9.8\!~\mathrm{\mu m}$. We also have the wave vector $k_{4}\!=2\pi/\lambda\approx6.0\!\cdot\!10^{6}\!~\mathrm{m}^{\!-1}$ and the intensities of the incoming waves $I_{1}\!=\!I_{2}\!=\!I_{3}\!=\!I/3\approx0.35\!\cdot\!10^{25}\!~\mathrm{Wm}^{\!-2}$ (where $\lambda$ and $I$ are the wavelength and intensity of the original beam). We are now ready to substitute these numerical values into Eqs.$\ $(6.22) and (6.23), and find out what the generated intensity will look like at different spots. To get a picture of the angular distribution of scattered radiation, the ratio
\begin{eqnarray}
	\frac{I_{g}\!\left(r,\theta,\phi\right)}{I_{g,res}\!\left(r\right)}\!\!\!\!&=&\!\!\!\!\frac{256}{k^{6}_{4}b^{4}L^{2}}\frac{\sin^{2}\left[k_{4}\frac{L}{4}\left(1-\cos\phi\sin\theta\right)\right]}{\left(1-\cos\phi\sin\theta\right)^{2}}\frac{\sin^{2}\left[k_{4}\frac{b}{2}\left(\sin\phi\sin\theta\right)\right]}{\left(\sin\phi\sin\theta\right)^{2}}\nonumber \\&&\!\!\!\!\times\frac{\sin^{2}\left[k_{4}\frac{b}{2}\left(\cos\theta\right)\right]}{\left(\cos\theta\right)^{2}}
\end{eqnarray}
(where we have substituted $\phi_{4}\!=0$) is in Figure 7.1 plotted as a function of $\theta$ and $\phi$, for the current values of the parameters $k_{4}$, $b$ and $L$. In the limit of fixing $\theta=\pi/2$, the ratio as a function of the deviation $\phi$ from the resonant direction becomes
\begin{eqnarray}
	\left.\frac{I_{g}\!\left(r,\theta,\phi\right)}{I_{g,res}\!\left(r\right)}\right|_{\theta=\frac{\pi}{2}}=\frac{64}{k^{4}_{4}b^{2}L^{2}}\frac{\sin^{2}\left[k_{4}\frac{L}{4}\left(1-\cos\phi\right)\right]}{\left(1-\cos\phi\right)^{2}}\frac{\sin^{2}\left[k_{4}\frac{b}{2}\left(\sin\phi\right)\right]}{\left(\sin\phi\right)^{2}},
\end{eqnarray}
%-------------------------------------------------------------------------------------------------
\begin{figure}
	\centering
		\includegraphics[width=0.8\textwidth]{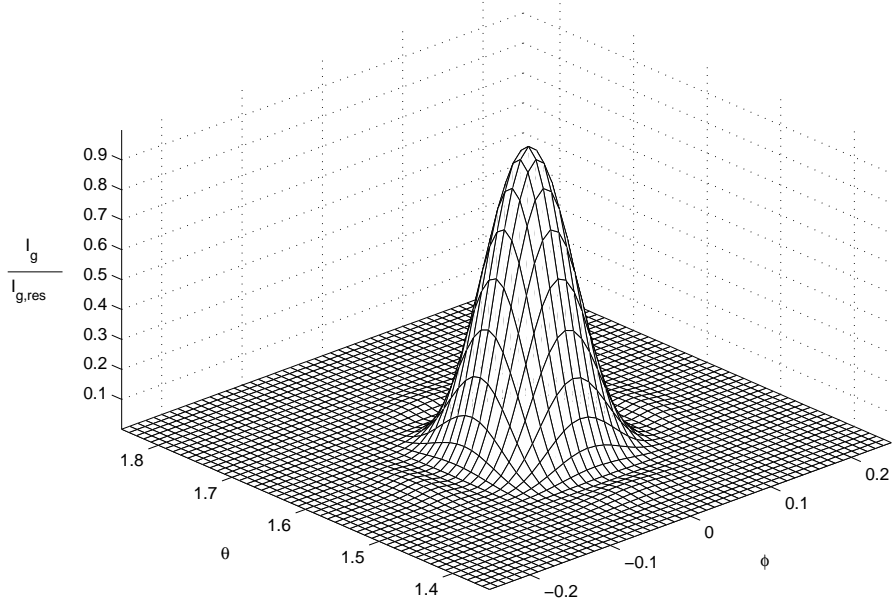}
	\caption{\emph{Angular distribution of the scattered intensity, Eq.$\ $(7.1), for the Vulcan laser}.}
	\label{fig:713graf1}
\end{figure}
%-------------------------------------------------------------------------------------------------
and is plotted in Figure 7.2, where the interference pattern might be seen somewhat clearer. Fixing $\phi=0$ instead, gives a dependence on the deviation $\vartheta\equiv\left(\theta-\pi/2\right)$ from resonance as
\begin{eqnarray}
	\left.\frac{I_{g}\!\left(r,\theta,\phi\right)}{I_{g,res}\!\left(r\right)}\right|_{\phi=0}=\frac{64}{k^{4}_{4}b^{2}L^{2}}\frac{\sin^{2}\left[k_{4}\frac{L}{4}\left(1-\cos\vartheta\right)\right]}{\left(1-\cos\vartheta\right)^{2}}\frac{\sin^{2}\left[k_{4}\frac{b}{2}\left(\sin\vartheta\right)\right]}{\left(\sin\vartheta\right)^{2}},
\end{eqnarray}
i.e.$\!$ of exactly the same form as the ratio (7.2), as expected from the symmetry of our chosen geometric configuration.
%--------------------------------------------------------------------
\begin{figure}[tbf]
	\centering
	\includegraphics[width=0.8\textwidth]{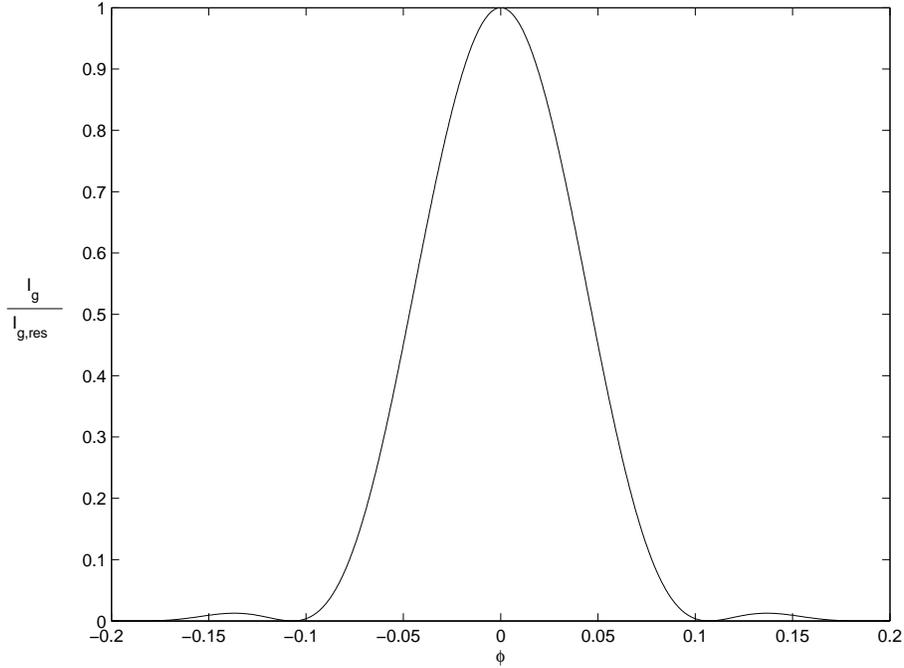}	
	\caption{\emph{Scattered intensity distribution as a function of the deviation $\phi$ from resonance, Eq.$\ $(7.2), for the Vulcan laser}.}
	\label{fig:713graf2}
\end{figure}
%--------------------------------------------------------------------
\subsection{Number of scattered photons}
The total generated power $P_{g}$ is found from integrating the intensity over a spherical shell (with radius $r$) centred at the origin, that is
\begin{eqnarray}
	P_{g}=\int^{2\pi}_{0}\int^{\pi}_{0}I_{g}\!\left(r,\theta,\phi\right)r^{2}\sin\theta ~d\theta ~d\phi=I_{g,res}\!\left(r\right)r^{2}\alpha^{2},
\end{eqnarray}
where
\begin{eqnarray}
	\alpha^{2}\!\!\!\!&\equiv&\!\!\!\!\int^{2\pi}_{0}\int^{\pi}_{0}\frac{I_{g}\!\left(r,\theta,\phi\right)}{I_{g,res}\!\left(r\right)}\sin\theta ~d\theta ~d\phi\nonumber \\ \!\!\!\!&=&\!\!\!\!\frac{256}{k^{6}_{4}b^{4}L^{2}}\int^{2\pi}_{0}\int^{\pi}_{0}\frac{\sin^{2}\left[k_{4}\frac{L}{4}\left(1-\cos\phi\sin\theta\right)\right]}{\left(1-\cos\phi\sin\theta\right)^{2}}\frac{\sin^{2}\left[k_{4}\frac{b}{2}\left(\sin\phi\sin\theta\right)\right]}{\left(\sin\phi\sin\theta\right)^{2}}\nonumber \\ &&\!\!\!\!\times\frac{\sin^{2}\left[k_{4}\frac{b}{2}\left(\cos\theta\right)\right]}{\left(\cos\theta\right)^{2}}\sin\theta ~d\theta ~d\phi.
\end{eqnarray}
The form Eq.$\ $(7.4) is written on tells us that about the same power would be generated by an intensity profile of constant magnitude $I_{g,res}$ spanning a square with the side of an angle $\alpha$.

A numerical solution of Eq.$\ $(7.5) gives the value $\alpha\!=\!0.096$ for our case of interest. The central interference peak in Figure 7.1 is definitely the most dominating, and integrating over just that region does not have any noticeable effects on $\alpha$. It is clear that all but a negligible part of the power is generated into the central peak directions. From Figures 7.1 and 7.2 we see that $\alpha$ is around half of the peak width, something which seems very reasonable. If we had not done the integration, this is probably what we would have guessed just from inspection of the plots.

Putting Eqs.$\ $(6.23) and (7.4) together yields
\begin{eqnarray}
	P_{g}=\frac{\xi^{2}}{16\pi^{4}c^{2}\epsilon^{4}_{0}}k^{4}_{4}b^{4}{L}^{2}\alpha^{2}{G}^{2}_{2d}I_{1}I_{2}I_{3}.
\end{eqnarray}
The time of interaction equals the pulse duration $\tau\!=\!L/c$, so the total radiated energy is given by
\begin{eqnarray}
	U_{g}=P_{g}\tau=\frac{\xi^{2}}{16\pi^{4}c^{3}\epsilon^{4}_{0}}k^{4}_{4}b^{4}{L}^{3}\alpha^{2}{G}^{2}_{2d}I_{1}I_{2}I_{3}.
\end{eqnarray}
Each generated photon carries an energy of $U_{ph}\!\!=\!\hbar ck_{4}$, and hence the estimated number of scattered photons per shot becomes
\begin{eqnarray}
	N=\frac{U_{g}}{U_{ph}}=\frac{\xi^{2}}{16\pi^{4}\hbar c^{4}\epsilon^{4}_{0}}k^{3}_{4}b^{4}{L}^{3}\alpha^{2}{G}^{2}_{2d}I_{1}I_{2}I_{3}.
\end{eqnarray}
\subsection{Optimization of the geometric factor}
All there is left to find is a value for the geometric factor $\textbf{G}_{2d}$ given by Eq.$\ $(5.8). Due to the one dimensional structure of the configuration, the optimization is in this case simple to perform. Since all waves travel along the x-axis, we can always arrange our coordinate system such that the optimal $\textbf{G}_{2d}$ lies along the z-axis. In fact, due to the symmetry of the configuration there must be geometric factors of exactly the same magnitudes pointing in every direction perpendicular to the x-axis. Anyway, we choose to maximize the magnitude of the z-component, which with the current angles imposed takes the form
\begin{eqnarray}
	\left(\textbf{G}_{2d}\right)_{z}\!\!\!\!&=&\!\!\!\!\frac{7}{4}\sin\gamma_{3}\sin\left(\gamma_{1}+\gamma_{2}\right)+\cos\gamma_{3}\cos\left(\gamma_{1}+\gamma_{2}\right)\nonumber\\&&\!\!\!\! +\frac{7}{4}\sin\gamma_{1}\sin\left(\gamma_{2}+\gamma_{3}\right)+\cos\gamma_{1}\cos\left(\gamma_{2}+\gamma_{3}\right)\nonumber\\ \!\!\!\!&=&\!\!\!\!\frac{3}{4}\sin\gamma_{3}\sin\left(\gamma_{1}+\gamma_{2}\right)+\frac{3}{4}\sin\gamma_{1}\sin\left(\gamma_{2}+\gamma_{3}\right)\nonumber\\&&\!\!\!\!+\cos\left(\gamma_{1}+\gamma_{2}-\gamma_{3}\right)+\cos\left(\gamma_{2}+\gamma_{3}-\gamma_{1}\right).
\end{eqnarray}
An upper limit is consequently $\left(\textbf{G}_{2d}\right)_{z}\!\!=\!3.5$, which in fact can be achieved by taking $\gamma_{1}\!=\!\gamma_{3}\!=\pi/2$ and $\gamma_{2}\!=0$. It is easy to check that the x- and y-components in this case vanish, as they should. Hence we conclude that we have found an optimal choice of polarization angles.
\subsection{Final estimations and practical adjustments}
Finally we plug in all the numerical values into Eq.$\ $(7.8), and find the estimated number of scattered photons per shot to be $N\!\approx22$, an amount which is definitely detectable.

It may be pointed out that in the earlier discussed case with a cubic interaction region, the width of the central peak becomes almost the same as above, and hence $\alpha$ does not change very much. Using Eq.$\ $(6.24) we conclude that the number of scattered photons is reduced by at least two orders of magnitude compared to the above result.

For practical reasons, the choice of $\phi_{1}\!=\!\phi_{3}\!=\pi$ and $\phi_{2}\!=\!\phi_{4}\!=0$ have to be slightly altered. We want the detector to be able to pick up all of the intensity generated into the central interference peak, i.e.$\!$ we have to make sure no other beams are present in that region. Following the discussion in Section 4.2, such a modified configuration is given by $\phi_{1}\!=\pi+2\delta$, $\ \phi_{2}\!=\delta$, $\ \phi_{3}\!=\pi+\delta$, $\ \phi_{4}\!=0$, $\ k_{1}\!=\!k_{4}\!\equiv k$, $\ k_{2}\!=\!k_{3}\!=\!k\cos\delta$, where $\delta$ has to be chosen as half the peak width. In this case it means $\delta\!=\!0.1$. Thus we must shift the frequencies of beam two and three by a factor of $\cos \ \!\!0.1$, giving us the new wavelengths $\lambda_{2}\!=\!\lambda_{3}\!=\!\lambda/\!\cos \ \!\!0.1
\!\approx\!1059\!~\mathrm{nm}$ and wave vectors $k_{2}\!=\!k_{3}\!=\!5.9\!\cdot\!10^{6}\!~\mathrm{m}^{\!-1}$. As before, we leave the technicalities of how to perform this shift to the experimentalists.

Since such small modifications are done, we will continue to use the same interaction model, leading to Eq.$\ $(7.8). All parameters are the same as earlier, except for ${G}^{2}_{2d}$ which has to be recalculated for the new angles. Putting $\phi_{1}\!=\pi+0.2$, $\ \phi_{2}\!=0.1$, $\ \phi_{3}\!=\pi+0.1$ and $\ \phi_{4}\!=0$ into Eq.$\ $(5.8), with $\gamma_{1}\!=\!\gamma_{3}\!=\pi/2$ and $\gamma_{2}\!=0$, gives $\textbf{G}_{2d}\!\approx\!3.47\ \!\hat{\textbf{z}}$. Actually this brings the number of scattered photons per shot down a bit, but the rounded off value will still be $N\!\approx22$.
\subsection{Suggestion of an experiment}
It is time to summarize the situation, and suggest a single experiment in which photon-photon scattering may be detected. The recipe goes as follows: Take the petawatt pulse of wavelength $1054~\mathrm{nm}$ produced by the Vulcan laser, and split it into three equal parts. Shift the wavelengths of the second and third pulses to $1059~\mathrm{nm}$. Let the pulses enter the x-y-plane at the angles $\phi_{1}\!=\pi+0.2$, $\ \phi_{2}\!=0.1$ and $\phi_{3}\!=\pi+0.1$, and with $\gamma_{1}\!=\!\gamma_{3}\!=\pi/2$ and $\gamma_{2}\!=0$, corresponding to polarization of pulse one and three in the x-y-plane while pulse two is polarized in the z-direction. Also synchronize the pulses to reach the origin at the same moment. On the x-axis, place a detector which at least spans the intervals $-0.1\leq\left(\theta\!-\!\pi/2\right)\leq0.1$ and $-0.1\leq\phi\leq0.1$.

For each shot, we expect several scattered photons to be detected. (More precisely our model predicts the average number to be 22, but we have to keep in mind that a lot of approximations have been made. What is important is that the amount of generated photons definitely is detectable.)
\section{X-ray free electron laser}
\subsection{Description}
Another interesting candidate to use for detection of photon-photon scattering is the XFEL (X-ray Free Electron Laser). Such a laser is planned as a part of the design of the superconducting electron-positron collider TESLA (TeV-Energy Superconducting Linear Accelerator), at DESY (Deutsches Electronen-Synchrotron), Hamburg \cite{XFEL}. The facility is scheduled to take up operation in 2012. TESLA mainly consists of a $33\!~\mathrm{km}$ long tunnel containing two linear accelerators, one for the electrons and one for the positrons, meeting at a collision site in the middle. The collider is capable of initially $500\!~\mathrm{GeV}$ total energy, but is extendable to $800\!~\mathrm{GeV}$. The first low energy part of the electron accelerator can also provide an electron beam for the XFEL.

The XFEL is based on the principle of SASE (Self-Amplified Spontaneous Emission), which can be summarized as follows: The high energy electrons coming from the accelerator is guided into an undulator, an arrangement of magnets forcing the electrons to follow a winding periodic slalom course. Due to these accelerations the electrons emit concentrated X-ray flashes, which will overtake the electrons ahead of them. The passing by of the photons causes the electrons to organize themselves into thin layers, and at the end of the undulator this structure is fully developed. The electrons in a certain layer will now emit radiation in a synchronized way, and hence produce high-intensity ultra-short X-ray pulses with laser light properties. This is quite different from how a traditional laser works, and one of the main advantages is that the wavelength easily may be altered by adjusting the electron acceleration.

The XFEL attracts scientists from many different disciplines including physics, chemistry, material science and structural biology. Here we will investigate how well it can be used for detection of photon-photon scattering.
\subsection{Laser capacity}
According to the predicted photon beam parameters, the XFEL will be able to produce $24\!~\mathrm{GW}$ pulses at wavelength $0.1\!~\mathrm{nm}$, with $100\!~\mathrm{fs}$ duration and a pulse width of $110\!~\mathrm{\mu m}$. The repetition rate will be as high as $30\ \!000$ pulses per second. However, the expected coherence time is only $0.3\!~\mathrm{fs}$, and hence it should be more accurate to model each pulse as a stream of subpulses, each with a length equal to the coherence length of $90\!~\mathrm{nm}$. The repetition rate may then be expressed as $10^{7}$ subpulses per second. 
\subsection{Estimated number of scattered photons}
Guided by the above properties, and on following the same method as for the Vulcan laser, we model the incoming subpulses as parallel epipeds with parameter values given by $L\!=\!90\!~\mathrm{nm}$, $b\!=\!110\!~\mathrm{\mu m}$, $k_{4}\!=6.3\!\cdot\!10^{10}\!~\mathrm{m}^{\!-1}$ and $I_{1}\!=\!I_{2}\!=\!I_{3}\!=0.66\!\cdot\!10^{18}\!~\mathrm{Wm}^{\!-2}$. Figure 7.3, a plot of the intensity distribution (7.2), guides us to assign the value $\alpha=\!10^{-6}$. Using the same geometrical factor as in the last section, Eq.$\ $(7.8) finally gives us $N\!\approx\!10^{\!-22}$. Taking the repetition rate into account, about $10^{\!-15}$ photons will be generated per second, and we may have to wait more than 30 million years to detect the first scattered photon. It looks like the high frequency and repetition rate of the XFEL are unable to compensate for its low power.
%-----------------------------------------------------------------------------------------------------
\begin{figure}
	\centering
	\includegraphics[width=0.8\textwidth]{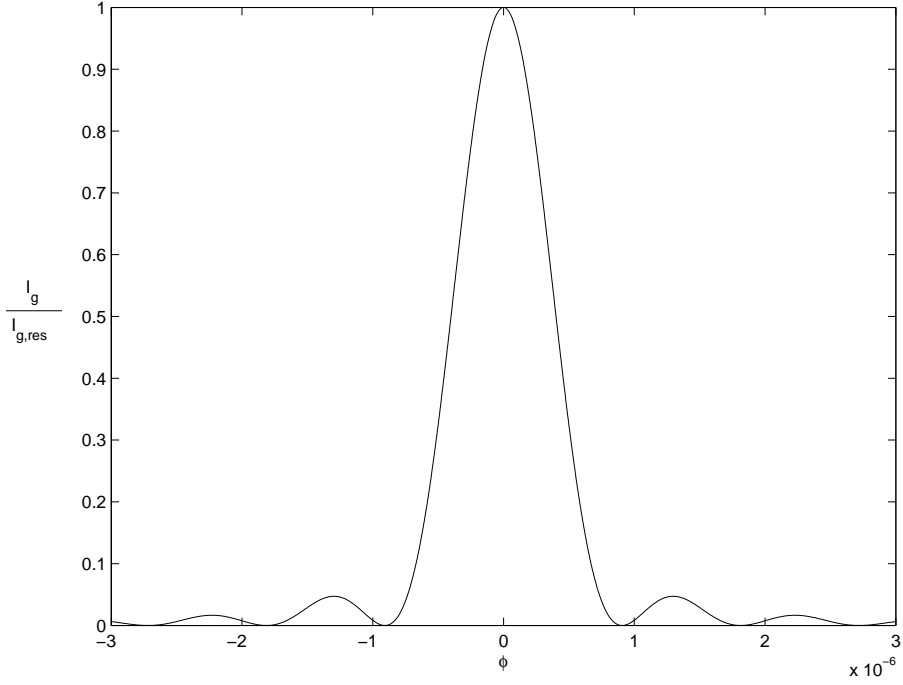}
	\caption{\emph{Scattered intensity distribution as a function of the deviation $\phi$ from resonance, Eq.$\ $(7.2), for the original beams of the XFEL}.}
	\label{fig:723graf1}
\end{figure}
%-----------------------------------------------------------------------------------------------------

The configuration used here is however not the optimal. Since we are dealing with pulses which are wider than long, we should instead consider the case with a cubic interaction region, as in Section 6.1. By performing the substitution $L/2\rightarrow b$ in Eqs.$\ $(7.1), (7.2) and (7.5) we get the correct formulas for the intensity distribution and $\alpha^{2}$ in this case. In fact a plot of the intensity distribution as a function of the deviation from the resonant direction will look almost exactly like Figure 7.3, and we may continue to use $\alpha=\!10^{-6}$. Hence we can estimate the scattering number for the optimal configuration from Eq.$\ $(6.24). Inserting ${G}_{2d_{cube}}^{2}\!\!=3.75$, and with the other parameter values as above, we see that the scattering number is increased by a factor $1.8\!\cdot\!10^{6}$ relative to the first result. Anyway, the scattering rate is still far to low, making it practically impossible to detect photon-photon scattering through this experiment.

Despite the negative results above, there is still some hope left. It may be possible, with future technology, to focus the width of the X-rays down to the theoretical diffraction limit of the order of one wavelength, i.e.$\!$ $b\!\sim\!\lambda$ \cite{Ringwald}. As a realistic value of the pulse width achievable in 2012 we take $b=1.1\!~\mathrm{nm}$, a compression of the original value by $10^{5}$ times. Correspondingly the intensity of each beam is increased by a factor $10^{10}$. From Figure 7.4, where Eq.$\ $(7.2) is plotted for these new values, we estimate $\alpha=0.05$. The scattered number of photons per subpulse predicted by our model, given by Eq.$\ $(7.8), becomes $N\!\approx0.003$. Hence we arrive at the incredible result of $30\ \!000$ scattered photons per second.
\begin{figure}
	\centering
	\includegraphics[width=0.8\textwidth]{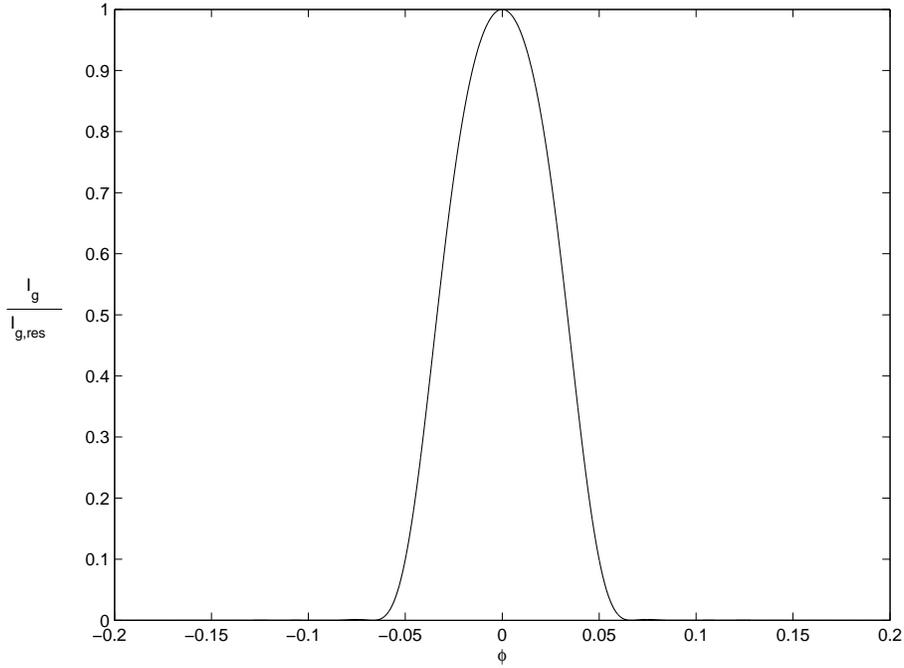}
	\caption{\emph{Scattered intensity distribution as a function of the deviation $\phi$ from resonance, Eq.$\ $(7.2), for the focused beams of the XFEL}.}
	\label{fig:focusedxfel}
\end{figure}
\section{Discussion}
\subsection{Improvements}
We cannot expect our simple model to predict any exact scattering numbers, but only estimate the order of magnitudes. In order to produce more precise numbers and intensity distributions to compare with experimental outcomes, we have to make some improvements. The modeling of the laser pulses as parallel epipeds of constant field strength serves our purpose of keeping the calculations analytical and simple well, but a more realistic approach would be to use pulses with a Gaussian-like distributed electric field strength, both in the radial direction as well as in the direction of propagation. This of course complicates the shape of the interaction region, and we have to rely on numerical calculations. No Gaussian models will be considered here, but such calculations would be an interesting follow up to this work. Moreover, we have not really taken the effects of the bandwidth into detailed account. This may especially be important for the XFEL, due to its short coherence time, and a deeper investigation would possibly lead to some modifications.

The improvements may however be overshadowed by practical errors, such as difficulties to tune the laser pulses exactly as we want them, or problems with creating a sufficiently good vacuum. There is really no meaning in calculating very precise scattering rates, as long as the experimental uncertainty widely exceeds the accuracy of the theory. In the first place we are really just asking ourselves whether some scattering can be measured at all, and the model used in this work should definitely be good enough to answer such a question. Anyway, what could be compared with experimental results is how the scattering number depends on the polarizations of the incoming beams. As long as the pulses in reality are symmetric around the axis of propagation, the ratios between the number of scattered photons for different sets of polarizations are independent of how the pulses are modeled. A sequence of experiments, each with different polarizations but otherwise the same, would hence work as a test of the results derived in Chapter 5.

\subsection{Other experiments}
Although this thesis has been pervaded by adjustments to suit the Vulcan laser and the XFEL well, there should be no big problems to modify the calculations to fit other situations as well. For instance it may be practically favourable to use different kinds of lasers, with different beam parameters, in the same experiment. We may not be able to follow exactly the same procedure of finding resonant configurations, and it may not be possible to model the interaction region in exactly the same way as here, but the basic approach will be the same, and the modifications needed should be quite easy to figure out.
%-----------------------------------------------------------------------------
\appendix
\chapter{The two dimensional geometric factor}
Here follows the detailed calculation of Eq.$\ $(5.7). We calculate the right hand side of Eq.$\ $(3.38) step by step, keeping only the resonant terms:\\
\\
\begin{eqnarray}
\left(E^{2}\!-B^{2}\right)\!\!\!\!&=&\!\!\!\!\left(\textbf{E}_{1}+\textbf{E}_{2}+\textbf{E}_{3}\right)^{2}-\left(\textbf{B}_{1}+\textbf{B}_{2}+\textbf{B}_{3}\right)^{2}\nonumber \\ \!\!\!\!&=&\!\!\!\! \left(E_{1}\sin\gamma_{1}\sin\phi_{1}+E_{2}\sin\gamma_{2}\sin\phi_{2}+E_{3}\sin\gamma_{3}\sin\phi_{3}\right)^{2}\nonumber \\ &&\!\!\!\!+ \left(E_{1}\sin\gamma_{1}\cos\phi_{1}+E_{2}\sin\gamma_{2}\cos\phi_{2}+E_{3}\sin\gamma_{3}\cos\phi_{3}\right)^{2}\nonumber \\ &&\!\!\!\!+\left(E_{1}\cos\gamma_{1}+E_{2}\cos\gamma_{2}+E_{3}\cos\gamma_{3}\right)^{2}\nonumber \\ &&\!\!\!\!-\left(E_{1}\cos\gamma_{1}\sin\phi_{1}+E_{2}\cos\gamma_{2}\sin\phi_{2}+E_{3}\cos\gamma_{3}\sin\phi_{3}\right)^{2}\nonumber \\ &&\!\!\!\!-\left(E_{1}\cos\gamma_{1}\cos\phi_{1}+E_{2}\cos\gamma_{2}\cos\phi_{2}+E_{3}\cos\gamma_{3}\cos\phi_{3}\right)^{2}\nonumber \\ &&\!\!\!\!-\left(E_{1}\sin\gamma_{1}+E_{2}\sin\gamma_{2}+E_{3}\sin\gamma_{3}\right)^{2}\nonumber \\ \!\!\!\!&=&\!\!\!\!E_{1}^{2}+E_{2}^{2}+E_{3}^{2}\nonumber \\ &&\!\!\!\!+2E_{1}E_{2}\left(\sin\gamma_{1}\sin\gamma_{2}\cos\left(\phi_{1}-\phi_{2}\right)+\cos\gamma_{1}\cos\gamma_{2}\right)\nonumber \\ &&\!\!\!\!+2E_{1}E_{3}\left(\sin\gamma_{1}\sin\gamma_{3}\cos\left(\phi_{1}-\phi_{3}\right)+\cos\gamma_{1}\cos\gamma_{3}\right)\nonumber \\ &&\!\!\!\!+2E_{2}E_{3}\left(\sin\gamma_{2}\sin\gamma_{3}\cos\left(\phi_{2}-\phi_{3}\right)+\cos\gamma_{2}\cos\gamma_{3}\right)\nonumber \\ &&\!\!\!\!-E_{1}^{2}-E_{2}^{2}-E_{3}^{2}\nonumber \\ &&\!\!\!\!-2E_{1}E_{2}\left(\cos\gamma_{1}\cos\gamma_{2}\cos\left(\phi_{1}-\phi_{2}\right)+\sin\gamma_{1}\sin\gamma_{2}\right)\nonumber \\ &&\!\!\!\!-2E_{1}E_{3}\left(\cos\gamma_{1}\cos\gamma_{3}\cos\left(\phi_{1}-\phi_{3}\right)+\sin\gamma_{1}\sin\gamma_{3}\right)\nonumber \\ &&\!\!\!\!-2E_{2}E_{3}\left(\cos\gamma_{2}\cos\gamma_{3}\cos\left(\phi_{2}-\phi_{3}\right)+\sin\gamma_{2}\sin\gamma_{3}\right)\nonumber \\ \!\!\!\!&=&\!\!\!\!4E_{1}E_{2}\cos\left(\gamma_{1}+\gamma_{2}\right)\sin^{2}\left(\frac{\phi_{1}-\phi_{2}}{2}\right)\nonumber \\ &&\!\!\!\!+4E_{1}E_{3}\cos\left(\gamma_{1}+\gamma_{3}\right)\sin^{2}\left(\frac{\phi_{1}-\phi_{3}}{2}\right)\nonumber \\ &&\!\!\!\!+4E_{2}E_{3}\cos\left(\gamma_{2}+\gamma_{3}\right)\sin^{2}\left(\frac{\phi_{2}-\phi_{3}}{2}\right)
\end{eqnarray}\\
%-----------E.B---------------------------------------------
\begin{eqnarray}
\left(\textbf{E}\cdot\textbf{B}\right)\!\!\!\!&=&\!\!\!\!\left(\textbf{E}_{1}+\textbf{E}_{2}+\textbf{E}_{3}\right)\cdot\left(\textbf{B}_{1}+\textbf{B}_{2}+\textbf{B}_{3}\right)\nonumber \\ \!\!\!\!&=&\!\!\!\! \left(E_{1}\sin\gamma_{1}\sin\phi_{1}+E_{2}\sin\gamma_{2}\sin\phi_{2}+E_{3}\sin\gamma_{3}\sin\phi_{3}\right)\nonumber \\ &&\ \ \ \!\!\!\!\times\left(E_{1}\cos\gamma_{1}\sin\phi_{1}+E_{2}\cos\gamma_{2}\sin\phi_{2}+E_{3}\cos\gamma_{3}\sin\phi_{3}\right)\nonumber \\ &&\!\!\!\!+ \left(E_{1}\sin\gamma_{1}\cos\phi_{1}+E_{2}\sin\gamma_{2}\cos\phi_{2}+E_{3}\sin\gamma_{3}\cos\phi_{3}\right)\nonumber \\ &&\ \ \ \!\!\!\!\times\left(E_{1}\cos\gamma_{1}\cos\phi_{1}+E_{2}\cos\gamma_{2}\cos\phi_{2}+E_{3}\cos\gamma_{3}\cos\phi_{3}\right)\nonumber \\ &&\!\!\!\!-  \left(E_{1}\cos\gamma_{1}+E_{2}\cos\gamma_{2}+E_{3}\cos\gamma_{3}\right)\nonumber \\ &&\ \ \ \!\!\!\!\times\left(E_{1}\sin\gamma_{1}+E_{2}\sin\gamma_{2}+E_{3}\sin\gamma_{3}\right)\nonumber \\ \!\!\!\!&=&\!\!\!\! \frac{1}{2}E_{1}^{2}\sin2\gamma_{1}+\frac{1}{2}E_{2}^{2}\sin2\gamma_{2}+\frac{1}{2}E_{3}^{2}\sin2\gamma_{3}\nonumber \\ &&\!\!\!\!+E_{1}E_{2}\sin\left(\gamma_{1}+\gamma_{2}\right)\cos\left(\phi_{1}-\phi_{2}\right)\nonumber \\ &&\!\!\!\!+ E_{1}E_{3}\sin\left(\gamma_{1}+\gamma_{3}\right)\cos\left(\phi_{1}-\phi_{3}\right)\nonumber \\ &&\!\!\!\!+ E_{2}E_{3}\sin\left(\gamma_{2}+\gamma_{3}\right)\cos\left(\phi_{2}-\phi_{3}\right)\nonumber \\ &&\!\!\!\!- \frac{1}{2}E_{1}^{2}\sin2\gamma_{1}-\frac{1}{2}E_{2}^{2}\sin2\gamma_{2}-\frac{1}{2}E_{3}^{2}\sin2\gamma_{3}\nonumber\\ &&\!\!\!\!-E_{1}E_{2}\sin\left(\gamma_{1}+\gamma_{2}\right)- E_{1}E_{3}\sin\left(\gamma_{1}+\gamma_{3}\right)-E_{2}E_{3}\sin\left(\gamma_{2}+\gamma_{3}\right)\nonumber \\ \!\!\!\!&=&\!\!\!\! -2E_{1}E_{2}\sin\left(\gamma_{1}+\gamma_{2}\right)\sin^{2}\left(\frac{\phi_{1}-\phi_{2}}{2}\right)\nonumber \\ &&\!\!\!\!- 2E_{1}E_{3}\sin\left(\gamma_{1}+\gamma_{3}\right)\sin^{2}\left(\frac{\phi_{1}-\phi_{3}}{2}\right)\nonumber \\ &&\!\!\!\!- 2E_{2}E_{3}\sin\left(\gamma_{2}+\gamma_{3}\right)\sin^{2}\left(\frac{\phi_{2}-\phi_{3}}{2}\right)
\end{eqnarray}\\
%-----------------------------(E^2-B^2)E---------------------------------------
\begin{eqnarray}
\left(E^{2}\!-B^{2}\right)\textbf{E}\!\!\!\!&=&\!\!\!\!\frac{\tilde{E}_{1}\tilde{E}_{2}\tilde{E}^{*}_{3}}{2}\left[\sin\gamma_{3}\sin\phi_{3}\cos\left(\gamma_{1}+\gamma_{2}\right)\sin^{2}\left(\frac{\phi_{1}-\phi_{2}}{2}\right)\right.\nonumber \\ &&\ \ \ \ \ \ \ \ \ \ + \sin\gamma_{2}\sin\phi_{2}\cos\left(\gamma_{1}+\gamma_{3}\right)\sin^{2}\left(\frac{\phi_{1}-\phi_{3}}{2}\right)\nonumber \\ &&\ \ \ \ \ \ \ \ \ \ + \left.\sin\gamma_{1}\sin\phi_{1}\cos\left(\gamma_{2}+\gamma_{3}\right)\sin^{2}\left(\frac{\phi_{2}-\phi_{3}}{2}\right)\right]e^{i\left(\textbf{k}_{4}\cdot\textbf{r}-\omega_{4}t\right)}\hat{\textbf{x}}\nonumber \\ &&\!\!\!\!- \frac{\tilde{E}_{1}\tilde{E}_{2}\tilde{E}^{*}_{3}}{2}\left[\sin\gamma_{3}\cos\phi_{3}\cos\left(\gamma_{1}+\gamma_{2}\right)\sin^{2}\left(\frac{\phi_{1}-\phi_{2}}{2}\right)\right.\nonumber \\ &&\ \ \ \ \ \ \ \ \ \ \ \ \ \!\!+ \sin\gamma_{2}\cos\phi_{2}\cos\left(\gamma_{1}+\gamma_{3}\right)\sin^{2}\left(\frac{\phi_{1}-\phi_{3}}{2}\right)\nonumber \\ &&\ \ \ \ \ \ \ \ \ \ \ \ \ \!\!+ \left.\sin\gamma_{1}\cos\phi_{1}\cos\left(\gamma_{2}+\gamma_{3}\right)\sin^{2}\left(\frac{\phi_{2}-\phi_{3}}{2}\right)\right]e^{i\left(\textbf{k}_{4}\cdot\textbf{r}-\omega_{4}t\right)}\hat{\textbf{y}}\nonumber \\ &&\!\!\!\!+ \frac{\tilde{E}_{1}\tilde{E}_{2}\tilde{E}^{*}_{3}}{2}\left[\cos\gamma_{3}\cos\left(\gamma_{1}+\gamma_{2}\right)\sin^{2}\left(\frac{\phi_{1}-\phi_{2}}{2}\right)\right.\nonumber \\ &&\ \ \ \ \ \ \ \ \ \ \ \ \ \!\!+ \cos\gamma_{2}\cos\left(\gamma_{1}+\gamma_{3}\right)\sin^{2}\left(\frac{\phi_{1}-\phi_{3}}{2}\right)\nonumber \\ &&\ \ \ \ \ \ \ \ \ \ \ \ \ \!\!+ \left.\cos\gamma_{1}\cos\left(\gamma_{2}+\gamma_{3}\right)\sin^{2}\left(\frac{\phi_{2}-\phi_{3}}{2}\right)\right]e^{i\left(\textbf{k}_{4}\cdot\textbf{r}-\omega_{4}t\right)}\hat{\textbf{z}}
\end{eqnarray}\\
%-------------------------(E^2-B^2)B-------------------------------
\begin{eqnarray}
\left(E^{2}\!-B^{2}\right)\textbf{B}\!\!\!\!&=&\!\!\!\!\frac{\tilde{E}_{1}\tilde{E}_{2}\tilde{E}^{*}_{3}}{2}\left[\cos\gamma_{3}\sin\phi_{3}\cos\left(\gamma_{1}+\gamma_{2}\right)\sin^{2}\left(\frac{\phi_{1}-\phi_{2}}{2}\right)\right.\nonumber \\ &&\ \ \ \ \ \ \ \ \ \ + \cos\gamma_{2}\sin\phi_{2}\cos\left(\gamma_{1}+\gamma_{3}\right)\sin^{2}\left(\frac{\phi_{1}-\phi_{3}}{2}\right)\nonumber \\ &&\ \ \ \ \ \ \ \ \ \ + \left.\cos\gamma_{1}\sin\phi_{1}\cos\left(\gamma_{2}+\gamma_{3}\right)\sin^{2}\left(\frac{\phi_{2}-\phi_{3}}{2}\right)\right]e^{i\left(\textbf{k}_{4}\cdot\textbf{r}-\omega_{4}t\right)}\hat{\textbf{x}}\nonumber \\ &&\!\!\!\!- \frac{\tilde{E}_{1}\tilde{E}_{2}\tilde{E}^{*}_{3}}{2}\left[\cos\gamma_{3}\cos\phi_{3}\cos\left(\gamma_{1}+\gamma_{2}\right)\sin^{2}\left(\frac{\phi_{1}-\phi_{2}}{2}\right)\right.\nonumber \\ &&\ \ \ \ \ \ \ \ \ \ \ \ \ \!\!+ \cos\gamma_{2}\cos\phi_{2}\cos\left(\gamma_{1}+\gamma_{3}\right)\sin^{2}\left(\frac{\phi_{1}-\phi_{3}}{2}\right)\nonumber \\ &&\ \ \ \ \ \ \ \ \ \ \ \ \ \!\!+ \cos\gamma_{1}\cos\phi_{1}\cos\left(\gamma_{2}+\gamma_{3}\right)\sin^{2}\left(\frac{\phi_{2}-\phi_{3}}{2}\right)]e^{i\left(\textbf{k}_{4}\cdot\textbf{r}-\omega_{4}t\right)}\hat{\textbf{y}}\nonumber \\ &&\!\!\!\!- \frac{\tilde{E}_{1}\tilde{E}_{2}\tilde{E}^{*}_{3}}{2}\left[\sin\gamma_{3}\cos\left(\gamma_{1}+\gamma_{2}\right)\sin^{2}\left(\frac{\phi_{1}-\phi_{2}}{2}\right)\right.\nonumber \\ &&\ \ \ \ \ \ \ \ \ \ \ \ \ \!\!+ \sin\gamma_{2}\cos\left(\gamma_{1}+\gamma_{3}\right)\sin^{2}\left(\frac{\phi_{1}-\phi_{3}}{2}\right)\nonumber \\ &&\ \ \ \ \ \ \ \ \ \ \ \ \ \!\!+  \left.\sin\gamma_{1}\cos\left(\gamma_{2}+\gamma_{3}\right)\sin^{2}\left(\frac{\phi_{2}-\phi_{3}}{2}\right)\right]e^{i\left(\textbf{k}_{4}\cdot\textbf{r}-\omega_{4}t\right)}\hat{\textbf{z}}
\end{eqnarray}\\
%------------------(E.B)E----------------------------------------
\begin{eqnarray}
\left(\textbf{E}\cdot\textbf{B}\right)\textbf{E}\!\!\!\!&=&\!\!\!\!-\frac{\tilde{E}_{1}\tilde{E}_{2}\tilde{E}^{*}_{3}}{4}\left[\sin\gamma_{3}\sin\phi_{3}\sin\left(\gamma_{1}+\gamma_{2}\right)\sin^{2}\left(\frac{\phi_{1}-\phi_{2}}{2}\right)\right.\nonumber \\ &&\ \ \ \ \ \ \ \ \ \ \ \ + \sin\gamma_{2}\sin\phi_{2}\sin\left(\gamma_{1}+\gamma_{3}\right)\sin^{2}\left(\frac{\phi_{1}-\phi_{3}}{2}\right)\nonumber \\ &&\ \ \ \ \ \ \ \ \ \ \ \ + \left.\sin\gamma_{1}\sin\phi_{1}\sin\left(\gamma_{2}+\gamma_{3}\right)\sin^{2}\left(\frac{\phi_{2}-\phi_{3}}{2}\right)\right]e^{i\left(\textbf{k}_{4}\cdot\textbf{r}-\omega_{4}t\right)}\hat{\textbf{x}}\nonumber \\ &&\!\!\!\!+ \frac{\tilde{E}_{1}\tilde{E}_{2}\tilde{E}^{*}_{3}}{4}\left[\sin\gamma_{3}\cos\phi_{3}\sin\left(\gamma_{1}+\gamma_{2}\right)\sin^{2}\left(\frac{\phi_{1}-\phi_{2}}{2}\right)\right.\nonumber \\ &&\ \ \ \ \ \ \ \ \ \ \ \ + \sin\gamma_{2}\cos\phi_{2}\sin\left(\gamma_{1}+\gamma_{3}\right)\sin^{2}\left(\frac{\phi_{1}-\phi_{3}}{2}\right)\nonumber \\ &&\ \ \ \ \ \ \ \ \ \ \ \ + \left.\sin\gamma_{1}\cos\phi_{1}\sin\left(\gamma_{2}+\gamma_{3}\right)\sin^{2}\left(\frac{\phi_{2}-\phi_{3}}{2}\right)\right]e^{i\left(\textbf{k}_{4}\cdot\textbf{r}-\omega_{4}t\right)}\hat{\textbf{y}}\nonumber \\ &&\!\!\!\!- \frac{\tilde{E}_{1}\tilde{E}_{2}\tilde{E}^{*}_{3}}{4}\left[\cos\gamma_{3}\sin\left(\gamma_{1}+\gamma_{2}\right)\sin^{2}\left(\frac{\phi_{1}-\phi_{2}}{2}\right)\right.\nonumber \\ &&\ \ \ \ \ \ \ \ \ \ \ \ + \cos\gamma_{2}\sin\left(\gamma_{1}+\gamma_{3}\right)\sin^{2}\left(\frac{\phi_{1}-\phi_{3}}{2}\right)\nonumber \\ &&\ \ \ \ \ \ \ \ \ \ \ \ + \left.\cos\gamma_{1}\sin\left(\gamma_{2}+\gamma_{3}\right)\sin^{2}\left(\frac{\phi_{2}-\phi_{3}}{2}\right)\right]e^{i\left(\textbf{k}_{4}\cdot\textbf{r}-\omega_{4}t\right)}\hat{\textbf{z}}
\end{eqnarray}\\
\newpage
%--------------(E.B)B--------------------------------------
\begin{eqnarray}
\left(\textbf{E}\cdot\textbf{B}\right)\textbf{B}\!\!\!\!&=&\!\!\!\!-\frac{\tilde{E}_{1}\tilde{E}_{2}\tilde{E}^{*}_{3}}{4}\left[\cos\gamma_{3}\sin\phi_{3}\sin\left(\gamma_{1}+\gamma_{2}\right)\sin^{2}\left(\frac{\phi_{1}-\phi_{2}}{2}\right)\right.\nonumber \\ &&\ \ \ \ \ \ \ \ \ \ \ \ + \cos\gamma_{2}\sin\phi_{2}\sin\left(\gamma_{1}+\gamma_{3}\right)\sin^{2}\left(\frac{\phi_{1}-\phi_{3}}{2}\right)\nonumber \\ &&\ \ \ \ \ \ \ \ \ \ \ \ + \left.\cos\gamma_{1}\sin\phi_{1}\sin\left(\gamma_{2}+\gamma_{3}\right)\sin^{2}\left(\frac{\phi_{2}-\phi_{3}}{2}\right)\right]e^{i\left(\textbf{k}_{4}\cdot\textbf{r}-\omega_{4}t\right)}\hat{\textbf{x}}\nonumber \\ &&\!\!\!\!+ \frac{\tilde{E}_{1}\tilde{E}_{2}\tilde{E}^{*}_{3}}{4}\left[\cos\gamma_{3}\cos\phi_{3}\sin\left(\gamma_{1}+\gamma_{2}\right)\sin^{2}\left(\frac{\phi_{1}-\phi_{2}}{2}\right)\right.\nonumber \\ &&\ \ \ \ \ \ \ \ \ \ \ \ + \cos\gamma_{2}\cos\phi_{2}\sin\left(\gamma_{1}+\gamma_{3}\right)\sin^{2}\left(\frac{\phi_{1}-\phi_{3}}{2}\right)\nonumber \\ &&\ \ \ \ \ \ \ \ \ \ \ \ + \left.\cos\gamma_{1}\cos\phi_{1}\sin\left(\gamma_{2}+\gamma_{3}\right)\sin^{2}\left(\frac{\phi_{2}-\phi_{3}}{2}\right)\right]e^{i\left(\textbf{k}_{4}\cdot\textbf{r}-\omega_{4}t\right)}\hat{\textbf{y}}\nonumber \\ &&\!\!\!\!+ \frac{\tilde{E}_{1}\tilde{E}_{2}\tilde{E}^{*}_{3}}{4}\left[\sin\gamma_{3}\sin\left(\gamma_{1}+\gamma_{2}\right)\sin^{2}\left(\frac{\phi_{1}-\phi_{2}}{2}\right)\right.\nonumber \\ &&\ \ \ \ \ \ \ \ \ \ \ \ + \sin\gamma_{2}\sin\left(\gamma_{1}+\gamma_{3}\right)\sin^{2}\left(\frac{\phi_{1}-\phi_{3}}{2}\right)\nonumber \\ &&\ \ \ \ \ \ \ \ \ \ \ \ + \left.\sin\gamma_{1}\sin\left(\gamma_{2}+\gamma_{3}\right)\sin^{2}\left(\frac{\phi_{2}-\phi_{3}}{2}\right)\right]e^{i\left(\textbf{k}_{4}\cdot\textbf{r}-\omega_{4}t\right)}\hat{\textbf{z}}
\end{eqnarray}\\
%------------------nabla . (E^2-B^2)E---------------------------
\begin{eqnarray}
\nabla\cdot\left[\left(E^{2}\!-B^{2}\right)\textbf{E}\right]\!\!\!\!&=&\!\!\!\!\frac{\tilde{E}_{1}\tilde{E}_{2}\tilde{E}^{*}_{3}}{2}ik_{4}\left[\sin\gamma_{3}\left(\sin\phi_{3}\cos\phi_{4}-\cos\phi_{3}\sin\phi_{4}\right)\right.\nonumber \\ &&\ \ \ \ \ \ \ \ \ \ \ \ \ \ \ \ \ \times \cos\left(\gamma_{1}+\gamma_{2}\right)\sin^{2}\left(\frac{\phi_{1}-\phi_{2}}{2}\right)\nonumber \\ &&\ \ \ \ \ \ \ \ \ \ \ \ \ \ + \sin\gamma_{2}\left(\sin\phi_{2}\cos\phi_{4}-\cos\phi_{2}\sin\phi_{4}\right)\nonumber \\ &&\ \ \ \ \ \ \ \ \ \ \ \ \ \ \ \ \ \times \cos\left(\gamma_{1}+\gamma_{3}\right)\sin^{2}\left(\frac{\phi_{1}-\phi_{3}}{2}\right)\nonumber \\ &&\ \ \ \ \ \ \ \ \ \ \ \ \ \ + \sin\gamma_{1}\left(\sin\phi_{1}\cos\phi_{4}-\cos\phi_{1}\sin\phi_{4}\right)\nonumber \\ &&\ \ \ \ \ \ \ \ \ \ \ \ \ \ \ \ \ \times \left.\cos\left(\gamma_{2}+\gamma_{3}\right)\sin^{2}\left(\frac{\phi_{2}-\phi_{3}}{2}\right)\right]e^{i\left(\textbf{k}_{4}\cdot\textbf{r}-\omega_{4}t\right)} \nonumber \\ \!\!\!\!&=&\!\!\!\! \frac{\tilde{E}_{1}\tilde{E}_{2}\tilde{E}^{*}_{3}}{2}ik_{4}\left[\sin\gamma_{3}\cos\left(\gamma_{1}+\gamma_{2}\right)\sin\left(\phi_{3}-\phi_{4}\right)\sin^{2}\left(\frac{\phi_{1}-\phi_{2}}{2}\right)\right.\nonumber \\ &&\ \ \ \ \ \ \ \ \ \ \ \ \ \ + \sin\gamma_{2}\cos\left(\gamma_{1}+\gamma_{3}\right)\sin\left(\phi_{2}-\phi_{4}\right)\sin^{2}\left(\frac{\phi_{1}-\phi_{3}}{2}\right)\nonumber \\ &&\ \ \ \ \ \ \ \ \ \ \ \ \ \ + \sin\gamma_{1}\cos\left(\gamma_{2}+\gamma_{3}\right)\nonumber \\ &&\ \ \ \ \ \ \ \ \ \ \ \ \ \ \ \ \ \times \left.\sin\left(\phi_{1}-\phi_{4}\right)\sin^{2}\left(\frac{\phi_{2}-\phi_{3}}{2}\right)\right]e^{i\left(\textbf{k}_{4}\cdot\textbf{r}-\omega_{4}t\right)}
\end{eqnarray}\\
%-----------------------d/dt (E^2-B^2)E------------------------------
\begin{eqnarray}
	\frac{\partial}{\partial t}\left[\left(E^{2}\!-B^{2}\right)\textbf{E}\right]=-i\omega_{4}\left[\left(E^{2}\!-B^{2}\right)\textbf{E}\right]
\end{eqnarray}\\
%------------------------nabla cross (E^2-B^2)B---------------------------
\begin{eqnarray}
\nabla\times\left[\left(E^{2}\!-B^{2}\right)\textbf{B}\right]\!\!\!\!&=&\!\!\!\!-\frac{\tilde{E}_{1}\tilde{E}_{2}\tilde{E}^{*}_{3}}{2}ik_{4}\left[\sin\gamma_{3}\cos\left(\gamma_{1}+\gamma_{2}\right)\sin\phi_{4}\sin^{2}\left(\frac{\phi_{1}-\phi_{2}}{2}\right)\right.\nonumber \\ &&\ \ \ \ \ \ \ \ \ \ \ \ \ \ \ \ + \sin\gamma_{2}\cos\left(\gamma_{1}+\gamma_{3}\right)\sin\phi_{4}\sin^{2}\left(\frac{\phi_{1}-\phi_{3}}{2}\right)\nonumber \\ &&\ \ \ \ \ \ \ \ \ \ \ \ \ \ \ \ + \left.\sin\gamma_{1}\cos\left(\gamma_{2}+\gamma_{3}\right)\sin\phi_{4}\sin^{2}\left(\frac{\phi_{2}-\phi_{3}}{2}\right)\right]e^{i\left(\textbf{k}_{4}\cdot\textbf{r}-\omega_{4}t\right)}\hat{\textbf{x}}\nonumber \\ &&\!\!\!\!+ \frac{\tilde{E}_{1}\tilde{E}_{2}\tilde{E}^{*}_{3}}{2}ik_{4}\left[\sin\gamma_{3}\cos\left(\gamma_{1}+\gamma_{2}\right)\cos\phi_{4}\sin^{2}\left(\frac{\phi_{1}-\phi_{2}}{2}\right)\right.\nonumber \\ &&\ \ \ \ \ \ \ \ \ \ \ \ \ \ \ \ + \sin\gamma_{2}\cos\left(\gamma_{1}+\gamma_{3}\right)\cos\phi_{4}\sin^{2}\left(\frac{\phi_{1}-\phi_{3}}{2}\right)\nonumber \\ &&\ \ \ \ \ \ \ \ \ \ \ \ \ \ \ \ + \left.\sin\gamma_{1}\cos\left(\gamma_{2}+\gamma_{3}\right)\cos\phi_{4}\sin^{2}\left(\frac{\phi_{2}-\phi_{3}}{2}\right)\right]e^{i\left(\textbf{k}_{4}\cdot\textbf{r}-\omega_{4}t\right)}\hat{\textbf{y}} \nonumber \\ &&\!\!\!\!- \frac{\tilde{E}_{1}\tilde{E}_{2}\tilde{E}^{*}_{3}}{2}ik_{4}\left[\cos\gamma_{3}\cos\left(\gamma_{1}+\gamma_{2}\right)\cos\left(\phi_{3}-\phi_{4}\right)\sin^{2}\left(\frac{\phi_{1}-\phi_{2}}{2}\right)\right.\nonumber \\ &&\ \ \ \ \ \ \ \ \ \ \ \ \ \ \ \ + \cos\gamma_{2}\cos\left(\gamma_{1}+\gamma_{3}\right)\cos\left(\phi_{2}-\phi_{4}\right)\sin^{2}\left(\frac{\phi_{1}-\phi_{3}}{2}\right)\nonumber \\ &&\ \ \ \ \ \ \ \ \ \ \ \ \ \ \ \ + \cos\gamma_{1}\cos\left(\gamma_{2}+\gamma_{3}\right)\nonumber \\ &&\ \ \ \ \ \ \ \ \ \ \ \ \ \ \ \ \ \ \ \times \left.\cos\left(\phi_{1}-\phi_{4}\right)\sin^{2}\left(\frac{\phi_{2}-\phi_{3}}{2}\right)\right]e^{i\left(\textbf{k}_{4}\cdot\textbf{r}-\omega_{4}t\right)}\hat{\textbf{z}}\!\!
\end{eqnarray}\\
%---------------------nabla cross (E.B)E---------------------------------------
\begin{eqnarray}
\nabla\times\left[\left(\textbf{E}\cdot\textbf{B}\right)\textbf{E}\right]\!\!\!\!&=&\!\!\!\!-\frac{\tilde{E}_{1}\tilde{E}_{2}\tilde{E}^{*}_{3}}{4}ik_{4}\left[\cos\gamma_{3}\sin\left(\gamma_{1}+\gamma_{2}\right)\sin\phi_{4}\sin^{2}\left(\frac{\phi_{1}-\phi_{2}}{2}\right)\right.\nonumber \\ &&\ \ \ \ \ \ \ \ \ \ \ \ \ \ \ \ + \cos\gamma_{2}\sin\left(\gamma_{1}+\gamma_{3}\right)\sin\phi_{4}\sin^{2}\left(\frac{\phi_{1}-\phi_{3}}{2}\right)\nonumber \\ &&\ \ \ \ \ \ \ \ \ \ \ \ \ \ \ \ + \left.\cos\gamma_{1}\sin\left(\gamma_{2}+\gamma_{3}\right)\sin\phi_{4}\sin^{2}\left(\frac{\phi_{2}-\phi_{3}}{2}\right)\right]e^{i\left(\textbf{k}_{4}\cdot\textbf{r}-\omega_{4}t\right)}\hat{\textbf{x}}\nonumber \\ &&\!\!\!\!+ \frac{\tilde{E}_{1}\tilde{E}_{2}\tilde{E}^{*}_{3}}{4}ik_{4}\left[\cos\gamma_{3}\sin\left(\gamma_{1}+\gamma_{2}\right)\cos\phi_{4}\sin^{2}\left(\frac{\phi_{1}-\phi_{2}}{2}\right)\right.\nonumber \\ &&\ \ \ \ \ \ \ \ \ \ \ \ \ \ \ \ + \cos\gamma_{2}\sin\left(\gamma_{1}+\gamma_{3}\right)\cos\phi_{4}\sin^{2}\left(\frac{\phi_{1}-\phi_{3}}{2}\right)\nonumber \\ &&\ \ \ \ \ \ \ \ \ \ \ \ \ \ \ \ + \left.\cos\gamma_{1}\sin\left(\gamma_{2}+\gamma_{3}\right)\cos\phi_{4}\sin^{2}\left(\frac{\phi_{2}-\phi_{3}}{2}\right)\right]e^{i\left(\textbf{k}_{4}\cdot\textbf{r}-\omega_{4}t\right)}\hat{\textbf{y}}\nonumber \\ &&\!\!\!\!+ \frac{\tilde{E}_{1}\tilde{E}_{2}\tilde{E}^{*}_{3}}{4}ik_{4}\left[\sin\gamma_{3}\sin\left(\gamma_{1}+\gamma_{2}\right)\cos\left(\phi_{3}-\phi_{4}\right)\sin^{2}\left(\frac{\phi_{1}-\phi_{2}}{2}\right)\right.\nonumber \\ &&\ \ \ \ \ \ \ \ \ \ \ \ \ \ \ \ + \sin\gamma_{2}\sin\left(\gamma_{1}+\gamma_{3}\right)\cos\left(\phi_{2}-\phi_{4}\right)\sin^{2}\left(\frac{\phi_{1}-\phi_{3}}{2}\right)\nonumber \\ &&\ \ \ \ \ \ \ \ \ \ \ \ \ \ \ \ + \sin\gamma_{1}\sin\left(\gamma_{2}+\gamma_{3}\right)\nonumber \\ &&\ \ \ \ \ \ \ \ \ \ \ \ \ \ \ \ \ \ \ \times \left.\cos\left(\phi_{1}-\phi_{4}\right)\sin^{2}\left(\frac{\phi_{2}-\phi_{3}}{2}\right)\right]e^{i\left(\textbf{k}_{4}\cdot\textbf{r}-\omega_{4}t\right)}\hat{\textbf{z}}
\end{eqnarray}\\
%---------------------nabla . (E.B)B------------------------------
\begin{eqnarray}
\nabla\cdot\left[\left(\textbf{E}\cdot\textbf{B}\right)\textbf{B}\right]\!\!\!\!&=&\!\!\!\!-\frac{\tilde{E}_{1}\tilde{E}_{2}\tilde{E}^{*}_{3}}{4}ik_{4}\left[\cos\gamma_{3}\left(\sin\phi_{3}\cos\phi_{4}-\cos\phi_{3}\sin\phi_{4}\right)\right.\nonumber \\ &&\ \ \ \ \ \ \ \ \ \ \ \ \ \ \ \ \ \ \ \times\sin\left(\gamma_{1}+\gamma_{2}\right)\sin^{2}\left(\frac{\phi_{1}-\phi_{2}}{2}\right)\nonumber \\ &&\ \ \ \ \ \ \ \ \ \ \ \ \ \ \ \ + \cos\gamma_{2}\left(\sin\phi_{2}\cos\phi_{4}-\cos\phi_{2}\sin\phi_{4}\right)\nonumber \\ &&\ \ \ \ \ \ \ \ \ \ \ \ \ \ \ \ \ \ \ \times \sin\left(\gamma_{1}+\gamma_{3}\right)\sin^{2}\left(\frac{\phi_{1}-\phi_{3}}{2}\right)\nonumber \\ &&\ \ \ \ \ \ \ \ \ \ \ \ \ \ \ \ + \cos\gamma_{1}\left(\sin\phi_{1}\cos\phi_{4}-\cos\phi_{1}\sin\phi_{4}\right)\nonumber \\ &&\ \ \ \ \ \ \ \ \ \ \ \ \ \ \ \ \ \ \ \times \left.\sin\left(\gamma_{2}+\gamma_{3}\right)\sin^{2}\left(\frac{\phi_{2}-\phi_{3}}{2}\right)\right]e^{i\left(\textbf{k}_{4}\cdot\textbf{r}-\omega_{4}t\right)}= \nonumber \\ &&\!\!\!\! -\frac{\tilde{E}_{1}\tilde{E}_{2}\tilde{E}^{*}_{3}}{4}ik_{4}\left[\cos\gamma_{3}\sin\left(\gamma_{1}+\gamma_{2}\right)\right.\nonumber \\ &&\ \ \ \ \ \ \ \ \ \ \ \ \ \ \ \ \ \ \ \times\sin\left(\phi_{3}-\phi_{4}\right)\sin^{2}\left(\frac{\phi_{1}-\phi_{2}}{2}\right)\nonumber \\ &&\ \ \ \ \ \ \ \ \ \ \ \ \ \ \ \ + \cos\gamma_{2}\sin\left(\gamma_{1}+\gamma_{3}\right)\nonumber \\ &&\ \ \ \ \ \ \ \ \ \ \ \ \ \ \ \ \ \ \ \times\sin\left(\phi_{2}-\phi_{4}\right)\sin^{2}\left(\frac{\phi_{1}-\phi_{3}}{2}\right)\nonumber \\ &&\ \ \ \ \ \ \ \ \ \ \ \ \ \ \ \ + \cos\gamma_{1}\sin\left(\gamma_{2}+\gamma_{3}\right)\nonumber \\ &&\ \ \ \ \ \ \ \ \ \ \ \ \ \ \ \ \ \ \ \left.\times\sin\left(\phi_{1}-\phi_{4}\right)\sin^{2}\left(\frac{\phi_{2}-\phi_{3}}{2}\right)\right]e^{i\left(\textbf{k}_{4}\cdot\textbf{r}-\omega_{4}t\right)}
\end{eqnarray}\\
%--------------------d/dt (E.B)B-----------------------------
\begin{eqnarray}
	\frac{\partial}{\partial t}\left[\left(\textbf{E}\cdot\textbf{B}\right)\textbf{B}\right]=-i\omega_{4}\left[\left(\textbf{E}\cdot\textbf{B}\right)\textbf{B}\right]
\end{eqnarray}\\
%-----------------------(N1)----------------------------------
\begin{eqnarray}
	(N1)\!\!\!\!&=&\!\!\!\!\xi \tilde{E}_{1}\tilde{E}_{2}\tilde{E}^{*}_{3}ik_{4}\left\{\ \left[\frac{7}{4}\cos\gamma_{3}\sin\left(\gamma_{1}+\gamma_{2}\right)-\sin\gamma_{3}\cos\left(\gamma_{1}+\gamma_{2}\right)\right]\right.\nonumber \\ &&\ \ \ \ \ \ \ \ \ \ \ \ \ \ \ \ \ \ \ \ \times\sin\left(\phi_{3}-\phi_{4}\right)\sin^{2}\left(\frac{\phi_{1}-\phi_{2}}{2}\right)\nonumber \\ &&\ \ \ \ \ \ \ \ \ \ \ \ \ \ \ \ + \left[\frac{7}{4}\cos\gamma_{2}\sin\left(\gamma_{1}+\gamma_{3}\right)-\sin\gamma_{2}\cos\left(\gamma_{1}+\gamma_{3}\right)\right]\nonumber \\ &&\ \ \ \ \ \ \ \ \ \ \ \ \ \ \ \ \ \ \ \ \times\sin\left(\phi_{2}-\phi_{4}\right)\sin^{2}\left(\frac{\phi_{1}-\phi_{3}}{2}\right)\nonumber \\ &&\ \ \ \ \ \ \ \ \ \ \ \ \ \ \ \ + \left[\frac{7}{4}\cos\gamma_{1}\sin\left(\gamma_{2}+\gamma_{3}\right)-\sin\gamma_{1}\cos\left(\gamma_{2}+\gamma_{3}\right)\right]\nonumber \\ &&\ \ \ \ \ \ \ \ \ \ \ \ \ \ \ \ \ \ \ \ \times\left.\sin\left(\phi_{1}-\phi_{4}\right)\sin^{2}\left(\frac{\phi_{2}-\phi_{3}}{2}\right)\right\}e^{i\left(\textbf{k}_{4}\cdot\textbf{r}-\omega_{4}t\right)}
\end{eqnarray}\\
\newpage
%------------------------xhat . (N234)-----------------------------------
\begin{eqnarray}
	\hat{\textbf{x}}\cdot\overline{(N234)}\!\!\!\!&=&\!\!\!\!-\xi \tilde{E}_{1}\tilde{E}_{2}\tilde{E}^{*}_{3}ik_{4}\left\{\sin\gamma_{3}\cos\left(\gamma_{1}+\gamma_{2}\right)\left(\sin\phi_{3}-\sin\phi_{4}\right)\sin^{2}\left(\frac{\phi_{1}-\phi_{2}}{2}\right)\right.\nonumber \\ &&\ \ \ \ \ \ \ \ \ \ \ \ \ \ \ \ \ \ + \sin\gamma_{2}\cos\left(\gamma_{1}+\gamma_{3}\right)\left(\sin\phi_{2}-\sin\phi_{4}\right)\sin^{2}\left(\frac{\phi_{1}-\phi_{3}}{2}\right)\nonumber \\ &&\ \ \ \ \ \ \ \ \ \ \ \ \ \ \ \ \ \ + \left.\sin\gamma_{1}\cos\left(\gamma_{2}+\gamma_{3}\right)\left(\sin\phi_{1}-\sin\phi_{4}\right)\sin^{2}\left(\frac{\phi_{2}-\phi_{3}}{2}\right)\right\}e^{i\left(\textbf{k}_{4}\cdot\textbf{r}-\omega_{4}t\right)}\nonumber \\ &&\!\!\!\!+ \frac{7}{4}\xi \tilde{E}_{1}\tilde{E}_{2}\tilde{E}^{*}_{3}ik_{4}\left\{\cos\gamma_{3}\sin\left(\gamma_{1}+\gamma_{2}\right)\left(\sin\phi_{3}-\sin\phi_{4}\right)\sin^{2}\left(\frac{\phi_{1}-\phi_{2}}{2}\right)\right.\nonumber \\ &&\ \ \ \ \ \ \ \ \ \ \ \ \ \ \ \ \ \ \ \ + \cos\gamma_{2}\sin\left(\gamma_{1}+\gamma_{3}\right)\left(\sin\phi_{2}-\sin\phi_{4}\right)\sin^{2}\left(\frac{\phi_{1}-\phi_{3}}{2}\right)\nonumber \\ &&\ \ \ \ \ \ \ \ \ \ \ \ \ \ \ \ \ \ \ \ + \left.\cos\gamma_{1}\sin\left(\gamma_{2}+\gamma_{3}\right)\left(\sin\phi_{1}-\sin\phi_{4}\right)\sin^{2}\left(\frac{\phi_{2}-\phi_{3}}{2}\right)\right\}e^{i\left(\textbf{k}_{4}\cdot\textbf{r}-\omega_{4}t\right)}\nonumber \\ \!\!\!\!&=&\!\!\!\! \xi \tilde{E}_{1}\tilde{E}_{2}\tilde{E}^{*}_{3}ik_{4}\left\{\ \left[\frac{7}{4}\cos\gamma_{3}\sin\left(\gamma_{1}+\gamma_{2}\right)-\sin\gamma_{3}\cos\left(\gamma_{1}+\gamma_{2}\right)\right]\right.\nonumber \\ &&\ \ \ \ \ \ \ \ \ \ \ \ \ \ \ \ \ \ \ \ \times\left(\sin\phi_{3}-\sin\phi_{4}\right)\sin^{2}\left(\frac{\phi_{1}-\phi_{2}}{2}\right)\nonumber \\ &&\ \ \ \ \ \ \ \ \ \ \ \ \ \ \ \ + \left[\frac{7}{4}\cos\gamma_{2}\sin\left(\gamma_{1}+\gamma_{3}\right)-\sin\gamma_{2}\cos\left(\gamma_{1}+\gamma_{3}\right)\right]\nonumber \\ &&\ \ \ \ \ \ \ \ \ \ \ \ \ \ \ \ \ \ \ \ \times\left(\sin\phi_{2}-\sin\phi_{4}\right)\sin^{2}\left(\frac{\phi_{1}-\phi_{3}}{2}\right)\nonumber \\ &&\ \ \ \ \ \ \ \ \ \ \ \ \ \ \ \ + \left[\frac{7}{4}\cos\gamma_{1}\sin\left(\gamma_{2}+\gamma_{3}\right)-\sin\gamma_{1}\cos\left(\gamma_{2}+\gamma_{3}\right)\right]\nonumber \\ &&\ \ \ \ \ \ \ \ \ \ \ \ \ \ \ \ \ \ \ \ \times\left.\left(\sin\phi_{1}-\sin\phi_{4}\right)\sin^{2}\left(\frac{\phi_{2}-\phi_{3}}{2}\right)\right\}e^{i\left(\textbf{k}_{4}\cdot\textbf{r}-\omega_{4}t\right)}
\end{eqnarray}\\
%-----------------------yhat . (N234)-----------------------------------
\begin{eqnarray}
	\hat{\textbf{y}}\cdot\overline{(N234)}\!\!\!\!&=&\!\!\!\!\xi \tilde{E}_{1}\tilde{E}_{2}\tilde{E}^{*}_{3}ik_{4}\left\{\sin\gamma_{3}\cos\left(\gamma_{1}+\gamma_{2}\right)\left(\cos\phi_{3}-\cos\phi_{4}\right)\sin^{2}\left(\frac{\phi_{1}-\phi_{2}}{2}\right)\right.\nonumber \\ &&\ \ \ \ \ \ \ \ \ \ \ \ \ \ \ \ + \sin\gamma_{2}\cos\left(\gamma_{1}+\gamma_{3}\right)\left(\cos\phi_{2}-\cos\phi_{4}\right)\sin^{2}\left(\frac{\phi_{1}-\phi_{3}}{2}\right)\nonumber \\ &&\ \ \ \ \ \ \ \ \ \ \ \ \ \ \ \ + \left.\sin\gamma_{1}\cos\left(\gamma_{2}+\gamma_{3}\right)\left(\cos\phi_{1}-\cos\phi_{4}\right)\sin^{2}\left(\frac{\phi_{2}-\phi_{3}}{2}\right)\right\}e^{i\left(\textbf{k}_{4}\cdot\textbf{r}-\omega_{4}t\right)}\nonumber \\ &&\!\!\!\!- \frac{7}{4}\xi \tilde{E}_{1}\tilde{E}_{2}\tilde{E}^{*}_{3}ik_{4}\left\{\cos\gamma_{3}\sin\left(\gamma_{1}+\gamma_{2}\right)\left(\cos\phi_{3}-\cos\phi_{4}\right)\sin^{2}\left(\frac{\phi_{1}-\phi_{2}}{2}\right)\right.\nonumber \\ &&\ \ \ \ \ \ \ \ \ \ \ \ \ \ \ \ \ \ \ \ + \cos\gamma_{2}\sin\left(\gamma_{1}+\gamma_{3}\right)\left(\cos\phi_{2}-\cos\phi_{4}\right)\sin^{2}\left(\frac{\phi_{1}-\phi_{3}}{2}\right)\nonumber \\ &&\ \ \ \ \ \ \ \ \ \ \ \ \ \ \ \ \ \ \ \ + \left.\cos\gamma_{1}\sin\left(\gamma_{2}+\gamma_{3}\right)\left(\cos\phi_{1}-\cos\phi_{4}\right)\sin^{2}\left(\frac{\phi_{2}-\phi_{3}}{2}\right)\right\}e^{i\left(\textbf{k}_{4}\cdot\textbf{r}-\omega_{4}t\right)}\nonumber \\ \!\!\!\!&=&\!\!\!\! -\xi \tilde{E}_{1}\tilde{E}_{2}\tilde{E}^{*}_{3}ik_{4}\left\{\ \left[\frac{7}{4}\cos\gamma_{3}\sin\left(\gamma_{1}+\gamma_{2}\right)-\sin\gamma_{3}\cos\left(\gamma_{1}+\gamma_{2}\right)\right]\right.\nonumber \\ &&\ \ \ \ \ \ \ \ \ \ \ \ \ \ \ \ \ \ \ \ \ \ \times\left(\cos\phi_{3}-\cos\phi_{4}\right)\sin^{2}\left(\frac{\phi_{1}-\phi_{2}}{2}\right)\nonumber \\ &&\ \ \ \ \ \ \ \ \ \ \ \ \ \ \ \ \ \ + \left[\frac{7}{4}\cos\gamma_{2}\sin\left(\gamma_{1}+\gamma_{3}\right)-\sin\gamma_{2}\cos\left(\gamma_{1}+\gamma_{3}\right)\right]\nonumber \\ &&\ \ \ \ \ \ \ \ \ \ \ \ \ \ \ \ \ \ \ \ \ \ \times\left(\cos\phi_{2}-\cos\phi_{4}\right)\sin^{2}\left(\frac{\phi_{1}-\phi_{3}}{2}\right)\nonumber \\ &&\ \ \ \ \ \ \ \ \ \ \ \ \ \ \ \ \ \ + \left[\frac{7}{4}\cos\gamma_{1}\sin\left(\gamma_{2}+\gamma_{3}\right)-\sin\gamma_{1}\cos\left(\gamma_{2}+\gamma_{3}\right)\right]\nonumber \\ &&\ \ \ \ \ \ \ \ \ \ \ \ \ \ \ \ \ \ \ \ \ \ \times\left.\left(\cos\phi_{1}-\cos\phi_{4}\right)\sin^{2}\left(\frac{\phi_{2}-\phi_{3}}{2}\right)\right\}e^{i\left(\textbf{k}_{4}\cdot\textbf{r}-\omega_{4}t\right)}
\end{eqnarray}\\
%-----------------------zhat . (N234)-----------------------------------
\begin{eqnarray}
	\hat{\textbf{z}}\cdot\overline{(N234)}\!\!\!\!&=&\!\!\!\!-2\xi \tilde{E}_{1}\tilde{E}_{2}\tilde{E}^{*}_{3}ik_{4}\left\{\cos\gamma_{3}\cos\left(\gamma_{1}+\gamma_{2}\right)\sin^{2}\left(\frac{\phi_{3}-\phi_{4}}{2}\right)\sin^{2}\left(\frac{\phi_{1}-\phi_{2}}{2}\right)\right.\nonumber \\ &&\ \ \ \ \ \ \ \ \ \ \ \ \ \ \ \ \ \ \ + \cos\gamma_{2}\cos\left(\gamma_{1}+\gamma_{3}\right)\sin^{2}\left(\frac{\phi_{2}-\phi_{4}}{2}\right)\sin^{2}\left(\frac{\phi_{1}-\phi_{3}}{2}\right)\nonumber \\ &&\ \ \ \ \ \ \ \ \ \ \ \ \ \ \ \ \ \ \ + \left.\cos\gamma_{1}\cos\left(\gamma_{2}+\gamma_{3}\right)\sin^{2}\left(\frac{\phi_{1}-\phi_{4}}{2}\right)\sin^{2}\left(\frac{\phi_{2}-\phi_{3}}{2}\right)\right\}e^{i\left(\textbf{k}_{4}\cdot\textbf{r}-\omega_{4}t\right)}\nonumber \\ &&\!\!\!\!- \frac{7}{4}2\xi \tilde{E}_{1}\tilde{E}_{2}\tilde{E}^{*}_{3}ik_{4}\left\{\sin\gamma_{3}\sin\left(\gamma_{1}+\gamma_{2}\right)\sin^{2}\left(\frac{\phi_{3}-\phi_{4}}{2}\right)\sin^{2}\left(\frac{\phi_{1}-\phi_{2}}{2}\right)\right.\nonumber \\ &&\ \ \ \ \ \ \ \ \ \ \ \ \ \ \ \ \ \ \ \ \ + \sin\gamma_{2}\sin\left(\gamma_{1}+\gamma_{3}\right)\sin^{2}\left(\frac{\phi_{2}-\phi_{4}}{2}\right)\sin^{2}\left(\frac{\phi_{1}-\phi_{3}}{2}\right)\nonumber \\ &&\ \ \ \ \ \ \ \ \ \ \ \ \ \ \ \ \ \ \ \ \ + \left.\sin\gamma_{1}\sin\left(\gamma_{2}+\gamma_{3}\right)\sin^{2}\left(\frac{\phi_{1}-\phi_{4}}{2}\right)\sin^{2}\left(\frac{\phi_{2}-\phi_{3}}{2}\right)\right\}e^{i\left(\textbf{k}_{4}\cdot\textbf{r}-\omega_{4}t\right)}\nonumber \\ \!\!\!\!&=&\!\!\!\! -2\xi \tilde{E}_{1}\tilde{E}_{2}\tilde{E}^{*}_{3}ik_{4}\left\{\ \left[\frac{7}{4}\sin\gamma_{3}\sin\left(\gamma_{1}+\gamma_{2}\right)+\cos\gamma_{3}\cos\left(\gamma_{1}+\gamma_{2}\right)\right]\right.\nonumber \\ &&\ \ \ \ \ \ \ \ \ \ \ \ \ \ \ \ \ \ \ \ \ \ \ \ \times\sin^{2}\left(\frac{\phi_{3}-\phi_{4}}{2}\right)\sin^{2}\left(\frac{\phi_{1}-\phi_{2}}{2}\right)\nonumber \\ &&\ \ \ \ \ \ \ \ \ \ \ \ \ \ \ \ \ \ \ \ + \left[\frac{7}{4}\sin\gamma_{2}\sin\left(\gamma_{1}+\gamma_{3}\right)+\cos\gamma_{2}\cos\left(\gamma_{1}+\gamma_{3}\right)\right]\nonumber \\ &&\ \ \ \ \ \ \ \ \ \ \ \ \ \ \ \ \ \ \ \ \ \ \ \ \times\sin^{2}\left(\frac{\phi_{2}-\phi_{4}}{2}\right)\sin^{2}\left(\frac{\phi_{1}-\phi_{3}}{2}\right)\nonumber \\ &&\ \ \ \ \ \ \ \ \ \ \ \ \ \ \ \ \ \ \ \ + \left[\frac{7}{4}\sin\gamma_{1}\sin\left(\gamma_{2}+\gamma_{3}\right)+\cos\gamma_{1}\cos\left(\gamma_{2}+\gamma_{3}\right)\right]\nonumber \\ &&\ \ \ \ \ \ \ \ \ \ \ \ \ \ \ \ \ \ \ \ \ \ \ \ \times\left.\sin^{2}\left(\frac{\phi_{1}-\phi_{4}}{2}\right)\sin^{2}\left(\frac{\phi_{2}-\phi_{3}}{2}\right)\right\}e^{i\left(\textbf{k}_{4}\cdot\textbf{r}-\omega_{4}t\right)}
\end{eqnarray}\\
%-------------nabla(N1)---------------------------------------
\begin{eqnarray}
	\nabla(N1)=ik_{4}(N1)\left(\cos\phi_{4}\hat{\textbf{x}}+\sin\phi_{4}\hat{\textbf{y}}\right)
\end{eqnarray}\\
%-------------d/dt (N234)---------------------------------------
\begin{eqnarray}
	\frac{\partial}{\partial t}\overline{(N234)}=-i\omega_{4}\overline{(N234)}
\end{eqnarray}\\
By finally plugging all this into the right hand side of Eq.$\ $(3.38), we end up with the generated field of Eq.$\ $(5.7).
\chapter{The three dimensional geometric factor}
The details of the calculation of Eq.$\ $(5.19) are shown here. On keeping only resonant terms, we evaluate the right hand side of Eq.$\ $(3.38) step by step:\\
\\
\begin{eqnarray}
\left(E^{2}\!-B^{2}\right)\!\!\!\!&=&\!\!\!\!\left(\textbf{E}_{1}+\textbf{E}_{2}+\textbf{E}_{3}\right)^{2}-\left(\textbf{B}_{1}+\textbf{B}_{2}+\textbf{B}_{3}\right)^{2}\nonumber \\ \!\!\!\!&=&\!\!\!\! \left(E_{2}\sin\gamma_{2}+E_{3}\cos\beta_{3}\right)^{2}+\left(E_{3}\sin\beta_{3}-E_{1}\sin\gamma_{1}\right)^{2}+\left(E_{1}\cos\gamma_{1}+E_{2}\cos\gamma_{2}\right)^{2}\nonumber \\ &&\!\!\!\!-\left(E_{2}\cos\gamma_{2}-E_{3}\sin\beta_{3}\right)^{2}-\left(E_{3}\cos\beta_{3}-E_{1}\cos\gamma_{1}\right)^{2}-\left(E_{1}\sin\gamma_{1}+E_{2}\sin\gamma_{2}\right)^{2}\nonumber \\ \!\!\!\!&=&\!\!\!\!E_{1}^{2}+E_{2}^{2}+E_{3}^{2}-E_{1}^{2}-E_{2}^{2}-E_{3}^{2}+2E_{1}E_{2}\left(\cos\gamma_{1}\cos\gamma_{2}-\sin\gamma_{1}\sin\gamma_{2}\right)\nonumber \\ &&\!\!\!\!+2E_{1}E_{3}\left(\cos\gamma_{1}\cos\beta_{3}-\sin\gamma_{1}\sin\beta_{3}\right)+2E_{2}E_{3}\left(\cos\gamma_{2}\sin\beta_{3}+\sin\gamma_{2}\cos\beta_{3}\right)\nonumber \\ \!\!\!\!&=&\!\!\!\!2E_{1}E_{2}\cos\left(\gamma_{1}+\gamma_{2}\right)+2E_{1}E_{3}\cos\left(\gamma_{1}+\beta_{3}\right)+2E_{2}E_{3}\sin\left(\gamma_{2}+\beta_{3}\right)
\end{eqnarray}\\
%-----------E.B---------------------------------------------
\begin{eqnarray}
\left(\textbf{E}\cdot\textbf{B}\right)\!\!\!\!&=&\!\!\!\!\left(\textbf{E}_{1}+\textbf{E}_{2}+\textbf{E}_{3}\right)\cdot\left(\textbf{B}_{1}+\textbf{B}_{2}+\textbf{B}_{3}\right)\nonumber \\ \!\!\!\!&=&\!\!\!\! \left(E_{2}\sin\gamma_{2}+E_{3}\cos\beta_{3}\right)\left(E_{2}\cos\gamma_{2}-E_{3}\sin\beta_{3}\right)\nonumber \\ &&\!\!\!\!+ \left(-E_{1}\sin\gamma_{1}+E_{3}\sin\beta_{3}\right)\left(-E_{1}\cos\gamma_{1}+E_{3}\cos\beta_{3}\right)\nonumber \\ &&\!\!\!\!+\left(E_{1}\cos\gamma_{1}+E_{2}\cos\gamma_{2}\right)\left(-E_{1}\sin\gamma_{1}-E_{2}\sin\gamma_{2}\right)\nonumber \\ \!\!\!\!&=&\!\!\!\! E_{1}E_{2}\left(-\cos\gamma_{1}\sin\gamma_{2}-\cos\gamma_{2}\sin\gamma_{1}\right)\nonumber \\ &&\!\!\!\!+ E_{1}E_{3}\left(-\sin\gamma_{1}\cos\beta_{3}-\sin\beta_{3}\cos\gamma_{1}\right)\nonumber \\ &&\!\!\!\!+ E_{2}E_{3}\left(-\sin\gamma_{2}\sin\beta_{3}+\cos\beta_{3}\cos\gamma_{2}\right)\nonumber \\ \!\!\!\!&=&\!\!\!\! -E_{1}E_{2}\sin\left(\gamma_{1}+\gamma_{2}\right)-E_{1}E_{3}\sin\left(\gamma_{1}+\beta_{3}\right)+E_{2}E_{3}\cos\left(\gamma_{2}+\beta_{3}\right)
\end{eqnarray}\\
\newpage
%-----------------------------(E^2-B^2)E---------------------------------------
\begin{eqnarray}
\left(E^{2}\!-B^{2}\right)\textbf{E}\!\!\!\!&=&\!\!\!\!\frac{\tilde{E}_{1}\tilde{E}_{2}\tilde{E}^{*}_{3}}{4}\left[\cos\beta_{3}\cos\left(\gamma_{1}+\gamma_{2}\right)+\sin\gamma_{2}\cos\left(\gamma_{1}+\beta_{3}\right)\right]e^{i\left(\textbf{k}_{4}\cdot\textbf{r}-\omega_{4}t\right)}\hat{\textbf{x}}\nonumber \\ &&\!\!\!\!+ \frac{\tilde{E}_{1}\tilde{E}_{2}\tilde{E}^{*}_{3}}{4}\left[\sin\beta_{3}\cos\left(\gamma_{1}+\gamma_{2}\right)-\sin\gamma_{1}\sin\left(\gamma_{2}+\beta_{3}\right)\right]e^{i\left(\textbf{k}_{4}\cdot\textbf{r}-\omega_{4}t\right)}\hat{\textbf{y}}\nonumber \\ &&\!\!\!\!+ \frac{\tilde{E}_{1}\tilde{E}_{2}\tilde{E}^{*}_{3}}{4}\left[\cos\gamma_{2}\cos\left(\gamma_{1}+\beta_{3}\right)+\cos\gamma_{1}\sin\left(\gamma_{2}+\beta_{3}\right)\right]e^{i\left(\textbf{k}_{4}\cdot\textbf{r}-\omega_{4}t\right)}\hat{\textbf{z}}\ \ \ \ \ \ \ \ 
\end{eqnarray}\\
%-------------------------(E^2-B^2)B-------------------------------
\begin{eqnarray}
\left(E^{2}\!-B^{2}\right)\textbf{B}\!\!\!\!&=&\!\!\!\!\frac{\tilde{E}_{1}\tilde{E}_{2}\tilde{E}^{*}_{3}}{4}\left[-\sin\beta_{3}\cos\left(\gamma_{1}+\gamma_{2}\right)+ \cos\gamma_{2}\cos\left(\gamma_{1}+\beta_{3}\right)\right]e^{i\left(\textbf{k}_{4}\cdot\textbf{r}-\omega_{4}t\right)}\hat{\textbf{x}}\nonumber \\ &&\!\!\!\!+ \frac{\tilde{E}_{1}\tilde{E}_{2}\tilde{E}^{*}_{3}}{4}\left[\cos\beta_{3}\cos\left(\gamma_{1}+\gamma_{2}\right)-\cos\gamma_{1}\sin\left(\gamma_{2}+\beta_{3}\right)\right]e^{i\left(\textbf{k}_{4}\cdot\textbf{r}-\omega_{4}t\right)}\hat{\textbf{y}}\nonumber \\ &&\!\!\!\!+ \frac{\tilde{E}_{1}\tilde{E}_{2}\tilde{E}^{*}_{3}}{4}\left[-\sin\gamma_{2}\cos\left(\gamma_{1}+\beta_{3}\right)-\sin\gamma_{1}\sin\left(\gamma_{2}+\beta_{3}\right)\right]e^{i\left(\textbf{k}_{4}\cdot\textbf{r}-\omega_{4}t\right)}\hat{\textbf{z}}\ \ \ \ \ \ \ \ 
\end{eqnarray}\\
%------------------(E.B)E----------------------------------------
\begin{eqnarray}
\left(\textbf{E}\cdot\textbf{B}\right)\textbf{E}\!\!\!\!&=&\!\!\!\!\frac{\tilde{E}_{1}\tilde{E}_{2}\tilde{E}^{*}_{3}}{8}\left[-\cos\beta_{3}\sin\left(\gamma_{1}+\gamma_{2}\right)-\sin\gamma_{2}\sin\left(\gamma_{1}+\beta_{3}\right)\right]e^{i\left(\textbf{k}_{4}\cdot\textbf{r}-\omega_{4}t\right)}\hat{\textbf{x}}\nonumber \\ &&\!\!\!\!+ \frac{\tilde{E}_{1}\tilde{E}_{2}\tilde{E}^{*}_{3}}{8}\left[-\sin\beta_{3}\sin\left(\gamma_{1}+\gamma_{2}\right)-\sin\gamma_{1}\cos\left(\gamma_{2}+\beta_{3}\right)\right]e^{i\left(\textbf{k}_{4}\cdot\textbf{r}-\omega_{4}t\right)}\hat{\textbf{y}}\nonumber \\ &&\!\!\!\!- \frac{\tilde{E}_{1}\tilde{E}_{2}\tilde{E}^{*}_{3}}{8}\left[-\cos\gamma_{2}\sin\left(\gamma_{1}+\beta_{3}\right)+\cos\gamma_{1}\cos\left(\gamma_{2}+\beta_{3}\right)\right]e^{i\left(\textbf{k}_{4}\cdot\textbf{r}-\omega_{4}t\right)}\hat{\textbf{z}}\ \ \ \ \ \ \ \ 
\end{eqnarray}\\
%--------------(E.B)B--------------------------------------
\begin{eqnarray}
\left(\textbf{E}\cdot\textbf{B}\right)\textbf{B}\!\!\!\!&=&\!\!\!\!\frac{\tilde{E}_{1}\tilde{E}_{2}\tilde{E}^{*}_{3}}{8}\left[\sin\beta_{3}\sin\left(\gamma_{1}+\gamma_{2}\right)-\cos\gamma_{2}\sin\left(\gamma_{1}+\beta_{3}\right)\right]e^{i\left(\textbf{k}_{4}\cdot\textbf{r}-\omega_{4}t\right)}\hat{\textbf{x}}\nonumber \\ &&\!\!\!\!+ \frac{\tilde{E}_{1}\tilde{E}_{2}\tilde{E}^{*}_{3}}{8}\left[-\cos\beta_{3}\sin\left(\gamma_{1}+\gamma_{2}\right)-\cos\gamma_{1}\cos\left(\gamma_{2}+\beta_{3}\right)\right]e^{i\left(\textbf{k}_{4}\cdot\textbf{r}-\omega_{4}t\right)}\hat{\textbf{y}}\nonumber \\ &&\!\!\!\!+ \frac{\tilde{E}_{1}\tilde{E}_{2}\tilde{E}^{*}_{3}}{8}\left[\sin\gamma_{2}\sin\left(\gamma_{1}+\beta_{3}\right)-\sin\gamma_{1}\cos\left(\gamma_{2}+\beta_{3}\right)\right]e^{i\left(\textbf{k}_{4}\cdot\textbf{r}-\omega_{4}t\right)}\hat{\textbf{z}}\ \ \ \ \ \ \ \ 
\end{eqnarray}\\
%------------------nabla . (E^2-B^2)E---------------------------
\begin{eqnarray}
\nabla\cdot\left[\left(E^{2}\!-B^{2}\right)\textbf{E}\right]\!\!\!\!&=&\!\!\!\!\frac{\tilde{E}_{1}\tilde{E}_{2}\tilde{E}^{*}_{3}}{4}ik\left[\left(\cos\beta_{3}+\sin\beta_{3}\right)\cos\left(\gamma_{1}+\gamma_{2}\right)\right.\nonumber \\ &&\ \ \ \ \ \ \ \ \ \ \ \ + \left(\sin\gamma_{2}-\frac{1}{2}\cos\gamma_{2}\right)\cos\left(\gamma_{1}+\beta_{3}\right)\nonumber \\ &&\ \ \ \ \ \ \ \ \ \ \ \ + \left.\left(-\sin\gamma_{1}-\frac{1}{2}\cos\gamma_{1}\right)\sin\left(\gamma_{2}+\beta_{3}\right)\right]e^{i\left(\textbf{k}_{4}\cdot\textbf{r}-\omega_{4}t\right)}\ \ \ \ \ \ \ \ 
\end{eqnarray}\\
%-----------------------d/dt (E^2-B^2)E------------------------------
\begin{eqnarray}
	\frac{\partial}{\partial t}\left[\left(E^{2}\!-B^{2}\right)\textbf{E}\right]=-\frac{3}{2}i\omega\left[\left(E^{2}\!-B^{2}\right)\textbf{E}\right]
\end{eqnarray}\\
%------------------------nabla cross (E^2-B^2)B---------------------------
\begin{eqnarray}
\nabla\times\left[\left(E^{2}\!-B^{2}\right)\textbf{B}\right]\!\!\!\!&=&\!\!\!\!\frac{\tilde{E}_{1}\tilde{E}_{2}\tilde{E}^{*}_{3}}{4}ik\left[-\sin\gamma_{2}\cos\left(\gamma_{1}+\beta_{3}\right)-\sin\gamma_{1}\sin\left(\gamma_{2}+\beta_{3}\right)\right.\nonumber \\ &&\ \ \ \ \ \ \ \ \ \ \ \ + \left.\frac{1}{2}\cos\beta_{3}\cos\left(\gamma_{1}+\gamma_{2}\right)-\frac{1}{2}\cos\gamma_{1}\sin\left(\gamma_{2}+\beta_{3}\right)\right]e^{i\left(\textbf{k}_{4}\cdot\textbf{r}-\omega_{4}t\right)}\hat{\textbf{x}}\nonumber \\ &&\!\!\!\!+ \frac{\tilde{E}_{1}\tilde{E}_{2}\tilde{E}^{*}_{3}}{4}ik\left[\frac{1}{2}\sin\beta_{3}\cos\left(\gamma_{1}+\gamma_{2}\right)-\frac{1}{2}\cos\gamma_{2}\cos\left(\gamma_{1}+\beta_{3}\right)\right.\nonumber \\ &&\ \ \ \ \ \ \ \ \ \ \ \ \ \ \ + \left.\sin\gamma_{2}\cos\left(\gamma_{1}+\beta_{3}\right)+\sin\gamma_{1}\sin\left(\gamma_{2}+\beta_{3}\right)\right]e^{i\left(\textbf{k}_{4}\cdot\textbf{r}-\omega_{4}t\right)}\hat{\textbf{y}} \nonumber \\ &&\!\!\!\!+ \frac{\tilde{E}_{1}\tilde{E}_{2}\tilde{E}^{*}_{3}}{4}ik\left[\sin\beta_{3}\cos\left(\gamma_{1}+\gamma_{2}\right)-\cos\gamma_{2}\cos\left(\gamma_{1}+\beta_{3}\right)\right.\nonumber \\ &&\ \ \ \ \ \ \ \ \ \ \ \ \ \ \ + \cos\beta_{3}\cos\left(\gamma_{1}+\gamma_{2}\right)\nonumber \\ &&\ \ \ \ \ \ \ \ \ \ \ \ \ \ \ \left.-\cos\gamma_{1}\sin\left(\gamma_{2}+\beta_{3}\right)\right]e^{i\left(\textbf{k}_{4}\cdot\textbf{r}-\omega_{4}t\right)}\hat{\textbf{z}}
\end{eqnarray}\\
%---------------------nabla cross (E.B)E---------------------------------------
\begin{eqnarray}
\nabla\times\left[\left(\textbf{E}\cdot\textbf{B}\right)\textbf{E}\right]\!\!\!\!&=&\!\!\!\!\frac{\tilde{E}_{1}\tilde{E}_{2}\tilde{E}^{*}_{3}}{8}ik\left[-\cos\gamma_{2}\sin\left(\gamma_{1}+\beta_{3}\right)+\cos\gamma_{1}\cos\left(\gamma_{2}+\beta_{3}\right)\right.\nonumber \\ &&\ \ \ \ \ \ \ \ \ \ \ \ \left.-\frac{1}{2}\sin\beta_{3}\sin\left(\gamma_{1}+\gamma_{2}\right)-\frac{1}{2}\sin\gamma_{1}\cos\left(\gamma_{2}+\beta_{3}\right)\right]e^{i\left(\textbf{k}_{4}\cdot\textbf{r}-\omega_{4}t\right)}\hat{\textbf{x}}\nonumber \\ &&\!\!\!\!+ \frac{\tilde{E}_{1}\tilde{E}_{2}\tilde{E}^{*}_{3}}{8}ik\left[\frac{1}{2}\cos\beta_{3}\sin\left(\gamma_{1}+\gamma_{2}\right)+\frac{1}{2}\sin\gamma_{2}\sin\left(\gamma_{1}+\beta_{3}\right)\right.\nonumber \\ &&\ \ \ \ \ \ \ \ \ \ \ \ \ \ \ +\left.\cos\gamma_{2}\sin\left(\gamma_{1}+\beta_{3}\right)-\cos\gamma_{1}\cos\left(\gamma_{2}+\beta_{3}\right)\right]e^{i\left(\textbf{k}_{4}\cdot\textbf{r}-\omega_{4}t\right)}\hat{\textbf{y}}\nonumber \\ &&\!\!\!\!+ \frac{\tilde{E}_{1}\tilde{E}_{2}\tilde{E}^{*}_{3}}{8}ik\left[\cos\beta_{3}\sin\left(\gamma_{1}+\gamma_{2}\right)+\sin\gamma_{2}\sin\left(\gamma_{1}+\beta_{3}\right)\right.\nonumber \\ &&\ \ \ \ \ \ \ \ \ \ \ \ \ \ \ - \sin\beta_{3}\sin\left(\gamma_{1}+\gamma_{2}\right)\nonumber \\ &&\ \ \ \ \ \ \ \ \ \ \ \ \ \ \ \left.-\sin\gamma_{1}\cos\left(\gamma_{2}+\beta_{3}\right)\right]e^{i\left(\textbf{k}_{4}\cdot\textbf{r}-\omega_{4}t\right)}\hat{\textbf{z}}
\end{eqnarray}\\
%---------------------nabla . (E.B)B------------------------------
\begin{eqnarray}
\nabla\cdot\left[\left(\textbf{E}\cdot\textbf{B}\right)\textbf{B}\right]\!\!\!\!&=&\!\!\!\!\frac{\tilde{E}_{1}\tilde{E}_{2}\tilde{E}^{*}_{3}}{8}ik\left[\left(\sin\beta_{3}-\cos\beta_{3}\right)\sin\left(\gamma_{1}+\gamma_{2}\right)\right.\nonumber \\ &&\ \ \ \ \ \ \ \ \ \ \ \ + \left(-\cos\gamma_{2}-\frac{1}{2}\sin\gamma_{2}\right)\sin\left(\gamma_{1}+\beta_{3}\right)\nonumber \\ &&\ \ \ \ \ \ \ \ \ \ \ \ + \left.\left(-\cos\gamma_{1}+\frac{1}{2}\sin\gamma_{1}\right)\cos\left(\gamma_{2}+\beta_{3}\right)\right]e^{i\left(\textbf{k}_{4}\cdot\textbf{r}-\omega_{4}t\right)}\ \ \ \ \ \ \ \ 
\end{eqnarray}\\
%--------------------d/dt (E.B)B-----------------------------
\begin{eqnarray}
	\frac{\partial}{\partial t}\left[\left(\textbf{E}\cdot\textbf{B}\right)\textbf{B}\right]=-\frac{3}{2}i\omega\left[\left(\textbf{E}\cdot\textbf{B}\right)\textbf{B}\right]
\end{eqnarray}\\
\newpage
%-----------------------(N1)----------------------------------
\begin{eqnarray}
	(N1)\!\!\!\!&=&\!\!\!\!-\xi\frac{\tilde{E}_{1}\tilde{E}_{2}\tilde{E}^{*}_{3}}{2}ik\ \left[\left(\cos\beta_{3}+\sin\beta_{3}\right)\cos\left(\gamma_{1}+\gamma_{2}\right)\right.\nonumber \\ &&\ \ \ \ \ \ \ \ \ \ \ \ \ \ \ \ \ +\left(\sin\gamma_{2}-\frac{1}{2}\cos\gamma_{2}\right)\cos\left(\gamma_{1}+\beta_{3}\right)\nonumber \\ &&\ \ \ \ \ \ \ \ \ \ \ \ \ \ \ \ \ +\left(-\sin\gamma_{1}-\frac{1}{2}\cos\gamma_{1}\right)\sin\left(\gamma_{2}+\beta_{3}\right)\nonumber \\ &&\ \ \ \ \ \ \ \ \ \ \ \ \ \ \ \ \ + \frac{7}{4}\left(\sin\beta_{3}-\cos\beta_{3}\right)\sin\left(\gamma_{1}+\gamma_{2}\right)\nonumber \\ &&\ \ \ \ \ \ \ \ \ \ \ \ \ \ \ \ \ +\frac{7}{4}\left(-\cos\gamma_{2}-\frac{1}{2}\sin\gamma_{2}\right)\sin\left(\gamma_{1}+\beta_{3}\right)\nonumber \\ &&\ \ \ \ \ \ \ \ \ \ \ \ \ \ \ \ \ \left.+\frac{7}{4}\left(-\cos\gamma_{1}+\frac{1}{2}\sin\gamma_{1}\right)\cos\left(\gamma_{2}+\beta_{3}\right)\right]e^{i\left(\textbf{k}_{4}\cdot\textbf{r}-\omega_{4}t\right)}
\end{eqnarray}\\
%------------------------xhat . (N234)-----------------------------------
\begin{eqnarray}
	\hat{\textbf{x}}\cdot\overline{(N234)}\!\!\!\!&=&\!\!\!\!\xi \frac{\tilde{E}_{1}\tilde{E}_{2}\tilde{E}^{*}_{3}}{2}ik\left\{\ \left[-\frac{1}{2}\sin\gamma_{2}\cos\left(\gamma_{1}+\beta_{3}\right)+\sin\gamma_{1}\sin\left(\gamma_{2}+\beta_{3}\right)\right.\right.\nonumber \\ &&\ \ \ \ \ \ \ \ \ \ \ \ \ \ \ \ \ \left.-2\cos\beta_{3}\cos\left(\gamma_{1}+\gamma_{2}\right)+\frac{1}{2}\cos\gamma_{1}\sin\left(\gamma_{2}+\beta_{3}\right)\right] \nonumber \\ &&\ \ \ \ \ \ \ \ \ \ \ \ \ \ +\frac{7}{4}\left[\frac{1}{2}\cos\gamma_{2}\sin\left(\gamma_{1}+\beta_{3}\right)+\cos\gamma_{1}\cos\left(\gamma_{2}+\beta_{3}\right)\right.\nonumber \\ &&\ \ \ \ \ \ \ \ \ \ \ \ \ \ \ \ \ \ \ -2\sin\beta_{3}\sin\left(\gamma_{1}+\gamma_{2}\right)\nonumber \\ &&\ \ \ \ \ \ \ \ \ \ \ \ \ \ \ \ \ \ \!\ \left.\left.-\frac{1}{2}\sin\gamma_{1}\cos\left(\gamma_{2}+\beta_{3}\right)\right]\ \right\}e^{i\left(\textbf{k}_{4}\cdot\textbf{r}-\omega_{4}t\right)}
\end{eqnarray}\\
%-----------------------yhat . (N234)-----------------------------------
\begin{eqnarray}
	\hat{\textbf{y}}\cdot\overline{(N234)}\!\!\!\!&=&\!\!\!\!\xi \frac{\tilde{E}_{1}\tilde{E}_{2}\tilde{E}^{*}_{3}}{2}ik\left\{\ \left[-2\sin\beta_{3}\cos\left(\gamma_{1}+\gamma_{2}\right)+\frac{1}{2}\cos\gamma_{2}\cos\left(\gamma_{1}+\beta_{3}\right)\right.\right.\nonumber \\ &&\ \ \ \ \ \ \ \ \ \ \ \ \ \ \ \ \ \left.-\sin\gamma_{2}\cos\left(\gamma_{1}+\beta_{3}\right)+\frac{1}{2}\sin\gamma_{1}\sin\left(\gamma_{2}+\beta_{3}\right)\right] \nonumber \\ &&\ \ \ \ \ \ \ \ \ \ \ \ \ \ +\frac{7}{4}\left[2\cos\beta_{3}\sin\left(\gamma_{1}+\gamma_{2}\right)+\frac{1}{2}\sin\gamma_{2}\sin\left(\gamma_{1}+\beta_{3}\right)\right.\nonumber \\ &&\ \ \ \ \ \ \ \ \ \ \ \ \ \ \ \ \ \ \ +\cos\gamma_{2}\sin\left(\gamma_{1}+\beta_{3}\right)\nonumber \\ &&\ \ \ \ \ \ \ \ \ \ \ \ \ \ \ \ \ \ \!\ \left.\left.+\frac{1}{2}\cos\gamma_{1}\cos\left(\gamma_{2}+\beta_{3}\right)\right]\ \right\}e^{i\left(\textbf{k}_{4}\cdot\textbf{r}-\omega_{4}t\right)}
\end{eqnarray}\\
\newpage
%-----------------------zhat . (N234)-----------------------------------
\begin{eqnarray}
	\hat{\textbf{z}}\cdot\overline{(N234)}\!\!\!\!&=&\!\!\!\!\xi \frac{\tilde{E}_{1}\tilde{E}_{2}\tilde{E}^{*}_{3}}{2}ik\left\{\ \left[-\sin\beta_{3}\cos\left(\gamma_{1}+\gamma_{2}\right)-\cos\beta_{3}\cos\left(\gamma_{1}+\gamma_{2}\right)\right.\right.\nonumber \\ &&\ \ \ \ \ \ \ \ \ \ \ \ \ \ \ \ \ \left.-\frac{1}{2}\cos\gamma_{2}\cos\left(\gamma_{1}+\beta_{3}\right)-\frac{1}{2}\cos\gamma_{1}\sin\left(\gamma_{2}+\beta_{3}\right)\right] \nonumber \\ &&\ \ \ \ \ \ \ \ \ \ \ \ \ \ +\frac{7}{4}\left[\cos\beta_{3}\sin\left(\gamma_{1}+\gamma_{2}\right)-\sin\beta_{3}\sin\left(\gamma_{1}+\gamma_{2}\right)\right.\nonumber \\ &&\ \ \ \ \ \ \ \ \ \ \ \ \ \ \ \ \ \ \ -\frac{1}{2}\sin\gamma_{2}\sin\left(\gamma_{1}+\beta_{3}\right)\nonumber \\ &&\ \ \ \ \ \ \ \ \ \ \ \ \ \ \ \ \ \ \!\ \left.\left.+\frac{1}{2}\sin\gamma_{1}\cos\left(\gamma_{2}+\beta_{3}\right)\right]\ \right\}e^{i\left(\textbf{k}_{4}\cdot\textbf{r}-\omega_{4}t\right)}
\end{eqnarray}\\
%-------------nabla(N1)---------------------------------------
\begin{eqnarray}
	\nabla(N1)=ik(N1)\left(\hat{\textbf{x}}+\hat{\textbf{y}}-\frac{1}{2}\hat{\textbf{z}}\right)
\end{eqnarray}\\
%-------------d/dt (N234)---------------------------------------
\begin{eqnarray}
	\frac{\partial}{\partial t}\overline{(N234)}=-\frac{3}{2}i\omega\overline{(N234)}
\end{eqnarray}\\
By plugging these expressions into Eq.$\ $(3.38), we finally arrive at Eq.$\ $(5.19).
%-----------------------------------------------------------------------------------------------
\chapter{Geometric factors in hep-ph/0510076}
Below follow the geometric factors refered to in the paper \emph{Using high-power lasers for detection of elastic photon-photon scattering} (hep-ph/0510076).
Due to somewhat different conventions, $\textbf{G}_{3d}$ here differs by a constant factor from the expression presented above in the MSc thesis.\\

For wave vectors confined to the x-y-plane, $\hat{\textbf{k}}_{j}\!\!=\!\cos\phi_{j}\hat{\textbf{x}}+\sin\phi_{j}\hat{\textbf{y}}$, and with $\gamma_{j}$ defined as the angle between the z-axis and the polarization vector of the $E_{j}$-field, in such a way it is positive in clockwise direction when looking in the $\hat{\textbf{k}}_{j}$-direction, the two dimensional geometric factor becomes
\begin{eqnarray}
	\textbf{G}_{2d}\!\!\!\!&=&\!\!\!\!+\frac{1}{2}\left\{\ \left[\frac{7}{4}\cos\gamma_{3}\sin\left(\gamma_{1}+\gamma_{2}\right)-\sin\gamma_{3}\cos\left(\gamma_{1}+\gamma_{2}\right)\right]\right.\nonumber \\ &&\ \ \ \ \ \ \times\left[\sin\left(\phi_{3}-\phi_{4}\right)\cos\phi_{4}-\left(\sin\phi_{3}-\sin\phi_{4}\right)\right]\sin^{2}\left(\frac{\phi_{1}-\phi_{2}}{2}\right)\nonumber \\ &&\ \ \ + \left[\frac{7}{4}\cos\gamma_{2}\sin\left(\gamma_{1}+\gamma_{3}\right)-\sin\gamma_{2}\cos\left(\gamma_{1}+\gamma_{3}\right)\right]\nonumber \\ &&\ \ \ \ \ \ \times\left[\sin\left(\phi_{2}-\phi_{4}\right)\cos\phi_{4}-\left(\sin\phi_{2}-\sin\phi_{4}\right)\right]\sin^{2}\left(\frac{\phi_{1}-\phi_{3}}{2}\right)\nonumber \\ &&\ \ \ + \left[\frac{7}{4}\cos\gamma_{1}\sin\left(\gamma_{2}+\gamma_{3}\right)-\sin\gamma_{1}\cos\left(\gamma_{2}+\gamma_{3}\right)\right]\nonumber \\ &&\ \ \ \ \ \ \times\left.\left[\sin\left(\phi_{1}-\phi_{4}\right)\cos\phi_{4}-\left(\sin\phi_{1}-\sin\phi_{4}\right)\right]\sin^{2}\left(\frac{\phi_{2}-\phi_{3}}{2}\right)\ \right\}\hat{\textbf{x}}\nonumber \\ &&\!\!\!\!+ \frac{1}{2}\left\{\ \left[\frac{7}{4}\cos\gamma_{3}\sin\left(\gamma_{1}+\gamma_{2}\right)-\sin\gamma_{3}\cos\left(\gamma_{1}+\gamma_{2}\right)\right]\right.\nonumber \\ &&\ \ \ \ \ \ \times\left[\sin\left(\phi_{3}-\phi_{4}\right)\sin\phi_{4}+\left(\cos\phi_{3}-\cos\phi_{4}\right)\right]\sin^{2}\left(\frac{\phi_{1}-\phi_{2}}{2}\right)\nonumber \\ &&\ \ \ + \left[\frac{7}{4}\cos\gamma_{2}\sin\left(\gamma_{1}+\gamma_{3}\right)-\sin\gamma_{2}\cos\left(\gamma_{1}+\gamma_{3}\right)\right]\nonumber \\ &&\ \ \ \ \ \ \times\left[\sin\left(\phi_{2}-\phi_{4}\right)\sin\phi_{4}+\left(\cos\phi_{2}-\cos\phi_{4}\right)\right]\sin^{2}\left(\frac{\phi_{1}-\phi_{3}}{2}\right)\nonumber \\ &&\ \ \ + \left[\frac{7}{4}\cos\gamma_{1}\sin\left(\gamma_{2}+\gamma_{3}\right)-\sin\gamma_{1}\cos\left(\gamma_{2}+\gamma_{3}\right)\right]\nonumber \\ &&\ \ \ \ \ \ \times\left.\left[\sin\left(\phi_{1}-\phi_{4}\right)\sin\phi_{4}+\left(\cos\phi_{1}-\cos\phi_{4}\right)\right]\sin^{2}\left(\frac{\phi_{2}-\phi_{3}}{2}\right)\ \right\}\hat{\textbf{y}}\nonumber \\ &&\!\!\!\!+ \left\{\ \left[\frac{7}{4}\sin\gamma_{3}\sin\left(\gamma_{1}+\gamma_{2}\right)+\cos\gamma_{3}\cos\left(\gamma_{1}+\gamma_{2}\right)\right]\right.\nonumber \\ &&\ \ \ \ \times\ \sin^{2}\left(\frac{\phi_{3}-\phi_{4}}{2}\right)\sin^{2}\left(\frac{\phi_{1}-\phi_{2}}{2}\right)\nonumber \\ &&\ + \left[\frac{7}{4}\sin\gamma_{2}\sin\left(\gamma_{1}+\gamma_{3}\right)+\cos\gamma_{2}\cos\left(\gamma_{1}+\gamma_{3}\right)\right]\nonumber \\ &&\ \ \ \ \times\ \sin^{2}\left(\frac{\phi_{2}-\phi_{4}}{2}\right)\sin^{2}\left(\frac{\phi_{1}-\phi_{3}}{2}\right)\nonumber \\ &&\ + \left[\frac{7}{4}\sin\gamma_{1}\sin\left(\gamma_{2}+\gamma_{3}\right)+\cos\gamma_{1}\cos\left(\gamma_{2}+\gamma_{3}\right)\right]\nonumber \\ &&\ \ \ \ \times\left.\ \sin^{2}\left(\frac{\phi_{1}-\phi_{4}}{2}\right)\sin^{2}\left(\frac{\phi_{2}-\phi_{3}}{2}\right)\ \right\}\hat{\textbf{z}}.
\end{eqnarray}

With wave vectors $\textbf{k}_{1}\!=k\hat{\textbf{x}}$, $\textbf{k}_{2}\!=k\hat{\textbf{y}}$, $\textbf{k}_{3}\!=\frac{k}{2}\hat{\textbf{z}}$, $\textbf{k}_{4}\!=k\hat{\textbf{x}}+k\hat{\textbf{y}}-\!\frac{1}{2}k\hat{\textbf{z}}$, and $\beta_{3}$ defined as the angle between the x-axis and the polarization vector of the $E_{3}$-field, in such way that it is positive in clockwise direction when looking in the $\hat{\textbf{k}}_{3}$-direction, we get the three dimensional geometric factor
\begin{eqnarray}
	\textbf{G}_{3d}\!\!\!\!&=&\!\!\!\!-\frac{2}{9}\left\{\ \left[\left(\frac{1}{2}\sin\beta_{3}-\cos\beta_{3}\right)\cos\left(\gamma_{1}+\gamma_{2}\right)\right.\right.\nonumber \\&&\ \ \ \ \ \ +\left(\frac{1}{8}\sin\gamma_{2}-\frac{1}{4}\cos\gamma_{2}\right)\cos\left(\gamma_{1}+\beta_{3}\right)\nonumber \\&&\ \ \ \ \ \ +\!\left.\left(\frac{1}{4}\sin\gamma_{1}+\frac{1}{8}\cos\gamma_{1}\right)\sin\left(\gamma_{2}+\beta_{3}\right)\right]\nonumber \\&&\ \ \ +\frac{7}{4}\left[\left(-\sin\beta_{3}-\frac{1}{2}\cos\beta_{3}\right)\sin\left(\gamma_{1}+\gamma_{2}\right)\right.\nonumber \\&&\ \ \ \ \ \ \ \  +\left(-\frac{1}{8}\cos\gamma_{2}-\frac{1}{4}\sin\gamma_{2}\right)\sin\left(\gamma_{1}+\beta_{3}\right)\nonumber \\&&\ \ \ \ \ \ \ \  +\!\!\left.\left.\left(\frac{1}{4}\cos\gamma_{1}-\frac{1}{8}\sin\gamma_{1}\right)\cos\left(\gamma_{2}+\beta_{3}\right)\right]\ \right\}\hat{\textbf{x}}\nonumber \\&&\!\!\!\! -\frac{2}{9}\left\{\ \left[\left(\frac{1}{2}\cos\beta_{3}-\sin\beta_{3}\right)\cos\left(\gamma_{1}+\gamma_{2}\right)\right.\right.\nonumber \\&&\ \ \ \ \ \ +\left(\frac{1}{8}\cos\gamma_{2}-\frac{1}{4}\sin\gamma_{2}\right)\cos\left(\gamma_{1}+\beta_{3}\right)\nonumber \\&&\ \ \ \ \ \  +\!\left.\left(-\frac{1}{4}\cos\gamma_{1}-\frac{1}{8}\sin\gamma_{1}\right)\sin\left(\gamma_{2}+\beta_{3}\right)\right]\nonumber \\&&\ \ \  +\frac{7}{4}\left[\left(\cos\beta_{3}+\frac{1}{2}\sin\beta_{3}\right)\sin\left(\gamma_{1}+\gamma_{2}\right)\right.\nonumber \\&&\ \ \ \ \ \ \ \  +\left(\frac{1}{8}\sin\gamma_{2}+\frac{1}{4}\cos\gamma_{2}\right)\sin\left(\gamma_{1}+\beta_{3}\right)\nonumber \\&&\ \ \ \ \ \ \ \  +\!\!\left.\left.\left(\frac{1}{4}\sin\gamma_{1}-\frac{1}{8}\cos\gamma_{1}\right)\cos\left(\gamma_{2}+\beta_{3}\right)\right]\ \right\}\hat{\textbf{y}}\nonumber \\&&\!\!\!\! +\frac{4}{9}\left\{\ \left[\frac{1}{2}\left(\cos\beta_{3}+\sin\beta_{3}\right)\cos\left(\gamma_{1}+\gamma_{2}\right)\right.\right.\nonumber \\&&\ \ \ \ \ \ +\frac{1}{8}\left(\sin\gamma_{2}+\cos\gamma_{2}\right)\cos\left(\gamma_{1}+\beta_{3}\right)\nonumber \\ &&\ \ \ \ \ \ \!\left.+\frac{1}{8}\left(\cos\gamma_{1}-\sin\gamma_{1}\right)\sin\left(\gamma_{2}+\beta_{3}\right)\right]\nonumber \\&&\ \ \ +\frac{7}{4}\left[\frac{1}{2}\left(\sin\beta_{3}-\cos\beta_{3}\right)\sin\left(\gamma_{1}+\gamma_{2}\right)\right.\nonumber \\&&\ \ \ \ \ \ \ \ +\frac{1}{8}\left(\sin\gamma_{2}-\cos\gamma_{2}\right)\sin\left(\gamma_{1}+\beta_{3}\right)\nonumber \\&&\ \ \ \ \ \ \ \ +\!\!\left.\left.\frac{1}{8}\left(-\cos\gamma_{1}-\sin\gamma_{1}\right)\cos\left(\gamma_{2}+\beta_{3}\right)\right]\ \right\}\hat{\textbf{z}}.
\end{eqnarray}
For this configuration the number of scattered photons per shot is estimated to
\begin{eqnarray}
	N_{3d}=1.31\eta^{2}G_{3d}^{2}\left(\frac{1\,{\rm \mu m}}{\lambda_{4}}\right)^{3}\left(\frac{L}{1\,{\rm\mu m}}\right)\left(\frac{P_{1}P_{2}P_{3}}{1\,{\rm PW^3}}\right),
\end{eqnarray}
where $P_{j}$ is the power of the incoming pulses, $L$ the pulse length, $\lambda_{4}$ the generated wavelength, and with $\eta^{2}$ given by
\begin{eqnarray} 
\eta^{2}\!\!\!\!&\equiv&\!\!\!\!\int^{2\pi}_{0}\int^{\pi}_{0}\frac{I_{g}\!\left(r,\theta,\phi\right)}{I_{g,res}\!\left(r\right)}\sin\theta ~d\theta ~d\phi\nonumber \\ \!\!\!\!&=&\!\!\!\!\frac{64}{k^{6}_{4}b^{6}}\int^{2\pi}_{0}\int^{\pi}_{0}\frac{\sin^{2}\left[k_{4}\frac{b}{2}\left(1-\cos\phi\sin\theta\right)\right]}{\left(1-\cos\phi\sin\theta\right)^{2}}\frac{\sin^{2}\left[k_{4}\frac{b}{2}\left(\sin\phi\sin\theta\right)\right]}{\left(\sin\phi\sin\theta\right)^{2}}\nonumber \\ &&\!\!\!\!\times\frac{\sin^{2}\left[k_{4}\frac{b}{2}\left(\cos\theta\right)\right]}{\left(\cos\theta\right)^{2}}\sin\theta ~d\theta ~d\phi,
\end{eqnarray}
where $I_{g}$ is the generated intensity, $I_{g,res}$ its maximum value in the resonant direction, and $b$ is the pulse width. 

An optimal choice of polarization angles for the three dimensional configuration is given by $\gamma_{1}=0$, $\gamma_{2}=\beta_{3}=\pi/2$, yielding $G_{3d}^{2}\!=0.77$.
%-----------------------------------------------------------------------------------------------


\begin{thebibliography}{99}
\thispagestyle{empty}
	\bibitem{Byron} F.W.$\!$ Byron and R.W.$\!$ Fuller, \textit{Mathematics of classical and quantum physics} (Dover publications, New York, 1992).
	\bibitem{Casimir} H.B.G.$\!$ Casimir, Proc.$\!$ Kon.$\!$ Ned.$\!$ Akad.$\!$ Wetenschap. B \textbf{52} 793 (1948).
	\bibitem{Vulcan} Central Laser Facility-Vulcan Glass Laser,\\ \texttt{http://www.clf.rl.ac.uk/Facilities/vulcan/index.htm}, 2005-01-24.
	\bibitem{Dittrich} W.$\!$ Dittrich and H.$\!$ Gies, \textit{Probing the quantum vacuum} (Springer, Berlin, 2000).
	\bibitem{Eriksson} D.$\!$ Eriksson, G.$\!$ Brodin, M.$\!$ Marklund and L.$\!$ Stenflo, \textit{Possibility to measure elastic photon-photon scattering in vacuum}, Phys.$\!$ Rev. A \textbf{70}, 013808 (2004).
	\bibitem{Greiner} W.$\!$ Greiner, B.$\!$ M$\mathrm{\ddot{u}}$ller and J.$\!$ Rafelski, \textit{Quantum electrodynamics of strong fields} (Springer, Berlin, 1985).
	\bibitem{Heisenberg} W.$\!$ Heisenberg and H.$\!$ Euler, Z.$\!$ Phys. \textbf{98} 714 (1936).
	\bibitem{Jackson} J.D.$\!$ Jackson, \textit{Classical electrodynamics} (Wiley, New York, 1999).
	\bibitem{Joshi} C.J.$\!$ Joshi and P.B.$\!$ Corkum, \textit{Interactions of ultra-intense laser light with matter}, Phys.$\!$ Today \textbf{48} 36 (1995).
	\bibitem{Mandl} F.$\!$ Mandl and G.$\!$ Shaw, \textit{Quantum field theory} (Wiley, Chichester, 1993).
	\bibitem{Mourou} G.A.$\!$ Mourou, C.P.J.$\!$ Barty and M.D.$\!$ Perry, \textit{Ultrahigh-intensity lasers: Physics at the extreme on a tabletop}, Phys.$\!$ Today \textbf{51} 22 (1998).
	\bibitem{Pukhov} A.$\!$ Pukhov, \textit{Strong field interaction of laser radiation}, Rep.$\!$ Prog.$\!$ Phys. \textbf{66} 47 (2003).
	\bibitem{XFEL} F.$\!$ Richard \textit{et al.}, \textit{Tesla XFEL technical design report} (DESY, Hamburg, 2002).
	\bibitem{Ringwald} A.$\!$ Ringwald, \textit{Pair production from vacuum at the focus of an X-ray free electron laser}, Phys.$\!$ Lett. B \textbf{510} 107 (2001).
	\bibitem{Schwinger}  J.$\!$ Schwinger, \textit{On gauge invariance and vacuum polarization}, Phys.$\!$ Rev. \textbf{82} 664 (1951).
	\bibitem{Shen} B.$\!$ Shen, M.Y.$\!$ Yu and X.$\!$ Wang, \textit{Photon-photon scattering in a plasma channel}, Phys.$\!$ Plasmas \textbf{10} 4570 (2003).
	\bibitem{Weiland} J.$\!$ Weiland and H.$\!$ Wilhelmsson, \textit{Coherent non-linear interaction of waves in plasmas} (Pergamon, Oxford, 1977).
	\bibitem{Zee} A.$\!$ Zee, \textit{Quantum field theory in a nutshell} (Princeton university press, Princeton, 2003).
\end{thebibliography}
\end{document}